\documentclass[a4paper,12pt,onehalfspacing,headrules,tools]{report}
\usepackage[utf8]{inputenc}
\usepackage[T1]{fontenc}
\usepackage[french]{babel}
\usepackage{color}
\usepackage{amssymb}
\usepackage{amsmath}
\usepackage{epsfig}
\usepackage{graphicx}
\usepackage[ruled,vlined,french,titlenumbered]{algorithm2e}
\usepackage{fancyhdr}
\usepackage{longtable}
\usepackage[francais]{minitoc} 
\usepackage{rotating} 

\usepackage{array}
\usepackage{multirow}
 \date{}
\setlongtables
\newtheorem{theorem}{Théorème}

\newtheorem{property}{Propriété}

\newtheorem{definition}{Définition}
\newtheorem{remarque}{Remarque}
\newtheorem{exemple}{Exemple}
\newtheorem{lemma}[theorem]{Lemme}

\newtheorem{proposition}{Proposition}

\fancypagestyle{plain}{\fancyhead{}
 }

\begin{document}

\begin{titlepage}
\center\footnotesize RÉPUBLIQUE TUNISIENNE\\
\vspace{0.1em}\footnotesize MINISTÈRE DE L'ENSEIGNEMENT SUPÉRIEUR ET DE LA RECHERCHE SCIENTIFIQUE\\
\vspace{0.1em}\footnotesize UNIVERSITÉ DE TUNIS EL MANAR\\
\vspace{0.1em}\footnotesize FACULTÉ DES SCIENCES DE TUNIS\vspace{0.1em}

\begin{figure}[!h]
\centering
\includegraphics[width=0.2\textwidth]{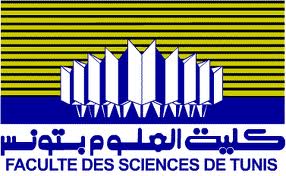}
\end{figure}

\vspace{2em}
{\bfseries\huge Mémoire de Mastère} \\
\vspace{0.25em}
{\slshape\normalsize Présenté en vue de l’obtention du diplôme de } \\
\vspace{0.25em}
{\textrm{\normalsize\textbf{Mastère de recherche en Informatique}}} \\ 
\vspace{1em}
{\slshape\normalsize Par } \\
\vspace{1em}
{\bfseries\large Seif Allah \textsc{Ben Chaabene}} \\
\vspace{3em}\hrule

{\textrm{\Huge\textbf{
Nouvelles représentations concises exactes des motifs rares
}}}

\hrule

\vspace{4em}
\textbf{Soutenu le 14/07/2010 devant le jury composé de :} \\ \vspace{1em}

\begin{table}[!ht]
\centering
\resizebox{0.9\linewidth}{!}{
\begin{tabular}{p{6cm}p{6.5cm}l}
\textbf{M. Zaher \textsc{Mahjoub}}&Professeur&Président\vspace{2px}\\
\textbf{Mme. Leila \textsc{Jemni}}& Maître-Assistant &Membre\vspace{2px}\\
\textbf{M. Sadok \textsc{Ben Yahia}}&Professeur &Directeur de mémoire\vspace{2px}
\end{tabular}
}
\end{table}

\vspace{1mm}
\textbf{Au sein du laboratoire : LIPAH}

\vspace{7.5em}

\center\large{2009-2010}
\end{titlepage}

\dominitoc 
\thispagestyle{empty}
\chapter*{Dédicace}
\bigskip \bigskip \bigskip
\bigskip \bigskip \bigskip
\bigskip \bigskip \bigskip
\bigskip \bigskip \bigskip
\begin{center}
Je dédie ce mémoire à mes parents pour leur grand amour, leur sacrifices et leur tendresse. Je le dédie également à mes frères, ma soeur, mes amis et mes proches.
Qu'ils trouvent ici le témoignage de ma gratitude et de mon respect.
\end{center}
\bigskip
\begin{flushright}\begin{small}\textit{Seif Allah Ben Chaabene}.\end{small}\end{flushright}
\newpage
\chapter*{Remerciements}
En premier lieu, je remercie Dieu le tout puissant pour ses faveurs et sa gratitude. Ensuite, je tiens à exprimer mes remerciements pour tous ceux qui m'ont aidé et encouragé à accomplir ce travail. En particulier, j'adresse mes remerciements à :\\

Mr. \textbf{Zaher Mahjoub}, Professeur à la Faculté des Sciences de Tunis, pour avoir accepté de présider le jury.\\

Mme. \textbf{Leila Jemni}, Maître de Conférences à la Faculté des Sciences de Tunis, pour avoir accepté de rapporter ce travail. Je la remercie pour l'attention avec laquelle elle a lu et évalué ce travail.\\

Mr. \textbf{Sadok Ben Yahia}, Maître de Conférences à la Faculté des Sciences de Tunis et directeur de ce mémoire, pour m'avoir accepté au sein de son équipe. J'ai spécialement apprécié ses qualités humaines indéniables, sa disponibilité et ses précieux conseils. Qu'il trouve ici l'expression de mon respect.\\

Mr. \textbf{Tarek Hamrouni}, Assistant à l'Institut Supérieur des Art Multimédias de Manouba, pour ses qualités humaines indéniables, son soutien, ses conseils, son aide tout au long de la préparation de ce mémoire et ses encouragements qui m'ont incité à perfectionner ce mémoire sans relâche.\\

Enfin, j'adresse un grand merci pour tous les membres du département des Sciences de l'Informatique de la Faculté des Sciences de Tunis. J'adresse aussi un remerciement particulier à tous mes collègues et tous ceux qui m'ont aidé dans ma vie.
\pagenumbering{roman}
\tableofcontents
\listoffigures
\listoftables
\newpage
\thispagestyle{empty}
\listofalgorithms
\thispagestyle{empty}
\bigskip
\bigskip \bigskip
\begin{center} \LARGE{\textbf{Notations}} \end{center}
\vfill
$\mathcal{MF} :$ Ensemble des motifs fréquents.\\
$\mathcal{MR} :$ Ensemble des motifs rares.\\
$\mathcal{MF}e :$ Ensemble des motifs fermés.\\
$\mathcal{MFF} :$ Ensemble des motifs fermés fréquents.\\
$\mathcal{MFR} :$ Ensemble des motifs fermés rares.\\
$\mathcal{GM} :$ Ensemble des motifs générateurs minimaux.\\
$\mathcal{GMR} :$ Ensemble des motifs générateur minimaux rares.\\
$\mathcal{GMF} :$ Ensemble des motifs générateur minimaux fréquents.\\
$\mathcal{MFM} :$ Ensemble des motifs fréquents maximaux.\\
$\mathcal{MRM} :$ Ensemble des motifs rares minimaux.\\
$\mathcal{MZ} :$ Ensemble des motifs zéros.\\
$\mathcal{GMZ} :$ Ensemble des générateurs minimaux des motifs zéros. \\
$\mathcal{MND} :$ Ensemble des motifs non-dérivables. \\
$\mathcal{MNDF} :$ Ensemble des motifs non-dérivables fréquents. \\
$\mathcal{MEF} :$ Ensemble des motifs essentiels fréquents.\\
\texttt{RCEMF} : Représentation concise exacte des motifs fréquents.\\
\texttt{RCEMR} : Représentation concise exacte des motifs rares.\\
\vfill
\pagenumbering{arabic}
\chapter*{Introduction générale}
\markboth{Introduction générale}{Introduction générale}
L'extraction de connaissances dans les bases de données (ECBD) $^{(}$\footnote{Le terme originel anglais ``Knowldege Discovery in Databases'' a été introduit par Piatetsky-Shapiro en 1989. Afin de simplifier l'écriture nous utilisons dans la suite le terme français ``ECBD''.}$^{)}$ est un domaine dont l'essor va de pair avec la croissance des collectes d'informations et l'augmentation des capacités de stockage de données. L'ECBD tire son origine dans la volonté d'appréhender de manière rigoureuse des phénomènes complexes et a pour objectif de découvrir des informations pertinentes à partir de données brutes \cite{PS87}. Cette discipline se situe au croisement des bases de données, des statistiques, de l'intelligence artificielle et de l'interface homme-machine. C'est un processus subtil composé de plusieurs phases \cite{PSU96} : la préparation des données, la fouille de données, l'interprétation et l'évaluation des connaissance extraites.
Ce travail contribue plus particulièrement à la fouille de données, une étape centrale dans le processus de découverte des connaissances. Il est bien connu que l'exploration de données caractérisées par un grand nombre d'attributs est un problème algorithmiquement complexe.
La fouille de donnée est un processus d'analyse, dont l'approche est différente de celle utilisée en statistiques. Cette dernière présuppose en général que l'on se fixe une hypothèse, que les données permettent ou non de confirmer. Au contraire, la fouille des données adopte une démarche sans à priori (approche pragmatique) et essaie ainsi de faire émerger, à partir des données brutes, des inférences que l'expérimentateur peut ne pas soupçonner, et dont il aura à valider la pertinence \cite{PS87}.




La fouille des données se propose de transformer en information, ou en connaissance, de grands volumes de données qui peuvent être stockées de manière diverse, dans des bases de données relationnelles, dans un (ou plusieurs) entrepôts de données. Les données qui peuvent aussi être récupérées de sources riches plus ou moins structurées comme l'Internet, ou encore en temps réel (appel à un centre d'appel, retrait d'argent dans un distributeur à billets, etc.). Lorsque la source n'est pas directement un entrepôt de données, il s'agit très souvent de construire une base de données ou un datamart dédié à l'analyse et aux analystes. Cela suppose d'avoir à sa disposition une palette d'outils de gestion de données. La fouille des données tente alors de réaliser un arbitrage entre validité scientifique, interprétabilité des résultats et facilité d'utilisation, dans un environnement professionnel où le temps d'étude joue un rôle majeur et où les analystes ne sont pas toujours des statisticiens.\\
L'objectif majeur de la fouille de données est d'identifier des relations cachées et intéressantes entre les motifs d'une base de données, plusieurs travaux depuis 1994 se sont focalisés à l'extraction des règles associatives pour des supports fréquents, autrement dit on ne conserve que les motifs, dont la fréquence d'apparition est supérieure à un seuil donné d'avance. Ensuite on génère, à partir de ces motifs,  des règles valides exactes ou approximatives \cite{Agra96} de la forme "Tous les étudiants qui suivent le cours Introduction à UNIX suivent en parallèle un cours Programmation en C" \cite{Han00}.
Depuis son introduction par Agrawal \emph{et al.} \cite{AIS93}, la plupart des études en fouille de données se sont intéressées à l'extraction des motifs fréquents et à la génération des règles d'associations à partir de ces motifs fréquents \cite{thesepasq}. Parfois les règles d'association qui sont générées à partir des motifs fréquents sont non intéressantes (ou non importantes), dans le sens où un comportement fréquent est en général un comportement normal dans un contexte de base de données \cite{BV07,KR06}. Toutefois ces dernières années, beaucoup de travaux se sont focalisés à l'exploitation des motifs rares, i.e., ceux qui ne sont pas fréquents. Ainsi, ils ont montré l'intérêt majeur de ces motifs dans le cas des bases réelles \cite{laszlo06}.
Dans pareille situation, les événements peu fréquents, communément appelés \textbf{\emph{rares}}, sont plus intéressants puisqu'ils apportent une information plus pertinente à l'utilisateur final. Par exemple, dans le cas d'une application de repérage des fraudes où ces dernières constituent des événements rares par rapport à la taille totale des données, l'extraction des événements fréquents d'une base de données des tentatives de connexions ne permet pas de détecter les connexions fraudes. Par contre, la fouille des motifs rares permet de trouver les événements rares (\emph{inattendus}) où plusieurs parmi eux sont à l'origine des attaques. Nous prenons un deuxième exemple qui simule une base de données médicale où
nous nous intéressons à l\textquoteright{}identification de la cause
des maladies cardio-vasculaires \emph{(MCV).} Une règle d\textquoteright{}association
fréquente telle que \textquotedblleft{}\emph{\{niveau élevé de cholestérol\} $\Rightarrow$ \{MCV\}}\textquotedblright{} peut valider l\textquoteright{}hypothèse
que les individus qui ont un fort taux de cholestérol ont un risque
élevé de \emph{MCV}. A l\textquoteright{}opposé, si nous avons un
nombre suffisant de végétariens dans notre base de données, alors
une règle d\textquoteright{}association rare \textquotedblleft{}\emph{\{végétarien\}
$\Rightarrow$ \{MCV\}}\textquotedblright{} peut valider l\textquoteright{}hypothèse
que les végétariens ont un risque faible de contracter une \emph{MCV}.
Dans ce cas, les motifs \emph{\{végétarien\}} et \emph{\{MCV\}} sont
tous les deux fréquents, mais le motif \emph{\{végétarien, MCV\}}\begin{large}                                                                 \end{large}
est rare.\\
Nous avons pris deux exemples issus de deux domaines différents où nous pouvons appliquer la fouille des motifs rares. Cependant, il faut mentionner que d'autres domaines sont aussi en liaison avec l'exploitation des motifs rares, dont nous citons : médecine, biologie, sécurité et audit des risques et traçage des comportements personnels.\\
Après une étude de la littérature, nous avons remarqué qu'aucune des méthodes n'a été menée pour extraire une représentation concise exacte des motifs rares, la plupart des travaux se sont focalisés à l'extraction d'un ensemble séparateur entre les motifs rares et ceux fréquents, d'autres travaux se sont focalisés à la génération des motifs rares à partir d'un ensemble séparateur. Ceci nous a poussé à proposer de nouvelles approches permettant d'extraire deux représentations concises exactes des motifs rares. La première représentation est basée sur les générateurs minimaux rares ainsi que la deuxième représentation est basée sur les motifs fermés rares.
\section*{Organisation du document}
Les résultats de nos travaux de recherche sont synthétisés dans ce mémoire qui est composé de quatre chapitres.\\
Le premier chapitre présente les fondements mathématiques qui seront utilisés dans ce mémoire.\\
Le deuxième chapitre détaille les approches de la littérature, qui permettent d'extraire les motifs rares ou bien un sous-ensemble de l'ensemble total des motifs rares.\\
Le troisième chapitre introduit deux nouvelles représentations concises exactes des motifs rares l'une basée sur les générateurs minimaux rares et l'autre sur les fermés rares. Nous présentons aussi, dans ce chapitre, une description détaillée des algorithmes ainsi qu'une étude théorique concernant l'étude de complexité théoriques et des preuves de complétude des deux algorithmes.\\
Le quatrième chapitre présente une étude expérimentale orientée selon deux axes. D'une part, nous évaluons la cardinalité de nos présentations concises exactes afin de les comparer avec celle des méthodes présentes dans la littérature. D'autre part, nous évaluons les performances temporelles de nos algorithmes et nous les comparons \textit{vs.} celles des algorithmes d'extraction des motifs rares.\\
Ce mémoire se termine par une conclusion, qui résume l'ensemble de nos travaux ainsi qu'il présente quelques perspectives futures de recherche.
\chapter{Notions mathématiques}
\minitoc
\section{Introduction}
L'extraction des motifs est une tâche qui a fait l'objet de nombreux travaux sur des données binaires (c'est à dire indiquant la présence ou l'absence d'un item dans une transaction donnée), ces données sont appelées souvent données de type ``\textit{caddie de supermarché}'' \cite{AIS93}. Cette extraction nécessite une base mathématique rigoureuse. Pour cette raison, nous nous intéressons dans ce chapitre aux principales notions mathématiques sur lesquelles le problème d'extraction des itemsets fréquents et rares se fonde. Par ailleurs, nous décrivons les notions de générateur minimal et de fermeture induite par l'opérateur de fermeture de Galois pour les deux cas ``rare'' et ``fréquent''.

\section{Analyse formelle de concepts}
\begin{definition} \cite{thesepasq} \textbf{Base de transactions ou contexte formel}\\
Un \textit{contexte formel} est un triplet
$\mathcal{K}=(\mathcal{O},\mathcal{I},\mathcal{R})$, décrivant
deux ensembles finis $\mathcal{O}$ et $\mathcal{I}$ et une
relation (d'incidence) binaire, $\mathcal{R}$, entre $\mathcal{O}$
et $\mathcal{I}$ tel que $\mathcal{R} \subseteq \mathcal{O} \times
\mathcal{I}$. L'ensemble $\mathcal{O}$ est habituellement appelé
ensemble d'\textit{objets} (ou \textit{transactions}) et
$\mathcal{I}$ est appelé ensemble d'\textit{items} (ou
\textit{attributs}). Chaque couple ($o$,$i$) $\in$ $\mathcal{R}$
désigne que l'objet $o$ $\in$ $\mathcal{O}$ possède l'item $i$
$\in$ $\mathcal{I}$ (noté $o~\mathcal{R}~i$).
\end{definition}
Une transaction $T$, avec un identificateur appelé
\textit{TID} (Tuple IDentifier), contient un ensemble, non vide,
d'items de $\mathcal{I}$. Un sous-ensemble $X$ de $\mathcal{I}$ ou
$k$ = $\vert $X$\vert $ est appelé un \textit{$k$-itemset} ou
simplement un \textit{itemset}, et $k$ représente \textit{la
longueur de $X$}. Le nombre de transactions de $\mathcal{K}$
contenant un itemset $X$, $\vert ${\{}$T$ $ \in $ $\mathcal{K}$
$\vert $ $X$ $ \subseteq $ $T$ {\}}$\vert $, est appelé
\textit{support absolu} de $X$. Le \textit{support relatif} de $X$
est le quotient de son support absolu par le nombre total de
transactions de $\mathcal{K}$, \textit{i.e.}, Supp($X$) =
$\frac{\vert {\{}T \in \mathcal{K} \vert X \subseteq T
{\}}\vert}{\vert \mathcal{O}\vert}$. Un itemset $X$ est dit
\textit{fréquent} si son support relatif est
 un seuil minimum \textit{minsupp} $^{(}$\footnote{\textit{minsupp}
définie le nombre minimal de transactions auxquelles doit
appartenir un itemset $\textit{X}$ pour être qualifié de \emph{fréquent}.}$^{)}$
spécifié par l'utilisateur.\\

\begin{exemple}
Soit le contexte d'extraction $\mathcal{K}$ illustré par le tableau \ref{DB1}. Ce contexte peut être considéré comme une base de transactions tel que $\mathcal{O}=\{1,2,3,4,5\}$ et $\mathcal{I}=\{A,B,C,D,E\}$.
\end{exemple}
\begin{table}[!htbp]
  \centering
 \begin{tabular}{|c||c|c|c|c|c|}

\hline
 & A & B & C & D & E\tabularnewline
\hline
\hline
1 & 1 & 0 & 1 & 1 & 0\tabularnewline
\hline
2 & 0 & 1 & 1 & 0 & 1\tabularnewline
\hline
3 & 1 & 1 & 1 & 0 & 1\tabularnewline
\hline
4 & 0 & 1 & 0 & 0 & 1\tabularnewline
\hline
5 & 1 & 1 & 1 & 0 & 1\tabularnewline
\hline
\end{tabular}
 \caption{Exemple d'un contexte formel $\mathcal{K}$.}\label{DB1}
\end{table}

Dans ce mémoire, nous nous intéressons aux itemsets comme classe de motifs. Un itemset est défini comme suit :
\begin{definition} \textbf{Un itemset ou un motif}\\
Un itemset ou un motif est un sous-ensemble de $\mathcal{I}$.
\end{definition}

Pour alléger l'écriture, un motif sera noté sous la forme d'une chaîne plutôt que sous forme ensembliste (\textit{i.e.} $AB$ au lieu de $\{A,B\}$).\\

Dans la définition qui suit, nous allons formaliser les différents supports relatifs à un motif quelconque.

\begin{definition} \cite{Cas05} \textbf{Supports d'un itemset}\\
Soit $\mathcal{K}=\mathcal{(O,I,R)}$ une base de transactions et \textit{I} un itemset. Nous distinguons principalement deux types de supports correspondant à un itemset \textit{I} :

\[
Supp(I)=\{o\in\mathcal{O} \mid \forall i \in I, (o,i) \in \mathcal{R}\}\]
\[
Supp(\vee I)=\{o\in\mathcal{O} \mid \exists i \in I, (o,i) \in \mathcal{R}\}\]
Ainsi,
\begin{itemize}
  \item $Supp(I)$ est le nombre de transactions qui contiennent tous les items de l'itemset I.
  \item $Supp(\vee I)$ est le nombre de transactions qui contiennent au moins un item de I.
\end{itemize}
\end{definition}
\begin{exemple}
Soit le contexte d'extraction donné par le tableau \ref{DB1}. Calculons les différents supports de l'itemset BC :\\
$Supp(BC)= 3; Supp(\vee BC)= 5$;\\
\end{exemple}

\begin{remarque}
Il est à noter que si nous omettons de préciser la nature du support, nous considérons le support conjonctif.
\end{remarque}
\begin{remarque}
Un ensemble d'items forme un motif dont la taille est le nombre d'items qui le composent. Le support d'un motif \textit{I} correspond au nombre d'objets contenant le motif. Un motif est dit \emph{fréquent} si son support est supérieur ou égal à un seuil de fréquence minimum donné (noté $\textit{minsupp}$).
Celà étant, un motif est dit rare ou non fréquent si son support est inférieur ou égal à un support maximum, noté \textit{maxsupp}. Dans ce qui suit, la valeur de $\textit{maxsupp}$ se calcule à partir de celle de \textit{minsupp}, à savoir $\textit{maxsupp} = \textit{minsupp}-1$ (ici \textit{maxsupp} et \textit{minsupp} sont donnés en valeur absolue).\\
La recherche de motifs rares consiste à engendrer tous les motifs, dont le support est inférieur ou égal au seuil \textit{maxsupp}.

Il peut exister un intervalle de valeurs entre \textit{minsupp} et \textit{maxsupp}. Cependant, dans ce mémoire, nous avons travaillé dans le cas général, c'est à dire, un motif est considéré rare s'il n'est pas fréquent. Celà implique l'existence d'une seule frontière entre les motifs rares et les motifs fréquents.
\end{remarque}
\begin{exemple} Soit le contexte d'extraction du tableau \ref{DB1}, les seuils  \textit{maxsupp} et \textit{minsupp} sont fixés respectivement à 2 et 3. Dans le tableau \ref{tab_typ}, nous présentons les supports de quelques motifs tout en distinguant les motifs fréquents de ceux rares.

\begin{table}[!htbp]
  \centering
 \begin{tabular}{|c|c|c|}
\hline
 & Support & Type\tabularnewline
\hline \hline
Supp($A$) & 3 & fréquent \tabularnewline
\hline
Supp($AB$) & 2 & rare \tabularnewline
\hline
Supp($BE$) & 4 & fréquent \tabularnewline
\hline
Supp($CD$)& 1 & rare \tabularnewline
\hline
Supp($ABE$)& 2 & rare \tabularnewline
\hline
\end{tabular}
 \caption{Nature des motifs pour $\textit{maxsupp}= 2$.}\label{tab_typ}
\end{table}
\end{exemple}

\begin{definition} \textbf{Contrainte}\\
Une contrainte binaire $q$ est une application qui renvoie une valeur dans $\{0,1\}$. On dira qu'un motif $I$ d'un contexte formel $\mathcal{K}$ est valide (\textit{resp.} non valide) sous la contrainte $q$ si :
\[ q(I,\mathcal{K})=1 ~~~~~~(resp.~~~q(I,\mathcal{K})=0)\]
On dira aussi que $I$ satisfait la contrainte $q$ (\textit{resp.} ne satisfait pas la contrainte)
\end{definition}
Plusieurs contraintes peuvent être définies à savoir la contrainte de fréquence ou bien la contrainte de rareté. Dans ce qui suit, nous définissons deux contraintes importantes dans le domaine de fouille de données à l'instar de la contrainte de monotonie et celle de l'anti-monotonie.

\begin{definition} \textbf{Contrainte monotone}\\
Une contrainte $q$ est dite monotone si pour un motif $I$ satisfaisant la contrainte, tout sur-ensemble de $I$ satisfait aussi la contrainte \[\forall I,J \subseteq \mathcal{I} ~~et~~ I\subseteq J \Rightarrow q(I,\mathcal{K})=1 \Rightarrow q(J,\mathcal{K})=1 \]
\end{definition}

Par exemple, la contrainte de rareté est monotone. En effet, si $I$ est rare, alors tous ses sur-ensembles sont aussi rares.

\begin{definition} \textbf{Contrainte anti-monotone}\\
Une contrainte $q$ est dite anti-monotone si pour un motif $I$ satisfaisant la contrainte, tout sous-ensemble de $I$ satisfait aussi la contrainte d'anti-monotonie, \textit{i.e.}, 
\[\forall I,J \subseteq \mathcal{I} ~~et~~ I \supseteq J \Rightarrow q(I,\mathcal{K})=1 \Rightarrow q(J,\mathcal{K})=1 \]
\end{definition}

Par exemple, la contrainte de fréquence est anti-monotone. En effet, si $I$ est fréquent, alors tous ses sous-ensembles le sont.

\begin{property} \cite{bay98} (Propriété de fermeture vers le bas).\\
Tous les sous-ensembles d'un motif fréquent sont fréquents.
\end{property}\label{pro_rare}
\begin{property}(Propriété de monotonie).\\
Tous les sur-motifs d'un motif rare sont aussi rares.
\end{property}

\begin{lemma} \cite{galm96} {\textbf{Identité d'inclusion-exclusion}}\\
Les identités d'inclusion-exclusion assurent le lien entre le support conjonctif et le support disjonctif.

\[
Supp(I)=\sum_{\emptyset \subset I'\subseteq I} (-1)^{|I'|-1} \ Supp(\vee I) \]
\[
Supp(\vee I)=\sum_{\emptyset \subset I'\subseteq I} (-1)^{|I'|-1} \ Supp(I) \]

\end{lemma}
Ce lemme démontre que la connaissance des supports disjonctifs (\textit{resp.} conjonctif) de tous les sous-ensembles d'un itemset \textit{I} nous permet de calculer le support conjonctif (\textit{resp.} disjonctif) de \textit{I}. 

\begin{definition} \cite{thesepasq} \textbf{Motif fréquent}\\
Soit un contexte $\mathcal{K}=(\mathcal{O},\mathcal{I},\mathcal{R})$. L'ensemble $\mathcal{MF}$ des itemsets fréquents dans $\mathcal{K}$ est défini comme suit :\\
\[ \mathcal{MF} = \{I \subseteq \mathcal{I}~~|~~Supp(I)\geq minsupp\} \]
\end{definition}

Dans la suite, nous définissons la propriété de l'ensemble des motifs rares, qui est le dual de l'ensemble des motifs fréquents.

\begin{definition} \cite{laszlo06} \label{motif_rare}\textbf{Motif rare}\\
Soit un contexte $\mathcal{K}=(\mathcal{O},\mathcal{I},\mathcal{R})$. L'ensemble $\mathcal{MR}$ des itemset rares dans $\mathcal{K}$ est défini comme suit :\\
\[ \mathcal{MR} = \{I \subseteq \mathcal{I}~~|~~Supp(I)< minsupp\} \]
Ou d'une manière équivalente :
\[ \mathcal{MR} = \{I \subseteq \mathcal{I}~~|~~Supp(I)\leq maxsupp  ~~~avec~~~ maxsupp=minsupp-1\}.\]
\end{definition}

\begin{definition} \cite{Liu08} \textbf{Bordure}\\
Les bordures sont les limites entre les motifs qui satisfont une contrainte (bordure positive) et ceux qui ne satisfont pas (bordure négative). Les bordures forment des représentations condensées de tous les motifs valides.
\end{definition}
Dans ce qui suit, nous définissons la bordure positive et la bordure négative d'une manière formelle.
\begin{definition} \textbf{Bordure positive}\\
La bordure positive est l'ensemble des motifs maximaux au sens de l'inclusion satisfaisant la contrainte. D'une manière formelle, la bordure positive d'un ensemble d'éléments $S$ ordonnés par l'inclusion ensembliste $\subseteq$ est définie par :
\[  \mathcal{BD}^{+}(S)=\{x \subseteq \mathcal{I} ~~|~~ x \in S ~\wedge~~\forall y \supset x,~ y\notin S\} \]
\end{definition}

\begin{definition} \textbf{Bordure négative}\\
Le bordure négative est l'ensemble des motifs maximaux au sens de l'inclusion satisfaisant la contrainte. D'une manière formelle, la bordure négative d'un ensemble d'éléments $S$ ordonnés par l'inclusion ensembliste $\subseteq$ est définie par :
\[  \mathcal{BD}^{-}(S)=\{x \subseteq \mathcal{I} ~~|~~x \notin S ~\wedge~~ \forall y \subset x,~ y \in S\} \]
\end{definition}

\begin{figure}[!htbp]
\begin{center}
  \includegraphics[scale=0.8]{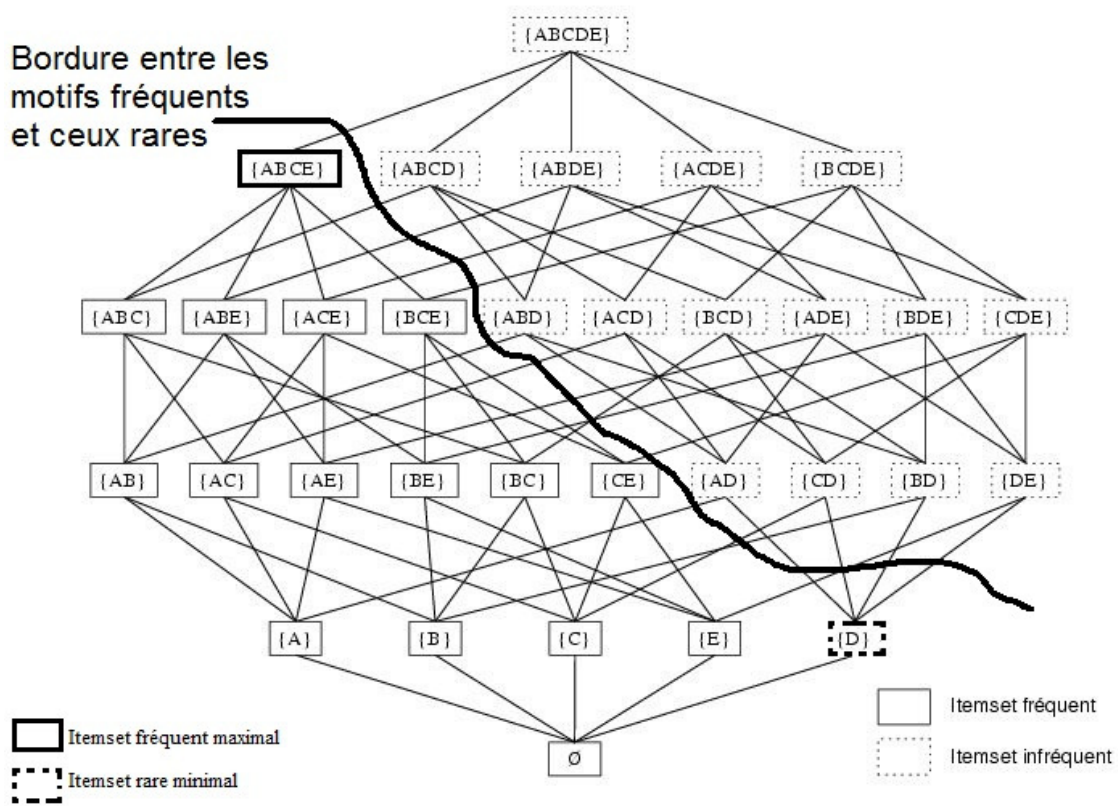}\\
  \caption{Treillis des itemsets associé au contexte $\mathcal{K}$ donné par le tableau \ref{DB1} pour \textit{minsupp} = 2.}\label{tr_itemset}
  \end{center}
\end{figure}

\begin{exemple}
Soit le treillis donné par la figure\ref{tr_itemset}. Pour \textit{minsupp=2}, les bordures positive et négative pour la contrainte fréquence sont comme suit :\\
$\mathcal{BD}^{+}(\mathcal{MF})=\{(ABCE,2)\}$.\\
$\mathcal{BD}^{-}(\mathcal{MF})=\{(D,1)\}$.
\end{exemple}
L'ensemble des motifs rares et l'ensemble des motifs fréquents ont tous les deux un sous-ensemble minimal générateur. \\
Dans le cas des motifs fréquents, ce sous-ensemble est l'ensemble des motifs fréquents maximaux $\mathcal{MFM}$. Un motif est dit motif \textit{fréquent maximal} s'il est fréquent et si tous ces sur-motifs ne le sont pas.\\
Dans le cas des motifs rares, l'ensemble générateur minimal est l'ensemble des motifs rares minimaux $\mathcal{MRM}$. Un motif rare minimal est un motif rare ayant tous ces sous-ensembles fréquents. L'ensemble des motifs rares minimaux forme un ensemble générateur minimal à partir duquel tous les motifs rares peuvent être retrouvés. D'une manière duale, l'ensemble des fréquents maximaux forme un ensemble générateur à partir duquel tous les motifs fréquents peuvent être retrouvés.

\begin{definition}\cite{bay98} \textbf{Ensemble des motifs fréquents maximaux}\\
Soit un contexte $\mathcal{K}=(\mathcal{O},\mathcal{I},\mathcal{R})$. L'ensemble $\mathcal{MFM}$ des itemsets maximaux fréquents dans $\mathcal{K}$ est défini comme suit :
\[ \mathcal{MFM} = \{I \subseteq \mathcal{I}~~|~~Supp(I)\geq minsupp \wedge \forall I' \supset I, ~Supp(I') < minsupp\} \]
\end{definition}
L'ensemble des motifs fréquents maximaux, noté $\mathcal{MFM}$, forme une bordure séparatrice entre les motifs fréquents et ceux rares dans le treillis des itemsets.\\
D'une manière duale, nous définissons l'ensemble des motifs rares minimaux.
\begin{definition}\cite{laszlo06} \textbf{Ensemble des motifs rares minimaux}\\
Soit un contexte $\mathcal{K}=(\mathcal{O},\mathcal{I},\mathcal{R})$. L'ensemble $\mathcal{MRM}$ des itemsets rares minimaux dans $\mathcal{K}$ est défini comme suit :
\[ \mathcal{MRM} = \{I \subseteq \mathcal{I}~~|~~Supp(I)\leq maxsupp \wedge \forall I' \subset I, ~Supp(I') > maxsupp\} \]
\end{definition}
\begin{remarque}
Un motif rare est minimal si tous ces sous-ensembles sont fréquents.
\end{remarque}

\begin{exemple}
Soit le treillis des itemsets illustré par la figure \ref{tr_itemset}. Ainsi, nous avons :\\
$\mathcal{MRM}=\{(D,1)\}$,\\
$\mathcal{MFM}=\{(ABCE,2)\}.$
\end{exemple}

\begin{remarque} \label{rq_bord}
Il faut signaler que l'ensemble des motifs rares minimaux et l'ensemble des motifs fréquents maximaux représentent respectivement la bordure positive et la bordure négative pour la contrainte fréquence. Aisni, nous avons :
\[\mathcal{MFM} \equiv \mathcal{BD}^{+}(\mathcal{MF}) \equiv \mathcal{BD}^{-}(\mathcal{MR})\]
\[\mathcal{MRM} \equiv \mathcal{BD}^{-}(\mathcal{MF}) \equiv \mathcal{BD}^{+}(\mathcal{MR})\]
\end{remarque}
\textbf{Notation.}\\
Dans la suite de ce mémoire, nous utilisons la notation $\mathcal{MFM}$ pour désigner la bordure positive de l'ensemble des motifs fréquents. Par opposition, nous utilisons la notation $\mathcal{MRM}$ pour désigner la bordure négative de l'ensemble des motifs fréquents.\\

L'ensemble des motifs rares se compose de deux sous-ensembles à savoir l'ensemble des motifs zéros et l'ensemble des motifs rares non zéros. Nous définissons l'ensemble des motifs zéros dans ce qui suit :
\begin{definition} \cite{LAP07} \textbf{Ensemble des motifs zéros}\\
Soit un contexte $\mathcal{K}=(\mathcal{O},\mathcal{I},\mathcal{R})$. L'ensemble $\mathcal{MZ}$ des itemsets zéros dans $\mathcal{K}$ est défini comme suit :\\
\[ \mathcal{MZ} = \{I \subseteq \mathcal{I}~~|~~Supp(I)= 0\} \]
\end{definition}
Il faut signaler qu'un motif rare non-zéro est un motif rare ayant un support différent de zéro (\textit{cf.} Définition \ref{motif_rare} de la page \pageref{motif_rare}).
\begin{exemple} \label{exp_mz}
Soit le treillis donné dans la figure \ref{tr_itemset}. Pour $\textit{maxsupp}=1$, l'ensemble des motifs zéros est comme suit :\\
\begin{small}
$\mathcal{MZ}=\{BD,DE,ABD,BCD,ADE,BDE,CDE,ABCD,ABDE,ACDE,BCDE,ABCDE\}$
\end{small}
\end{exemple}

D'une manière analogue, l'ensemble des motifs zéros admet un ensemble générateur minimal à partir duquel nous pouvons retrouver tous les motifs zéros. Dans la suite de ce mémoire, cet ensemble est noté $\mathcal{GMZ}$.
\begin{definition} \cite{LAP07} \textbf{Générateur de motif zéro}\\
Soit un contexte $\mathcal{K}=(\mathcal{O},\mathcal{I},\mathcal{R})$. L'ensemble $\mathcal{GMZ}$ des motifs zéros minimaux dans $\mathcal{K}$ est défini comme suit :\\
\[ \mathcal{GMZ} = \{I \subseteq \mathcal{I}~~|~~Supp(I)= 0 \wedge \forall I' \subset I, ~Supp(I')\neq 0\} \]

\end{definition}
\begin{property}\label{pro_zero}(Propriété de monotonie des motifs zéros).\\
Tous les sur-motifs d'un motif zéro sont des motifs zéros.
\end{property}

\begin{exemple}
Soit le treillis donné par la figure \ref{tr_itemset}. Pour $\textit{maxsupp}=1$, l'ensemble des motifs zéros minimaux est égal à :
$\mathcal{GMZ}=\{BD,DE\}$.\\
La distinction entre les motifs rares zéros et ceux rares non-zéros est importante puisque le nombre de motifs zéros est très élevé. Ainsi, nous nous sommes intéressés par les motifs dont le support dépasse $0$. Comme mentionné plus haut, nous ne générons pas les motifs zéros à cause de leur grand nombre. En se référant à l'exemple \ref{exp_mz}, seuls les motifs appartenant à l'ensemble $\mathcal{GMZ}$ seront maintenus le reste de ces motifs peut être dérivé en utilisant la propriété \ref{pro_zero} (\textit{cf.} page \pageref{pro_zero}).
\end{exemple}
Dans ce qui suit, nous présentons un autre type des motifs clés. Ainsi, ces motifs ne permettent pas une distinction entre les motifs rares et ceux fréquents, comme c'était le cas avec les motifs déjà présentés, mais ils permettent de faire la distinction entre les classes d'équivalences. Pour cela, nous commençons par définir une classe d'équivalence.

L'idée de base de la définition d'une classe d'équivalence est de regrouper les motifs considérés comme équivalents. Il s'agit d'un principe puissant, qui permet de réunir dans des classes d'équivalences les motifs correspondants aux mêmes objets.\\

\begin{definition} \cite{thesepasq} \textbf{Opérateurs de fermeture}\\
Soit un ensemble partiellement ordonné $(E,\leq)$. Une application $h$ de $(E,\leq)$ dans $(E,\leq)$ est un opérateur de fermeture si et seulement si elle possède les trois propriétés suivantes pour tous sous-ensembles $S,S' \in E$ :\\
\begin{enumerate}
  \item Isotonie : $S \leq S' \Rightarrow h(S) \leq h(S')$,
  \item Extensivité : $S \leq h(S)$,
  \item Idempotence : $h(h(S))=h(S) $
\end{enumerate}
Étant donné un opérateur de fermeture $h$ sur un ensemble partiellement ordonné $(E,\leq)$, un élément $x \in E$ est un élément fermé si l'image de $x$ par l'opérateur est lui même : $h(x)=x$

\end{definition}

\begin{definition}  \label{cx_galois}\cite{thesepasq} \textbf{Correspondance de Galois}\\
Soit un contexte d'extraction $\mathcal{K}$ =($\mathcal{O}$,$\mathcal{I}$,$\mathcal{R}$). Soit l'application
$f$, de l'ensemble des parties de
$\mathcal{O}$ $^{(}$\footnote{L'ensemble des parties d'un ensemble
d'éléments $\mathcal{O}$, constitué de tous les sous-ensembles de
$\mathcal{O}$, est noté 2$^{\mathcal{O}}$.}$^{)}$ dans l'ensemble
des parties de $\mathcal{I}$, qui associe à un ensemble d'objets
$O$ $\subseteq \mathcal{O}$ l'ensemble des items $i$ $\in$
$\mathcal{I}$ communs à tous les objets $o$ $\in$ $O$
\cite{ganter99}:
\begin{center}
$f: 2^\mathcal{O}{} \rightarrow 2^\mathcal{I}{}$\\
$f(O)=\{i \in \mathcal{I} | \forall o\in O \wedge
o\mathcal{R}i$ $\}$
\end{center}

Soit l'application $g$, de l'ensemble des parties de
$\mathcal{I}$ dans l'ensemble des parties de $\mathcal{O}$, qui
associe à tout ensemble d'items (communément appelé itemset) $I$
$\subseteq \mathcal{I}$ l'ensemble des objets $o$ $\subseteq
\mathcal{O}$ contenant tous les items $i$ $\in$ $I$
\cite{ganter99}:

\begin{center}
$g:2^\mathcal{I}{} \rightarrow 2^\mathcal{O}{}$\\
$g(I)=\{o\in \mathcal{O} | \forall i\in I \wedge o\mathcal{R}i$
$\}$
\end{center}

Le couple d'applications ($f$,$g$) est une \textit{correspondance
de Galois} entre l'ensemble des parties de $\mathcal{O}$ et
l'ensemble des parties de $\mathcal{I}$~\cite{ganter99}. Étant
donné une correspondance de Galois, les propriétés
suivantes sont vérifiées quelques soient $I$, $I_{_{1}}$,
$I_{_{2}}$ $\subseteq $ $\mathcal{I}$ et $O$, $O_{_{1}}$,
$O_{_{2}}$ $\subseteq$ $\mathcal{O}$ \cite{ganter99}:

1. $I_{_{1}}$ $\subseteq $ $I_{_{2}}$ $\Rightarrow $
$g$($I_{_{2}}$) $\subseteq $ $g$($I_{_{1}}$);

2. $O_{_{1}}$ $\subseteq $ $O_{_{2}}$ $\Rightarrow $
$f$($O_{_{2}}$) $ \subseteq $ $f$($O_{_{1}}$);

3. $O$ $ \subseteq $ $g$($I$) $ \Leftrightarrow $ $I$ $
\subseteq $ $f$($O$) $ \Leftrightarrow $ ($I$,$O$) $ \in $ $\mathcal{R}$.\\
\end{definition}

\begin{remarque}
L'opérateur de fermeture $h$ est le résultat de l'application consécutive de deux fonctions $g$ et $f$ sur un ensemble de motifs.\\
\begin{center}
$h(x)=f \circ g(x)$ où $x \in \mathcal{I}$
\end{center}
\end{remarque}

\begin{definition} \label{def_ce} \cite{thesepasq} \textbf{Relation d'équivalence}\\
Soient $P,Q$ deux motifs. La relation d'équivalence $\theta$ est définie comme suit :\\
\[ P\theta Q \Leftrightarrow h(P)=h(Q).\]
La classe d'équivalence de P est donnée par :
\[ [P]= \{Q \subseteq \mathcal{P} | P \theta Q\} \]
$h$ est l'opérateur de fermeture de Galois.
\end{definition}

\begin{definition} \cite{PBT99} \textbf{Motif fermé}\\
Étant donné l'opérateur de fermeture de la correspondance de Galois
$h$, un itemset $I$ $\subseteq \mathcal{I}$ tel que
$h$($I$) = $I$ est appelé \textit{itemset fermé}. Un itemset
fermé est donc un ensemble maximal d'items communs à un ensemble
d'objets.
\end{definition}
\textbf{Notation. } Dans la suite de ce mémoire, nous désignons par $\mathcal{MF}e$ l'ensemble des motifs fermés, par $\mathcal{MFF}$ l'ensemble des motifs fermés fréquents et par $\mathcal{MFR}$ l'ensemble des motifs fermés rares.
\begin{definition} \cite{PBT99} \textbf{Ensemble des motifs fermés fréquents}
\[
\mathcal{MFF}=\{I\subseteq \mathcal{I}\,|\, h(I)=I ~|~ Supp(I)\geq minsupp\}\]
\end{definition}
D'une manière duale, nous définissons un motif fermé rare comme suit.

\begin{definition} \textbf{Ensemble des motifs fermés rares}

\[
\mathcal{MFR}=\{I\subseteq \mathcal{I}\,|\, h(I)=I ~|~ 0\leq Supp(I)\leq maxsupp\}\]
où $h$ est l'opérateur de fermeture.
\end{definition}

\begin{exemple}
L'ensemble $\mathcal{MFR}$ des itemsets fermés rares du contexte $\mathcal{K}$ pour un seuil maximal de support égal à $3$ est présenté dans la table \ref{tabferm}. L'itemset ${BCE}$ est un itemset fermé rare pour $\textit{maxsupp}= 3$ car $supp({BCE})=|{2,3,5}| = 3 \leq \textit{minsupp}$. Dans une base de données médicale, cela signifie que $60\%$ (c'est-à-dire 3 cas sur 5) des patients ont les maladies B, C et E.
\end{exemple}
\begin{table}[!htbp]
  \centering
  \begin{tabular}{|c|c|}
    \hline
    Itemset fermé rare & Support\\
    \hline     \hline
    $C$ & 4 \\
    $AC$ & 3 \\
    $BE$ & 4 \\
    $ACD$ & 1 \\
    $BCE$ & 3 \\
    $ABCE$ & 1 \\
    \hline
  \end{tabular}
  \caption{Itemsets fermés rares extraits du contexte $\mathcal{K}$ pour $\textit{minsupp}= 3$.}\label{tabferm}
\end{table}

Dans ce qui suit, nous définissons les propriétés de l'idéal d'ordre et de filtre d'ordre pour un ensemble donné.
\begin{definition} \cite{GS05} \textbf{Idéal d'ordre}\\
Un ensemble S est un idéal d'ordre s'il vérifie les propriétés suivantes :
\begin{itemize}
  \item Si $\textit{x} \in S$, alors $\forall \textit{y} \subseteq \textit{x}$, $y \in S$.
  \item Si $\textit{x} \notin S$, alors $\forall \textit{x} \subseteq \textit{y}$, $\textit{y} \notin S$.
\end{itemize}
\end{definition}
\bigskip
\begin{definition} \cite{GS05} \textbf{Filtre d'ordre}\\
Un ensemble S est un filtre d'ordre s'il vérifie les propriétés suivantes :
\begin{itemize}
  \item Si $\textit{x} \in S$, alors  $\forall \textit{y} \supseteq \textit{x}$, $\textit{y} \in S$.
  \item Si $\textit{x} \notin S$, alors $\forall \textit{x} \supseteq \textit{y}$, $\textit{y} \notin S$.
\end{itemize}
\end{definition}

\begin{property}\label{pro_fre}
L'ensemble des motifs fréquent forme un idéal d'ordre dans $(2^{n},\subseteq)$ où n est le nombre d'items.
\end{property}
\begin{property}\label{pro_rare}
L'ensemble des motifs rares forme un filtre d'ordre dans $(2^{n},\subseteq)$ où n est le nombre d'items :
\begin{itemize}
\item Tous les sur-ensembles d'un motif rare sont rares.
\item Tous les sous-ensembles d'un motif non rare sont aussi non rares.
\end{itemize}
\end{property}

\begin{exemple}
Soit le treillis des itemsets donné par la figure \ref{tr_itemset}. Pour $\textit{minsupp}=2$, tous les sous-ensembles du motif $ABCE$ (fréquent) sont fréquents. D'une manière duale, tous les sur-ensembles du motif $D$ (rare) sont rares.
\end{exemple}

\begin{figure}[!htbp]
\begin{center}
  \includegraphics[scale=0.8]{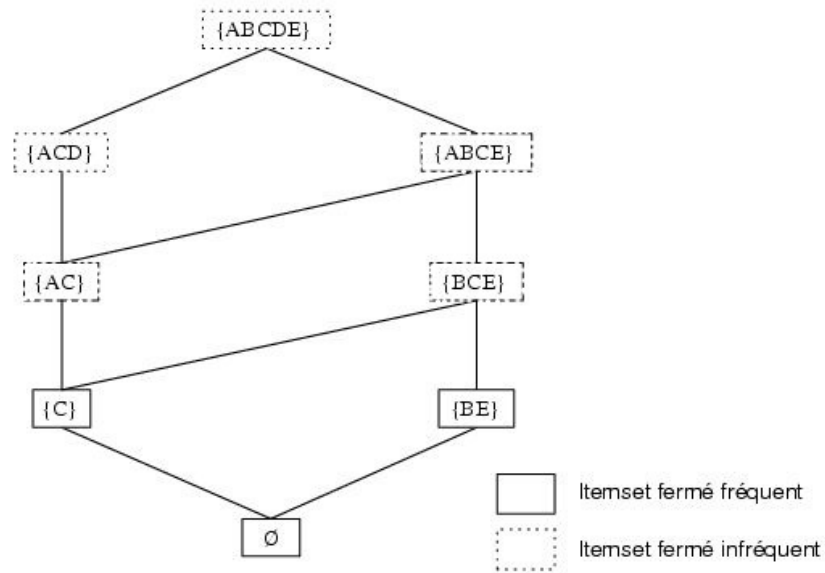}\\
  \caption{Treillis des itemsets fermés associé au contexte $\mathcal{K}$.}\label{tr_ferme}
  \end{center}
\end{figure}

\begin{definition} \cite{PBT99} \textbf{Treillis des itemsets fermés}\\
Soit $\mathcal{MF}e$ l'ensemble des itemsets fermés dans un contexte $\mathcal{K}=(\mathcal{O},\mathcal{I},\mathcal{R})$ et la relation d'ordre partiel $\subseteq$ sur les éléments de $\mathcal{MF}e$. Pour chaque couple d'éléments $l_{1}, l_{2} \in \mathcal{MF}e$ nous avons $l_{1} \subseteq l_{2}$. Nous appelons alors $l_{1}$ \textit{sous-ensemble fermé} de $l_{2}$ et $l_{2}$ \textit{sur-ensemble fermé} de $l_{1}$. L'ensemble partiellement ordonné $(\mathcal{MF}e,\subseteq)$ forme un treillis complet puisque pour tout sous-ensemble $S \subseteq \mathcal{MF}e$ il existe un plus petit majorant et un plus grand minorant \cite{ganter99,Wille99}.
\end{definition}
\begin{exemple}
Le treillis des itemsets fermés associé au contexte $\mathcal{K}$, du tableau \ref{DB1}, est représenté dans la figure \ref{tr_ferme}. Ce treillis contient 8 itemsets.
\end{exemple}
\begin{remarque} Le treillis des itemsets fermés d'une relation binaire finie (le contexte d'extraction) est isomorphe au treillis formé par les composantes intensions des concepts du treillis de concept \cite{ganter99,Wille99}, également appelé treillis de Galois, de cette relation binaire. \end{remarque}

Agrawal \textit{et al.} ont introduit dans \cite{AIS93}, les deux
propriétés suivantes relatives aux supports des itemsets
fréquents:
\begin{enumerate}
    \item Tous les sous-ensembles d'un itemset fréquent sont fréquents.
    \item Tous les sur-ensembles d'un itemset infréquent sont infréquents.
\end{enumerate}

Ces propriétés restent applicables dans le cas des itemsets
\textit{fermés} fréquents \cite{thesepasq}. Ainsi, nous avons
\begin{enumerate}
    \item Tous les sous-ensembles d'un itemset fermé fréquent sont fréquents.
    \item Tous les sur-ensembles d'un itemset fermé infréquent sont infréquents.
\end{enumerate}
Sachant que le support d'un itemset (fréquent) $I$ est égal au
support de sa fermeture $h(I)$ qui est le plus petit itemset
fermé contenant $I$: $Supp(I) = Supp(h(I))$ \cite{thesepasq}.\\
Dans ce qui suit, nous définissons les motifs minimaux d'une classe d'équivalence communément appelés générateurs minimaux \cite{Liu08}.
\begin{definition} \cite{Liu08} \textbf{Générateur minimal}\\
Un motif $I$ est dit un générateur minimal si et seulement si n'existe aucun motif $J$ tel que $J \subset I$ et $Supp(I)=Supp(J)$.
Selon la définition, l'ensemble vide est un générateur minimal de toute la base de donnée.
\end{definition}
\begin{definition} \cite{Liu08} \textbf{Ensemble des générateurs minimaux}
\[ \mathcal{GM}=\{ I \in \mathcal{I}~~|~~ \nexists J ~~tel ~~que~~ J \subset I ~~et~~ Supp(I)=Supp(J)\}\]
\end{definition}

\begin{definition} \cite{Liu08} \textbf{Générateurs minimaux fréquents}
\[
\mathcal{GMF}=\{I \in \mathcal{GM} ~~|~~ Supp(I)> maxsupp\}\]
\end{definition}

\begin{definition} \textbf{Générateurs minimaux rares}
\[
\mathcal{GMR}=\{I \in \mathcal{GM} ~~|~~ Supp(I)\leq maxsupp\}\]
\end{definition}

\begin{property} \cite{Liu08} \label{pro_gm}
L'ensemble des générateurs minimaux forme un idéal d'ordre dans $(2^{n},\subseteq)$ où n est le nombre d'items.
\[\forall I \in \mathcal{I} ~~|~~ I \notin GM \Rightarrow J \supseteq I, J \notin GM \]
\end{property}
\begin{remarque}
L'inclusion de l'ensemble des générateurs minimaux fréquents et de l'ensemble des générateurs minimaux rares forment l'ensemble des générateurs minimaux total du contexte puisque $maxsupp=minsupp-1$. D'une manière formelle :
\[ \mathcal{GM} = \mathcal{GMR} \cup \mathcal{GMF}.\]
De même pour l'ensemble des fermés, nous pouvons écrire :
\[ \mathcal{MF}e = \mathcal{MFF} \cup \mathcal{MFR}.\]
\end{remarque}

\begin{figure}[!htbp]
\begin{center}
  \includegraphics[scale=0.5]{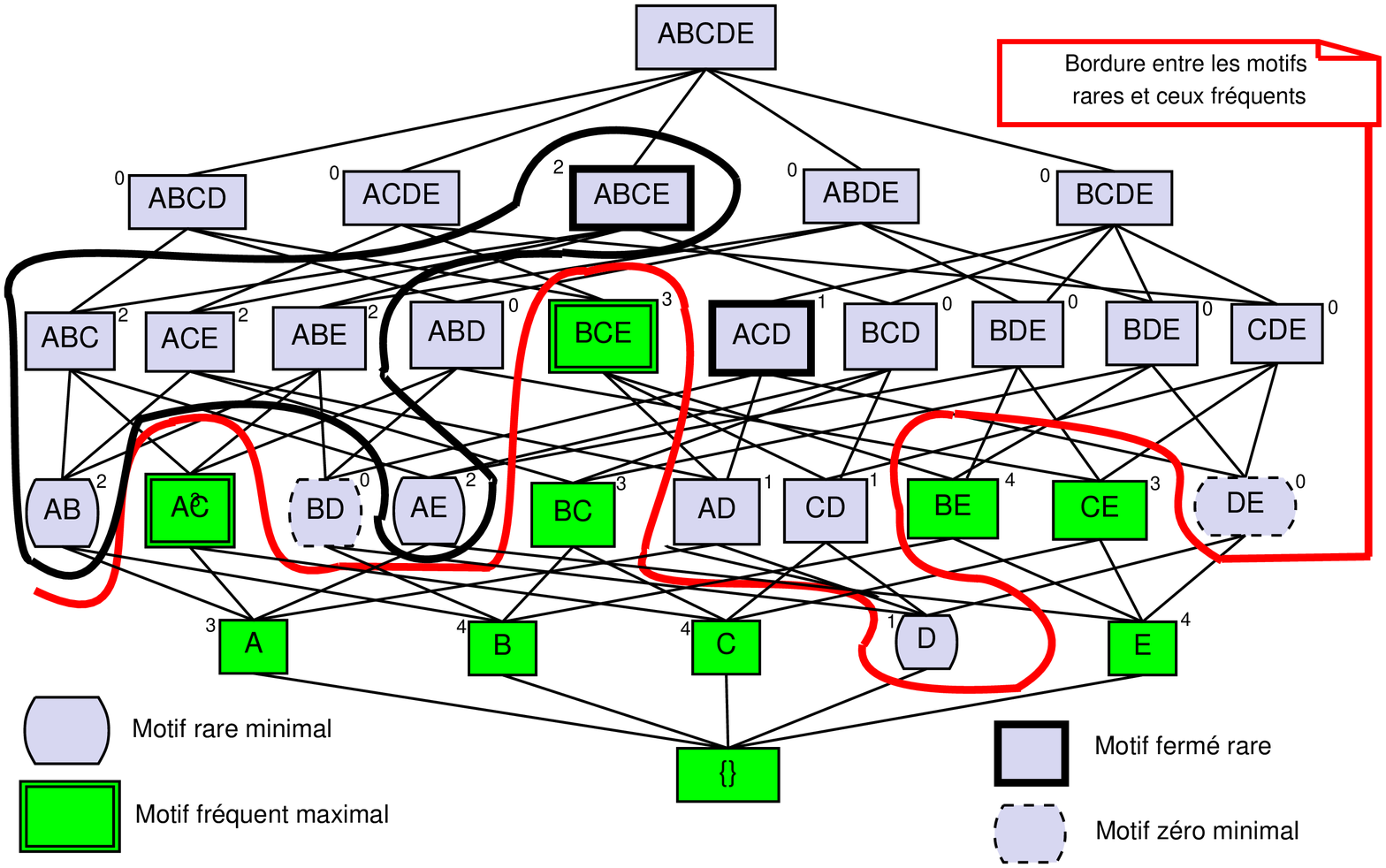}\\
  \caption{Treillis des itemsets associé au contexte $\mathcal{K}$ pour \textit{minsupp} = 3.}\label{tr}
  \end{center}
\end{figure}

\begin{figure}[!htbp]
\begin{center}
  \includegraphics[scale=0.5]{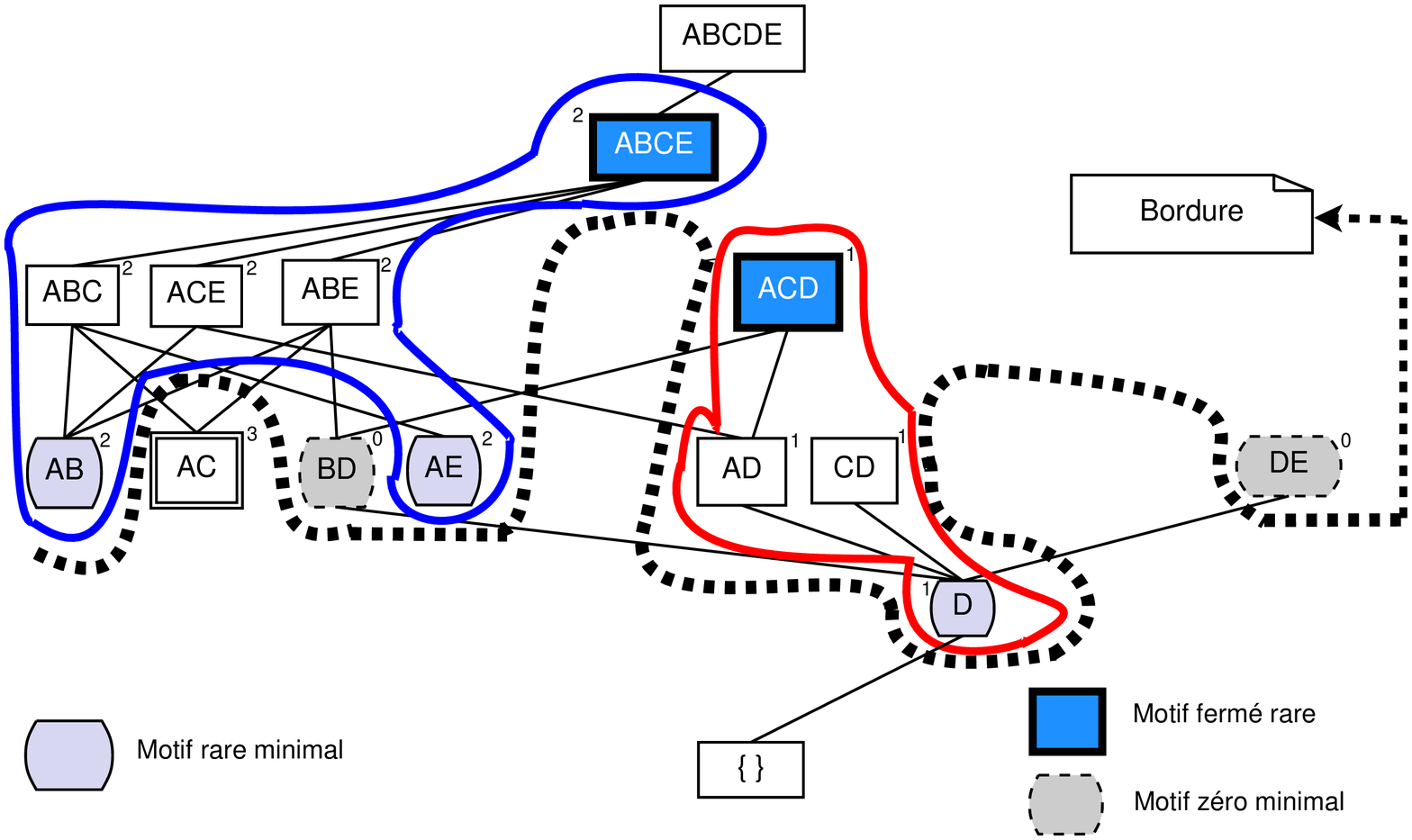}\\
  \caption{Une partie du treillis agrandie pour montrer les classes d'équivalence du contexte $\mathcal{K}$ pour \textit{minsupp} = 3.}\label{zoom_tr}
  \end{center}
\end{figure}

\begin{exemple}
Les itemsets fermés rares dans le contexte $\mathcal{K}$ pour un seuil minimal de support de $3$ sont représentés dans la figure \ref{tr_ferme}.
\end{exemple}
\begin{exemple}
Considérons le treillis des itemset défini par la figure \ref{tr} et sa partie agrandie donnée par la figure \ref{zoom_tr} associé au contexte  $\mathcal{K}$ pour $minsupp=3$. Nous remarquons qu'a partir de l'ensemble des générateurs minimaux rares $\mathcal{GMR}=\{(AB,2),(AE,2),(D,1)\}$ et de l'ensemble des générateurs zéros minimaux  $\mathcal{GMZ}=\{DE,BD\}$, nous pouvons distinguer les motifs fréquents de ceux rares et nous déduisons, pour les motifs rares, leurs supports exacts (\textit{cf.} la figure \ref{zoom_tr}).\\
par ailleurs, nous pouvons utiliser l'ensemble des fermés rares $\mathcal{MFR}=\{(ABCE,2),\\(ACD,2)\}$ et l'ensemble des minimaux rares, noté $\mathcal{MRM}=\{AB,AE,D\}$, pour déduire l'ensemble total des motifs rares munis de leurs supports exacts.
\end{exemple}

\section{Règles associatives}

La formalisation du problème d'extraction des règles associatives
a été introduite par Agrawal \textit{et al.} en
1993~\cite{AIS93}. Une règle associative $r$ est une relation
entre itemsets de la forme $r$: $X$ $\Rightarrow$ $(Y$-$X)$, dans
laquelle $X$ et $Y$ sont des itemsets fréquents, tel que $X
\subset Y$. Les itemsets $X$ et $(Y$-$X)$ sont appelés,
respectivement, \textit{prémisse} et \textit{conclusion} de la
règle $r$. La génération des règles associatives
est réalisée à partir d'un ensemble $\mathcal{MF}$
d'itemsets fréquents dans un contexte d'extraction
$\mathcal{K}$, pour un seuil minimal de support \textit{minsupp}.
Les règles associatives valides sont celles dont la mesure de
confiance, $Conf(r)$ = $\frac{Supp(Y)}{Supp(X)}$, est
supérieure ou égale à un seuil minimal de confiance,
défini par l'utilisateur et qui sera noté dans la suite
\textit{minconf}. Si $Conf(r) = 1$ alors $r$ est appelée
\textit{règle associative exacte}, sinon elle est appelée
\textit{règle associative approximative} \cite{PBT99}.

\bigskip

Ainsi, chaque règle associative, $X$ $\Rightarrow$ $(Y$-$X)$, est
caractérisée par:
\begin{enumerate}
    \item \textbf{Le niveau de support}: il correspond au nombre de fois où
    l'association est présente, rapportée au nombre de transactions
    comportant l'ensemble des items de Y. Le niveau de support
    permet de mesurer la fréquence de l'association \cite{LV93}.
    \item \textbf{Le niveau de confiance}: il correspond au nombre de fois où
    l'association est présente, rapportée au nombre de présence de $X$. Le niveau de confiance permet de mesurer la force de
    l'association \cite{LV93}.
\end{enumerate}

\bigskip

Ainsi, étant donné un contexte d'extraction $\mathcal{K}$, le
problème de l'extraction des règles associatives dans
$\mathcal{K}$ consiste à déterminer l'ensemble des règles
associatives dont le support et la confiance sont au moins égaux
respectivement à \textit{minsupp} et \textit{minconf}. Ce problème
peut être décomposé en deux sous-problèmes comme suit
\cite{AIS93}

1. Déterminer l'ensemble des itemsets fréquents dans
$\mathcal{K}$, \textit{i.e.}, les itemsets dont le support est
supérieur ou égal à \textit{minsupp}.

2. Pour chaque itemset fréquent $I_{_{1}}$, générer toutes les
règles associatives de la forme $r$: $I_{_{2}} \Rightarrow
I_{_{1}}$ tel que $I_{_{2}} \subset I_{_{1}}$ et dont la confiance
est supérieure ou égale à \textit{minconf}.

\bigskip

Ces deux problèmes sont résolus par un algorithme fondamental dans
la fouille de données, à savoir \textsc{Apriori} \cite{AIS93}. Le
premier sous-problème a une complexité exponentielle en fonction
du nombre d'itemsets. En effet, étant donné un ensemble d'items de
taille $n$, le nombre d'itemsets fréquents potentiels est égal à
$2^{n}$. Le deuxième sous-problème est exponentiel en la taille
des itemsets fréquents. En effet, pour un itemset fréquent $I$, le
nombre de règles associatives non triviales qui peuvent être
générées est 2$^{\vert I \vert}$ - 1. Toutefois, la génération des
règles associatives à partir des itemsets fréquents ne nécessite
aucun balayage de la base de données et les temps de calcul de
cette génération sont faibles comparés aux temps nécessaires pour
la découverte des itemsets fréquents \cite{thesepasq}. Néanmoins,
le problème de la pertinence et de l'utilité des
règles extraites est d'une première importance étant
donné que dans la plupart des bases de transactions
réelles, des milliers et m\^{e}me des millions de règles
associatives sont générées \cite{ST01,zaki04}. Or,
il a été constaté que dans la pratique, plusieurs règles étaient
redondantes \cite{BYahia04}.

\section{Conclusion}
En se basant sur les notions qui ont été énoncées dans ce chapitre, plusieurs efforts de recherche ont été entamés dans l'objectif de l'extraction et l'exploitation de l'ensemble des motifs rares d'une base de transaction ainsi que leurs supports.\\
Dans le chapitre qui suit, nous nous proposons de présenter les méthodes proposées dans la littérature ainsi qu'une étude comparative entre ces méthodes.
\chapter{Découverte des motifs rares : état de l'art}
\minitoc

\linesnumbered
\section{Introduction}
La plupart des études en fouille de données se sont intéressées à l'extraction
de motifs fréquents et à la génération des règles d'associations
à partir de ces motifs fréquents. Ces dernière années, plusieurs travaux de recherche
se focalisent à l'exploitation des motifs rares,
ainsi qu'ils montrent l'intérêt majeur de ces motifs dans le cas des
bases réelles \cite{MNS05,LAP07}. Le nombre de motifs rares pouvant être extraits est très grand ceci rend leur exploitation et leur manipulation quasi-impossible par des experts humains. Ainsi, le besoin de définir des ensembles de motifs, de taille réduite, à partir desquels nous pouvons générer avec exactitude l'ensemble total des motifs munis de leurs supports est né avec les motifs fréquents. De tels ensembles pour les motifs fréquents sont appelées \emph{représentations concises exactes}. Dans le cas des motifs rares, aucun travail n'a était mené pour extraire une représentation concise exacte.
Dans ce chapitre, nous présentons en premier lieu les principaux
algorithmes d'extraction des motifs rares. Par la suite, nous exposons les travaux qui extraient une représentation concise exacte des motifs fréquents.

\section{Motivations}
L'objectif majeur de la fouille de données est d'identifier
des relations cachées et intéressantes entre les motifs d'une
base de données gigantesque. Plusieurs travaux depuis 1996 sont focalisés
à l'extraction des règles d'associations
fréquentes issues des motifs fréquents. Autrement dit, nous ne conservons que les motifs,
dont la fréquence d'apparition est supérieure à un seuil donné d'avance
puis on génère à partir de ceux-ci des règles exactes ou
approximatives \cite{Agra96} de la forme ``Tous les
étudiants qui suivent le cours \emph{Introduction à UNIX} suivent
en parallèle un cours \emph{Programmation en C}'' \cite{Han00}. Jusqu'à présent la plupart des travaux se sont intéressés
à l'extraction des motifs fréquents et les règles
associatives fréquentes. Parfois les règles d'association
qui sont générées à partir des motifs fréquents sont non intéressantes,
dans le sens où un comportement fréquent est en général un comportement
normal dans un contexte d'extraction. Plusieurs études ont prouvés
l'importance majeure de la notion de la rareté dans
plusieurs domaines. Dans ce qui suit, nous citons l'intérêt
de la notion des motifs rares dans des contextes réels. En effet,
la découverte des motifs rares peut se révéler très intéressante dans
les domaines suivants : médecine, biologie, sécurité et audit des
risques et traçage des comportements personnels.\\
Prenons un exemple qui simule une base de données médicale où
nous nous intéressons à l'identification de la cause
des maladies cardio-vasculaires \emph{(MCV).} Une règle d'association
fréquente telle que ``\emph{\{niveau élevé de cholestérol\}
$\Rightarrow$ \{MCV\}}'' peut valider l'hypothèse
que les individus qui ont un fort taux de cholestérol ont un risque
élevé de \emph{MCV}. À l'opposé, si nous avons un
nombre suffisant de végétariens dans notre base de données, alors
une règle d'association rare ``\emph{\{végétarien\}
$\Rightarrow$ \{MCV\}}'' peut valider l'hypothèse
que les végétariens ont un risque faible de contracter une \emph{MCV}.
Dans ce cas, les motifs \emph{\{végétarien\}} et \emph{\{MCV\}} sont
tous les deux fréquents, mais le motif \emph{\{végétarien, MCV\}} est rare.

Un autre exemple issu du domaine de la pharmacovigilance,
qui est une branche de la pharmacologie dédiée à la détection et l'étude
des effets indésirables des médicaments. L'extraction des motifs rares dans une base de
données des effets indésirables des médicaments pourrait contribuer
à un suivi plus efficace des effets indésirables graves et servir ensuite
à prévenir les accidents mortels qui aboutissent au retrait de certains
médicaments (par exemple, le retrait de la cérivastatine, médicament
hypolipémiant, en août 2001) \cite{MNS05}.\\

De même, un troisième exemple étudié par \cite{MNS05,Siest98}, qui se base sur des données réelles de la cohorte \texttt{STANISLAS}. Cette étude a montré l'intérêt
de l'extraction des motifs rares pour la fouille de
données des cohortes supposées saines. Cette cohorte est composée
d'un millier de familles Françaises présumées saines.
Son principal objectif est de mettre en évidence l'influence
des facteurs génétiques et environnementaux sur la variabilité des
risques cardio-vasculaires. Une information intéressante à extraire
de cette base de données pour l'expert dans ce domaine
consiste en des profils qui associent des données génétiques à des
valeurs extrêmes ou limites de paramètres biologiques. Cependant,
ces types d'associations sont plutôt rares dans les
cohortes saines. Dans ce contexte, l'extraction de
motifs rares pourrait être très utile pour atteindre les objectifs
de l'expert.\\
Finalement, nous prenons un exemple issu du domaine de la sécurité informatique. Étant donné un
fichier log, qui représente les tentatives de connexions effectuées sur un serveur web d'authentification. Nous supposons l'organisation suivante du fichier log : adresse \texttt{IP}, temps de connexion, port, nombre de tentatives, etc. Le fait d'extraire
les motifs fréquents ainsi que les règles d'associations fréquentes
associées permet de connaître la plupart des adresses connectant au serveur, les services les plus utilisés et les pages les plus demandées. Cependant, les informations qui sont liées aux attaques à savoir; l'origine des attaques, les ports les plus attaqués et les services les plus visés sont quasiment impossible à dégager  à l'aide des algorithmes qui extraient des relations entre les motifs fréquents.
Ainsi, si le fichier log contient quatre champs à savoir : l'\texttt{IP} de l'utilisateur connecté, le port à travers lequel la connexion s'établie, la date et un dernier champ représente l'état de la connexion (établie avec succès ou bien échouée). Le nombre de transactions du contexte représente le nombre de connexions par jour, il est clair qu'une recherche des motifs rares dans un tel contexte nous conduise à trouver les connexions qui échouent puisqu'ils forment une minorité par rapport aux autres. Ces connexions, qui sont mal finies, peuvent être dûes à une simple faute de frappe, mais dans pas mal des cas se sont à l'origine des attaques.\\
\section{Algorithmes de la littérature}

Dans la littérature, plusieurs travaux s'intéressent aux motifs rares, que nous pouvons regrouper dans deux classes :

\begin{itemize}
\item Les algorithmes dédiés à l'extraction de l'ensemble total des motifs rares.
\item Les algorithmes dédiés à l'extraction d'une partie de l'ensemble des motifs rares.
\end{itemize}

\subsection{Algorithmes d'extraction d'une partie de l'ensemble des motifs rares}

Ci-dessous la liste exhaustive des travaux qui s'intéressent à l'extraction des motifs rares tout en introduisant des nouvelles mesures de qualité \cite{Duval07} autres que \emph{minsupp} et \emph{minconf} :

\begin{itemize}

\item \textsc{MSApriori} \cite{Yun03},

\item \textsc{AICluster} \cite{Koh08},

\item \textsc{Rsaa} \cite{Yun03},

\item \textsc{Morsa} \cite{Suk04}.
\end{itemize}
Dans la suite, nous détaillons quelques approches.

\subsubsection{Algorithme \textsc{MSApriori}}
\textsc{MSApriori} est un algorithme basé sur l'algorithme \textsc{Apriori}, Il a utilisé d'autres mesures de qualités. Dans le cas d'\textsc{Apriori}, un seul seuil de fréquence est utilisé pour tous les travaux associés, de même une seule valeur \emph{minconf} est appliquée pour toutes les règles. Cependant, dans des cas réels, cette unicité peut être non significative vu la diversité des items dans un contexte de données \cite{Yun03}. Afin de pallier ces problèmes, Liu \textit{et al.} en 1999, cinq ans après l'apparition de l'algorithme \textsc{Apriori}, ont pensé à des nouvelles mesures de qualité tout en utilisant les principes de l'algorithme Agrawal \textit{et al}. La suppression d'un seul seuil  \emph{minsupp} et l'introduction d'une liste des seuils étaient le point de différence entre ces deux travaux. \textsc{MSApriori} est l'abréviation de \texttt{M}ulti \texttt{S}upport \texttt{A}priori, c'est le premier algorithme utilisant la notion de plusieurs supports et l'affectation d'un \textit{minsupp} pour chaque item de la base
de données. Le résultat de l'algorithme est l'ensemble des motifs fréquents
par rapport à une liste des seuils. Cependant, cet ensemble peut inclure des motifs rares
qui ne peuvent pas être extraits à l'aide d'une méthode classique
utilisant un seul seuil comme le cas de l'algorithme \textsc{Apriori}. Il faut signaler que l'algorithme \textsc{MSApriori} n'est pas conçu pour extraire des motifs rares mais l'utilisation de plusieurs supports a permis d'avoir une partie de l'ensemble des motifs rares dans le résultat final.

\subsubsection{Algorithme \textsc{Rsaa}}
L'algorithme \textsc{Rsaa} est l'abréviation de \texttt{R}elative \texttt{S}upport \texttt{A}priori \texttt{A}lgorithm \cite{Yun03}. L'algorithme \textsc{Rsaa} est une amélioration de \textsc{MSApriori} basée sur \textsc{Apriori}. En 2003, Yun \emph{et al.} ont présenté les inconvénients majeurs
dus à l'utilisation de plusieurs seuils à savoir : les critères de choix de \emph{minsupp} pour chaque item, le ralentissement du programme à cause d'un nombre énorme de seuils, un coût supplémentaire pour l'évaluation de support des motifs composé de plusieurs items où l'algorithme a besoin de calculer un seuil minimal
global de tous les seuils. 
Ainsi, ces problèmes ont poussé Yun \emph{et al.} à présenter une nouvelle approche qui utilise deux seuils : le premier seuil
sert à extraire les motifs fréquents et le deuxième sert à extraire les motifs rares du contexte. Le
résultat de l'algorithme contient une partie des motifs fréquents
et une partie des motifs rares.

Au même fil d'idées, les autres approches sont à l'origine d'une amélioration de l'aspect réel du résultat à travers des nouvelles mesures de qualité tout en se basant sur l'algorithme \textsc{Apriori}. Par ailleurs, Il est à signaler que l'algorithme ``\textsc{Morsa}'' est conçu pour extraire des motifs rares à partir des bases de données non-binaires.\\

Toutes ces approches, déjà citées, se rassemblent au niveau du résultat à savoir l'extraction d'une partie des motifs rares.
En outre, les approches qui vont être citées dans ce qui suit, possèdent le même résultat à savoir l'ensemble total des motifs rares.

\subsection{Algorithmes d'extraction de l'ensemble total des motifs rares }

D'une manière chronologique, \textsc{Apriori-rare} est le premier algorithme permettant
d'extraire les motifs rares (\textit{cf.} Définition \ref{motif_rare}, page \pageref{motif_rare}) à partir d'un contexte d'extraction. Ensuite, plusieurs algorithmes ont été développés, dont nous citons :

\begin{itemize}
\item \textsc{MRG-Exp} de Szathmary \emph{et al.} \cite{LAP07},
\item \textsc{Afrim} de Adda \emph{et al.} \cite{AMV07},
\item \textsc{Minit} de Haglin et Manning \cite{Haglin07}.
\end{itemize}

\subsubsection{Algorithme \textsc{Apriori-rare} :}

Input : Le contexte d'extraction, \emph{minsupp}.

Output : L'ensemble des rares minimaux.

\begin{description}
\item [{Description}]~
\end{description}
L'algorithme \textsc{Apriori-rare} est inspiré de l'algorithme \textsc{Apriori}
\cite{AIS93}. L'idée est d'utiliser l'algorithme \textsc{Apriori}
(qui génère les motifs fréquents) avec une modification au niveau de la méthode d'élagage.
En effet, les motifs rares minimaux peuvent être trouvés simplement à
l'aide de l'algorithme \textsc{Apriori}. \textsc{Apriori} est basé sur deux principes (\textit{cf.} les propriétés \ref{pro_fre} et \ref{pro_rare}, page \pageref{pro_rare}). \textsc{Apriori} est conçu pour trouver les motifs fréquents, mais a pour ``effet collatéral'' d'explorer également les motifs rares minimaux. Quand \textsc{Apriori} trouve un motif rare, il ne
génèrera plus tard aucun de ses sur-motifs car ils sont de manière sûre infréquents. Puisque
\textsc{Apriori} explore le treillis des motifs niveau par niveau du bas vers le haut, il comptera aussi le
support des motifs rares minimaux. Ces motifs seront élagués, et plus tard l'algorithme peut
remarquer qu'un candidat a un sous-motif rare (en fait \textsc{Apriori} vérifie si tous les $(k - 1)$-sous-motifs
d'un $k$-candidat sont fréquents). Si l'un d'entre eux n'est pas fréquent, alors son
candidat est rare. En outre, cela signifie que le candidat a un sous-motif rare minimal. Grâce
à cette technique d'élagage, \textsc{Apriori} peut réduire significativement l'espace de recherche dans
le treillis des motifs.
Une légère modification d'\textsc{Apriori} suffit pour conserver les $\textsc{MRM}s$ $^{(}$\footnote{$\mathcal{MRM}$ est l'ensemble des \textbf{M}otifs \textbf{R}ares \textbf{M}inimaux.}$^{)}$. Si le support d'un
candidat est inférieur au support minimum \emph{maxsupp}, l'algorithme  l'enregistrera dans l'ensemble des motifs rares minimaux avant de l'élaguer (\textit{cf.} Algorithme \ref{algoApriori-rare}). Les motifs retenus avant l'élagage forment un sous-ensemble des motifs rares minimaux. Cet ensemble contient les motifs infréquents (inférieur à \emph{minsupp}) ainsi que ses sous-ensembles sont fréquents. En effet, cet ensemble forme une frontière entre les motifs fréquents et ceux rares (voir figure \ref{tr_itemset} page \pageref{tr_itemset}). Cette bordure a une spécification puisque à partir de celle ci nous pouvons retrouver tous les motifs rares en utilisant la contrainte de monotonie \cite{LAP07} (les sur-ensembles d'un motif rare sont rares).\bigskip \\
\noindent\fbox{\parbox{\linewidth}{
\emph{Pseudo code de L'algorithme :}\\
\'Etape 1: L'algorithme calcule les supports des motifs de taille
$k$.\\
\'Etape 2: Les motifs non fréquents sont enregistrés dans l'ensemble $\mathcal{MRM}$ avant de les élaguer.\\
\'Etape 3: Génération des itemsets de taille $k+1$ à partir de ceux
de taille $k$ et passage à l'étape 1. \\
L'algorithme se termine lorsqu'il ne reste aucun motif à générer.
}}
\begin{algorithm}[!htbp]
{
    \SetVline
    \setnlskip{-3pt}


\Donnees {\begin{enumerate}
   \item Contexte d'extraction $\mathcal{K}$.
  \item Seuil \emph{maxsupp}.
\end{enumerate} }

\Res{\begin{enumerate}
    \item Ensemble des motifs rares minimaux $\mathcal{MRM}$ et l'ensemble des motifs zéros.
\end{enumerate}}

    \Deb
{
    $\mathcal{C}_{1} \leftarrow \{l\_motif\}$\;
    $ i \leftarrow 1 $\;
    \Tq{$\mathcal{C}_{i}\neq\emptyset$}
    {
        SuppCount($\mathcal{C}_{i}$)\;
        $\mathcal{R}_{i} \leftarrow \{r\in \mathcal{C}_{i} |support(r) < \textit{minsupp}\} $\;
        $\mathcal{F}_{i} \leftarrow \{f\in \mathcal{C}_{i} |support(r) \geq \textit{minsupp}\} $\;
        $\mathcal{C}_{i+1} \leftarrow Apriori-Gen(\mathcal{F}_{i})$\;
        $i \leftarrow i + 1$\;
    }
    $\mathcal{MRM}\leftarrow\bigcup_{i} \mathcal{R}_{i}$ \;
    $\mathcal{GF}\leftarrow\bigcup_{i} \mathcal{F}_{i}$ \;
    \textbf{retourner} $\mathcal{MRM}$
}
}
  \caption{\textsc{Apriori-rare}}
  \label{algoApriori-rare}
\end{algorithm}

\begin{exemple} \textbf{Déroulement de l'algorithme \textsc{Apriori-rare}}\\
Soit le contexte $\mathcal{K}$ définie par le tableau \ref{DB1} (page \pageref{DB1}) pour $\textit{minsupp}=3$ ce qui signifie que $\textit{maxsupp}=2$. En première étape, un balayage de du contexte d'extraction permet l'extraction
des motifs de taille 1 munis de leur supports. Les motifs
dont le support est inférieur à un seuil \emph{minsupp} (dans notre exemple $minsupp=2$) vont être
ajoutés à l'ensemble $\mathcal{MRM}$ avant de les élaguer de la liste des candidats. Dans notre exemple, la première étape consiste à calculer les supports des motifs $A,B,C,D,E$ et élaguer les motifs non fréquents
(rares) c'est le cas du motif $D$ avec un support égal à $1$.
En deuxième étape, la génération des motifs de taille 2 à partir
de ceux de taille 1 est effectuée. À ce stade, l'ensemble $\mathcal{MRM}$ contient les motifs: ${D,AB,AE}$ munis de leurs supports $\{1,2,2\}$. L'algorithme s'arrête lorsque \emph{BCE} est généré, puisque c'est
le dernier motif à générer. En effet, l'algorithme retourne l'ensemble
des fréquents munis de leur supports et l'ensemble des motifs rares minimaux
munis de leurs supports. Dans notre exemple, l'ensemble
$\mathcal{MRM}$ contient $\{(D,1);(AB,2);(AE,2)\}$.

\end{exemple}

\begin{table}[tb]
  \centering
 \begin{tabular}{|c|c|c|c|c|c|}
\hline
 & A & B & C & D & E\tabularnewline
\hline
1 & 1 & 0 & 1 & 1 & 0\tabularnewline
\hline
2 & 0 & 1 & 1 & 0 & 1\tabularnewline
\hline
3 & 1 & 1 & 1 & 0 & 1\tabularnewline
\hline
4 & 0 & 1 & 0 & 0 & 1\tabularnewline
\hline
5 & 1 & 1 & 1 & 0 & 1\tabularnewline
\hline
\end{tabular}
 \caption{Contexte d'extraction}\label{base1}
\end{table}

\begin{table}[!htbp]
\fbox{\makebox[15cm]{
\begin{tabular}{|c|c|}
\hline
A & 3\tabularnewline
\hline
B & 4\tabularnewline
\hline
C & 4\tabularnewline
\hline
D & 1\tabularnewline
\hline
E & 4\tabularnewline
\hline
\end{tabular}\hfill{}\begin{tabular}{|c|c|}
\hline
A & 3\tabularnewline
\hline
B & 4\tabularnewline
\hline
C & 4\tabularnewline
\hline
E & 4\tabularnewline
\hline
\end{tabular}
}}

\caption{Calcul du support des motifs de taille 1 et élagage des motifs fréquents.}
\end{table}

\begin{table}[!htbp]
\fbox{\makebox[15cm]{
\begin{tabular}{|c|c|}
\hline
AB & 2\tabularnewline
\hline
AC & 3\tabularnewline
\hline
AE & 2\tabularnewline
\hline
BC & 3\tabularnewline
\hline
BE & 4\tabularnewline
\hline
CE & 3\tabularnewline
\hline
\end{tabular}\hfill{}\begin{tabular}{|c|c|}
\hline
AC & 3\tabularnewline
\hline
BC & 3\tabularnewline
\hline
BE & 4\tabularnewline
\hline
CE & 3\tabularnewline
\hline
\end{tabular}
}}
\caption{Calcul du support des motifs de taille 2 et élagage des motifs fréquents.}
\end{table}
\begin{table}[!htbp]
\fbox{\makebox[15cm]{
\begin{tabular}{|c|c|}
\hline
A & 3\tabularnewline
\hline
B & 4\tabularnewline
\hline
C & 4\tabularnewline
\hline
D & 1\tabularnewline
\hline
E & 4\tabularnewline
\hline
\end{tabular}
\hfill{}$\overset{(1)}{\Longrightarrow}$\hfill{}
\begin{tabular}{|c|c|}
\hline
A & 3\tabularnewline
\hline
B & 4\tabularnewline
\hline
C & 4\tabularnewline
\hline
E & 4\tabularnewline
\hline
\end{tabular}
\hfill{}$\overset{(2)}{\Longrightarrow}$\hfill{}
\begin{tabular}{|c|c|}
\hline
AB & 2\tabularnewline
\hline
AC & 3\tabularnewline
\hline
AE & 2\tabularnewline
\hline
BC & 3\tabularnewline
\hline
BE & 4\tabularnewline
\hline
CE & 3\tabularnewline
\hline
\end{tabular}
\hfill{}$\overset{(3)}{\Longrightarrow}$\hfill{}
\begin{tabular}{|c|c|}
\hline
AC & 3\tabularnewline
\hline
BC & 3\tabularnewline
\hline
BE & 4\tabularnewline
\hline
CE & 3\tabularnewline
\hline
\end{tabular}
}}
\begin{tabular}{|c|}
\hline
$(1)$ Calcul du support et élagage. \hspace{1.3cm} $(2)$ Génération des candidats de taille 2.\tabularnewline
(3) Élagage des motifs rares.\tabularnewline
\hline
\end{tabular}
\caption{Déroulement de l'algorithme \textsc{Apriori-rare}.}
\end{table}


À partir de l'ensemble $\mathcal{MRM}$, nous pouvons retrouver
tous les motifs rares (cette phase est assurée par l'algorithme \textsc{Arima}
\cite{LAP07}), puisque tous les sur-ensembles d'un motif rare le sont aussi.

L'algorithme \textsc{Arima} permet de retrouver tous
les motifs rares dans le contexte d'extraction à partir d'un ensemble des
rares minimaux tout en évitant les motifs zéros (pas de génération des sur-motifs d'un d'un motif zéro).

Dans ce qui suit, nous détaillons l'algorithme \textsc{Arima} qui utilise l'ensemble des $\mathcal{MRM}$s pour générer l'ensemble total des motifs rares munis de leurs supports.\\
\subsubsection*{L'algorithme ARIMA}
\textsc{Arima} est un algorithme qui prend en entrée, contrairement à tous les algorithmes de fouilles de données, un ensemble des motifs rares minimaux et fournit comme résultat l'ensemble total des motifs rares munis de leurs supports. L'algorithme \textsc{Arima}, n'a pas besoin ni d'un contexte d'extraction, ni d'un seuil \emph{minsupp} puisque son entrée n'est que le résultat d'un autre algorithme à l'instar d'\textsc{Apriori-rare}.

En première étape, \textsc{Arima} génère tous les sur-ensembles d'un motif rare minimal. Dans notre cas, l'algorithme commence par la génération des sur-ensembles
du plus petit itemset qui sont : \emph{{AD},{BD},{CD},{DE}}. Après un passage sur le contexte, leurs supports sont calculés. Puisque $\emph{\{BD\}}$, $\emph{\{DE\}}$ sont des motifs nuls,
alors ils vont être copiés dans la liste des $\mathcal{GMZ}$ (générateurs des motifs
zéros) et les motifs de supports non nuls vont être stockés dans un ensemble $\mathcal{R}$ qui servira à générer de nouveaux motifs. L'ensemble
$\mathcal{GMZ}$ sert à réduire le nombre de motifs à générer (on
ne génère pas les sur-motifs d'un motif nul). Dans notre exemple, tous
les sur-ensembles de \emph{\{BD\}} et \emph{\{DE\}} sont élagués.

Dans ce qui suit, nous présentons la deuxième algorithme de la littérature qui fournit l'ensemble total des motifs rares.

\begin{algorithm}[!tb]
{

    \SetVline
    \setnlskip{-3pt}


\Donnees {\begin{enumerate}
    \item Ensemble de motifs rares minimaux $\mathcal{MRM}$.
\end{enumerate} }

\Res{\begin{enumerate}
    \item Ensemble de motifs rares $\mathcal{MR}$.
\end{enumerate}}

    \Deb
{
    $\mathcal{GMZ} \leftarrow \emptyset $\;
    $S \leftarrow \{~tous~les~items~de~\mathcal{K}~\}$\;
    $i \leftarrow \{~longueur~de~plus~petit~\mathcal{MRM}~\} $\;
    $ \mathcal{C}_{i} \leftarrow { i - \mathcal{MRM} } $\;// Les motifs de $\mathcal{MRM}$ de taille $i$\\
    $ \mathcal{GMZ} \leftarrow \mathcal{GMZ} \bigcup_{z}\in \mathcal{C}_{i}|Supp(z) =0 $\;
    $\mathcal{R}_{i}\leftarrow\{ r \in \mathcal{C}_{i}|supp(r)>0\}$\;
    \Tq{$\mathcal{R}_{i}\neq \emptyset$}
    {
        \Tq{$\mathcal{R}_{i}(r)\neq \emptyset$}
        {
            $Cand \leftarrow \{ tous\;sur\;les~éléments\;de\;r\;utilisant\;S\}$\;
            \Tq{Cand(c)}
            {
                \Si{c~est~un~sous-motif~dans~$\mathcal{GMZ}$}
                {
                    Supprimer c de Cand\;
                }
                $\mathcal{C}_{i+1}  \leftarrow \mathcal{C}_{i+1} \bigcup Cand$\;
                $Cand \leftarrow \emptyset$\;
            }
        }
        $SupportCount(\mathcal{C}_{i}+1)$\;
        $\mathcal{GMZ} \leftarrow \{\mathcal{GMZ} \bigcup {z\in \mathcal{C}_{i} | supp(z) = 0}\}$;
        $\mathcal{R}_{i+1}  \leftarrow \{ r\in C_{i} | supp(r)>0\}$\;
        $i \leftarrow i+1 $\;
    }
    $\mathcal{MR}\leftarrow \bigcup \mathcal{R}_{i}$ \;
    \textbf{retourner} $\mathcal{MR}$
}
}
  \caption{\textsc{ARIMA}}
  \label{algoARIMA}
\end{algorithm}

\subsubsection{Algorithme \textsc{MRG-Exp}}
Input : Contexte d'extraction et un seuil \emph{minsupp}.

Output : l'ensemble des motifs rares minimaux $\mathcal{MRM}$.\\
L'algorithme \textsc{MRG-Exp}, comme son prédécesseur \textsc{Apriori-rare}, se base sur l'algorithme \textsc{Apriori} pour extraire les \emph{motifs rares minimaux} notés $\mathcal{MRM}$.
\begin{description}
\item [{Description}]~
\end{description}
L'algorithme \textsc{MRG-Exp} \cite{LAP07} est inspiré de
deux algorithmes \textsc{Apriori-rare} \cite{laszlo06} et \textsc{Apriori}
\cite{AIS93} tout en minimisant le nombre de motifs à générer et
à visiter. L'idée de \textsc{MRG-Exp} est la même qu'\textsc{Apriori-rare}
puisqu'il s'agit d'une exploitation en largeur de l'espace de recherche,
calcul de support des itemsets, génération des motifs basés sur
les motifs fréquents et conservation des motifs rares dans un ensemble
formant le résultat final avant de les élaguer. Cette dernière phase de l'algorithme
(l'élagage) est le seul point de différence entre les deux algorithmes.
L'algorithme \textsc{MRG-Exp} introduit un nouveau critère d'élagage basé sur le support d'un motif $I$ ainsi que les supports de tous
ces sous-ensembles directs. Deux critères d'élagage sont définis pour
minimiser les motifs visités par l'algorithme \textsc{MRG-Exp}. Le premier critère, fait partie
de l'algorithme \textsc{Apriori} et de son successeur \textsc{Apriori-rare;
}un critère qui est primordial dans la plupart des algorithmes de
fouille de données à savoir l'utilisation d'un seuil fixé d'avance
pour permettre de filtrer les motifs et d'utiliser la notion de monotonie. Le deuxième critère, est utilisé pour la première fois dans les algorithmes d'extraction des motifs rares et fréquemment dans les approches d'extraction des motifs fréquents à savoir l'utilisation de la notion de l'idéal d'ordre des générateurs minimaux. En effet,
l'algorithme commence par calculer le support des motifs de taille
1, élaguer les motifs non fréquents puis les conserver dans
un ensemble résultat, enfin générer les motifs de
taille 2 à partir de ceux de taille 1 non élagués et le même processus
se répète pour la génération des motifs de taille supérieure
à 2. Après la génération des motifs de taille $k+1$, à partir de ceux de taille k, l'algorithme teste
s'il y a un motif dont le support est égal à l'un de ces sous-ensembles
directs. Si c'est le cas, alors le motif sera élagué puisqu'il fait partie de la même classe d'équivalence de l'un de ces sous-ensembles et la génération de ses sur-ensembles n'a aucun intérêt.\bigskip \\
\noindent\fbox{\parbox{\linewidth}{
\emph{Pseudo code de l'algorithme \textsc{MRG-Exp} :}\\
\'Etape 1: Le calcul de support des motifs de taille $k$.\\
\'Etape 2: La copie des motifs rares dans un ensemble noté $\mathcal{RG}$ avant de
les élaguer. \\
\'Etape 3: La copie des motifs fréquents tout en respectant la notion de l'idéal d'ordre des générateurs minimaux.\\
\'Etape 4: Génération des motifs à partir de l'ensemble $\mathcal{FG}$ et retour
à l'étape 1.\\
L'algorithme se termine lorsqu'il ne reste aucun motif à générer.\\
}}

\begin{center}
\begin{table}[!htbp]
\fbox{
\hfill{}
\begin{tabular}{|c|c|}
\hline
A & 3\tabularnewline
\hline
B & 4\tabularnewline
\hline
C & 4\tabularnewline
\hline
D & 1\tabularnewline
\hline
E & 4\tabularnewline
\hline
\end{tabular}
\hfill{}
$\overset{(1)}{\Longrightarrow}$
\hfill{}
\begin{tabular}{|c|c|}
\hline
A & 3\tabularnewline
\hline
B & 4\tabularnewline
\hline
C & 4\tabularnewline
\hline
E & 4\tabularnewline
\hline
\end{tabular}
\hfill{}
$\overset{(2)}{\Longrightarrow}$
\hfill{}
\begin{tabular}{|c|c|}
\hline
AB & 2\tabularnewline
\hline
AC & 3\tabularnewline
\hline
AE & 2\tabularnewline
\hline
BC & 3\tabularnewline
\hline
BE & 4\tabularnewline
\hline
CE & 3\tabularnewline
\hline
\end{tabular}
\hfill{}
$\overset{(3)}{\Longrightarrow}$
\hfill{}
\begin{tabular}{|c|c|}
\hline
BC & 3\tabularnewline
\hline
BE & 4\tabularnewline
\hline
CE & 3\tabularnewline
\hline
\end{tabular}
\hfill{}
$\overset{(4)}{\Longrightarrow}$
\hfill{}
\begin{tabular}{|c|c|}
\hline
BCE & 3\tabularnewline
\hline
\end{tabular}
\hfill{}
$\overset{(5)}{\Longrightarrow}$
\hfill{}
$\textrm{\O}$.
}
\begin{tabular}{|c|}
\hline
$(1)$ Phase d'élagage. \hspace{0.5cm}$(2)$ Génération des candidats de taille 2.\tabularnewline
$(3)$ Phase d'élagage.   \hspace{0.22cm} $(4)$ Génération du candidat. \hspace{0.20cm} $(5)$ Aucun candidats à générer.\tabularnewline
\hline
\end{tabular}

\caption{Trace d'exécution de l'algorithme \textsc{MRG-Exp}.}
\end{table}
\end{center}

Nous remarquons la similitude de fonctionnement des deux algorithmes \textsc{MRG-Exp} et \textsc{Apriori-rare} tout en minimisant le nombre
des motifs à générer. Dans notre cas, l'algorithme calcule les supports
des motifs de taille 1 et ne garde que les motifs rares. Dans l'exemple, le motif \emph{\{D\}} va être copié dans l'ensemble des
$\mathcal{MRM}$ et $\emph{\{AB,AC,AE,BC,BE,CE\}}$ seront générés à partir de
ceux fréquents de taille 1. Un passage sur la base permet de connaître
le support de chacun d'eux. L'algorithme élague les motifs non fréquents, c'est à dire,
pas de génération de sur-ensembles d'un motif rare, élagage de motifs dont le support est égal au minimum de ces sous-ensembles directs. En effet, $\emph{\{AB,AE\}}$ sont élagués puisqu'ils sont rares et $\emph{\{AC\}}$
est élagué puisque son support est égal à celui de $\emph{A}$. De même, $\emph{BCE}$ est élagué puisque son support est égal au support de \emph{BC}.
\begin{algorithm}
{
    \SetVline
    \setnlskip{-3pt}


\Donnees {\begin{enumerate}
    \item Contexte d'extraction.
    \item Seuil \emph{minsupp}.
\end{enumerate} }

\Res{\begin{enumerate}
    \item Ensemble des motifs rares minimaux $\mathcal{MRM}$.
\end{enumerate}}

    \Deb
{
    $\mathcal{CG}_{l} \leftarrow \{  l-itemsets\} $\;
    $SupportCount(\mathcal{CG}_{l})$\;
    \Tq{$\mathcal{CG}_{l}(c)\neq \emptyset $}
    {
        $c.pred\_supp \leftarrow \emptyset.supp $ \;
        \eSi{$c.pred\_supp = c.pred\_supp$}
        {
        $c.key \leftarrow faux $\;
        }
        {
        $c.key \leftarrow vrai $ \;
        }
    }
    $\mathcal{RG}_{l}\leftarrow\{r\in \mathcal{CG}_{l}|(r.key = vrai) \ et \ (r.supp< \textit{minsupp})\}$\;
    $\mathcal{FG}_{l}\leftarrow\{f\in \mathcal{CG}_{l}|(r.key = vrai) \ et \ (r.supp\geq \textit{minsupp})\}$\;
    \Tq{$\mathcal{CG}_{l}(c)\neq \emptyset $}
    {
        $\mathcal{CG}_{i+1}\leftarrow GenCandidate(\mathcal{FG}_{i})$\;
        $SuppCount(\mathcal{\mathcal{CG}}_{i+1})$\;
        \Tq{$\mathcal{CG}_{l+1}(c)\neq \emptyset $}
        {
            \Si{$c.pred\_sup \neq c.pred\_supp$}
            {
                \eSi{c.Supp<\textit{minsupp}}
                {
                    $\mathcal{RG}_{i+1} \leftarrow \mathcal{RG}_{i+1} \bigcup \{c\}$ \;
                }
                {
                    $\mathcal{FG}_{i+1} \leftarrow \mathcal{FG}_{i+1} \bigcup \{c\} $\;
                }
            }
        }
    }

    $\mathcal{GF}\leftarrow\bigcup_{i}\mathcal{FG}_{i}$ // Générateur fréquents \;
    $\mathcal{MRM}\leftarrow\bigcup_{i}\mathcal{RG}_{i}$ // Générateur des rares minimaux \;
    \textbf{retourner} $\mathcal{MRM}$
}
}
  \caption{\textsc{MRG-Exp}}
  \label{algoMRG-EXP}
\end{algorithm}

\subsubsection{Algorithme \textsc{Afrim}}

Input : Base de données et seuil \emph{minsupp}.

Output : Ensemble des motifs rares.

\begin{description}
\item [{Description}]~
\end{description}
L'originalité de \textsc{Afrim} \cite{AMV07} réside au niveau d'exploitation du treillis. L'algorithme exploite l'ensemble des itemsets
à partir de l'itemset le plus grand en terme de nombre
d'items qui le compose. Dans son travail, Adda \textit{et al.}
ont présenté les avantages d'un parcours descendant (de haut en bas, c'est-à-dire par taille décroissante des candidats) à partir du motif admettant le plus grand nombre d'items dans sa
cardinalité vers l'ensemble vide pour qu'il puisse
utiliser la notion de monotonie qui assure l'affirmation suivante ;
\emph{Si un motif X est fréquent alors tous les motifs qui y sont inclus
le sont}. Ainsi, c'est l'unique critère d\textquoteright{}élagage de l'algorithme pour réduire le nombre de motifs à visiter. Puisque le parcours était
descendant alors l'algorithme génère les itemsets d'un niveau $k$ à partir
des itemsets de niveau $k+1$, les motifs fréquents seront élagués ainsi que tous ces sous-ensembles.\\

L'avantage de \textsc{Afrim} réside dans le fait que cet algorithme bénéficie d'un critère d'élagage qui n'était pas utilisé dans ses prédécesseurs
(\textsc{Apriori-rare} et \textsc{MRG-Exp}). Son critère d'élagage se base sur
une règle d'anti-monotonie : ``\textit{tous les sous-ensembles d'un motif fréquent
le sont}''. En se basant sur cette règle d'anti-monotonie, \textsc{Afrim} élague
à chaque itération tous les motifs fréquents ainsi que leurs descendants.
Théoriquement, l'idée est très intéressante. En effet, c'est plus logique
d'élaguer les motifs qui ne vont pas apparaître dans le résultat que
d'élaguer les motifs qui forment le résultat final et d'appliquer un
deuxième algorithme pour les récupérer.\bigskip \\
\noindent\fbox{\parbox{\linewidth}{
\emph{Pseudo code de l'algorithme :}\\
Étape 1 (initialisation): l'algorithme commence par calculer le support
de l'itemset de plus grande taille $n$ (en général il est nul), puis il
génère les motifs de tailles ($n-1$) et calcule leurs supports.\\
Étape 2 : tester et élaguer les motifs dont le support est supérieur à
\emph{minsupp} (motif fréquent) ainsi que tous ses sous-ensembles.\\
Étape 3 : génération des motifs de taille $k$ à partir de ceux de
taille ($k+1$) par l'intersection des motifs. Passer à l'étape 2.\\
L'algorithme s'arrête lorsqu'il ne reste aucun motif à générer.
}}

\begin{table}[!htbp]
\fbox{\makebox[15.8cm]{
\begin{tabular}{|c|c|}
\hline
ABCD & 0\tabularnewline
\hline
ABCE & 2\tabularnewline
\hline
ABDE & 0\tabularnewline
\hline
ACDE & 0\tabularnewline
\hline
BCDE & 0\tabularnewline
\hline
\end{tabular}\hfill{}\begin{tabular}{|c|c||c|c|}
\hline
ABC & 2 & ACE & 2\tabularnewline
\hline
ABD & 0 & BCE & 3\tabularnewline
\hline
ACD & 1 & ADE & 0\tabularnewline
\hline
BCD & 0 & BDE & 0\tabularnewline
\hline
ABE & 2 & CDE & 0\tabularnewline
\hline
\end{tabular}\hfill{}\begin{tabular}{|c|c||c|c|}
\hline
AB & 2 & CD & 1\tabularnewline
\hline
AC & 3 & AE & 2\tabularnewline
\hline
AD & 1 & DE & 0\tabularnewline
\hline
BD & 0 & - & \tabularnewline
\hline
\end{tabular}\hfill{}\begin{tabular}{|c|c|}
\hline
D & 1\tabularnewline
\hline

\end{tabular}
}}
\caption{Trace d'exécution de l'algorithme \textsc{Afrim}.}
\label{tabaf}

\end{table}

\begin{exemple} \textbf{Déroulement de l'algorithme \textsc{Afrim}}\\
Nous utilisons toujours le même contexte d'extraction cité au début dans le tableau \ref{base1}.
Dans l'exemple du tableau \ref{tabaf}, l'algorithme commence par l'itemset \emph{\{ABCDE\}}
de support nul, ensuite l'algorithme génère cinq motifs \emph{\{ABCD,ABCE,ABDE,ACDE,BCDE\}}
qui sont obtenus en supprimant à chaque fois un item à partir de
\emph{\{ABCDE\}}. Puisque les cinq motifs sont non fréquents ($minsupp=3$)
alors on conserve tous les éléments de cet ensemble. Les 5 itemsets vont donner
naissance à 10 motifs par l'intersection deux à deux. Par exemple,
\emph{ABC} est obtenu grâce à l'intersection de \emph{\{ABCD\}} et
\emph{\{ABCE\}}. Après un passage sur le contexte d'extraction, nous trouvons
les supports des différents motifs (tableau 2 à gauche), l'unique
itemset fréquent c'est \emph{BCE} alors l'algorithme élague ce dernier
ainsi que tous ces sous-ensembles et génère les motifs de
taille 2 à partir de ceux de taille 3. L'algorithme
\textsc{Afrim} génère à cette phase seulement 7 motifs au lieu de
dix puisque les motifs \emph{\{BC,BE,CE\}} sont élagués. En dernière
étape, l'algorithme génère seulement D et calcule son support. S'il n'y a aucun motif à générer, l'algorithme retourne l'ensemble des motifs rares munis de leurs supports.

Dans la figure \ref{tabaf}, nous présentons les différentes étapes d'exécution de l'algorithme \textsc{Afrim}.
\end{exemple}

\begin{algorithm}[!tb]
{

    \SetVline
    \setnlskip{-3pt}


\Donnees {\begin{enumerate}
    \item Un contexte d'extraction
    \item Seuil \textit{maxsupp}
    \item Seuil de la cardinalité maximum des motifs à extraire \emph{maxc}.
\end{enumerate} }

\Res{\begin{enumerate}
    \item L'ensemble des motifs rares dont la cardinalité est inférieure à \emph{maxc} noté $\mathcal{GR}$.
\end{enumerate}}

    \Deb
{
    $I\leftarrow \{ l'ensemble~des~motifs~du~contexte~\mathcal{K}\}$\;
    $ N \leftarrow |I| $\;
    $\mathcal{F}_n \leftarrow CandTest(Fn,D) $ \;
    $\mathcal{F}_n \leftarrow \{ F\backslash i;~f\in Fn~et~i\in I\} $ \;
    $\mathcal{F}_{n-1}\leftarrow CandTest(Fn-1,D) $ \;
    $k\leftarrow n-2$\;

    \Tq{$F_{k+1}\neq \emptyset $}
    {
        $\mathcal{F}_{k}\leftarrow GenCondidat(\mathcal{F}_{k+1})$ // Génération des $k$-candidats par l'intersection de k+1 condidats\;
        $\mathcal{GR}\leftarrow CandTest (\mathcal{F}_{k},\mathcal{K})$// Conservation des motifs rares\;
        $k$ $\leftarrow$ $k$ - $1$\;
    }
    \textbf{retourner} $\mathcal{GR}$
}
}
  \caption{\textsc{Afrim}}
  \label{algoAfRIM}
\end{algorithm}

\subsubsection{Algorithme \textsc{Minit}}

Input : Un contexte d'extraction $\mathcal{K}$, vecteur $V$ de taille $L$, seuil $\textit{minsupp}$, seuil $\textit{maxc}$.

Output : L'ensemble des motifs rares minimaux, dont la cardinalité
est inférieure ou égale à $\textit{maxc}$.

\begin{description}
\item [{Description}]~
\end{description}
\textsc{Minit} \cite{Haglin07} est un algorithme appartenant à la famille des algorithmes ``Diviser pour Régner''. Cette stratégie se base sur la décomposition du problème en sous-problèmes pour les rendre plus simples à résoudre. De ce fait, l'algorithme $\textsc{Minit}$ décompose le contexte initial en sous-contextes, traite les sous-contextes et re-formule le résultat final à partir des résultats locaux des sous-contextes. L'algorithme prend en paramètre un contexte
d'extraction $\mathcal{K}$, un vecteur $V$ contenant des valeurs booléennes et la taille des motifs
rares minimaux extraite noté \emph{maxc}. Le vecteur $V$ contient $m$ valeur où $m$ est le nombre
d'items dans le contexte $\mathcal{K}$, $V$ servira à marquer les
items visités. En première étape, l'algorithme
ordonne les motifs de taille $1$ de manière ascendante. La liste \emph{L}
des motifs ordonnés est traitée élément par élément, chaque élément
\emph{x} de \emph{L} sert à produire un sous contexte de $\mathcal{K}$ où \emph{x} se trouve dans toutes les transactions. Autrement dit, l'algorithme conserve seulement les transactions contenant le motif \emph{x}. À ce stade, l'algorithme
extrait toutes les transactions contenant \emph{x}, et à chaque création d'un nouveau contexte, l'algorithme
marque l'item qui a donné naissance à ce contexte,
le marquage se fait au niveau du vecteur $V$. Chaque valeur de $V$ correspond à un item parmi les items du contexte. Au début, les valeurs du vecteur valent 0,
le passage d'une valeur 0 à 1 à une position $i$ du vecteur
signifie qu'un sous-contexte, contenant le motif correspondant à
l'emplacement $i$ dans le vecteur, a été extrait. Autrement dit, l'extraction
des transactions d'un contexte qui contiennent un motif \emph{A} entraîne
la modification de la valeur 0 par 1 à la position numéro $1$ du vecteur
$V$. L'algorithme \textsc{Minit} est récursif. Il s\textquoteright{}arrête lorsque la variable \emph{maxc} est égale à 1, lorsqu'il s'arrête il retourne l'ensemble des motifs présents au dernier sous-contexte.

Dans la dernière étape, l'algorithme concatène les
items marqués dans le vecteur $V$ avec la liste des itemsets retournés lorsque \emph{maxc} est égal à 1.
Le résultat sera l'ensemble des motifs rares minimaux qui ont une cardinalité inférieur ou égale à $\textit{maxc}$.\\
Dans ce qui suit, nous présentons un exemple qui illustre les différents étapes d'exécution de l'algorithme.\\

\begin{table}[!htbp]
\begin{tabular}{|p{15.8cm}|}
\hline
\smallskip
\begin{tabular}{|c|c|}
\hline
A & 3\tabularnewline
\hline
B & 4\tabularnewline
\hline
C & 4\tabularnewline
\hline
D & 1\tabularnewline
\hline
E & 4\tabularnewline
\hline
\end{tabular}\hfill{}\begin{tabular}{|c|c|}
\hline
A & 3\tabularnewline
\hline
B & 4\tabularnewline
\hline
C & 4\tabularnewline
\hline
E & 4\tabularnewline
\hline
\end{tabular}
\smallskip
\tabularnewline
\hline
\smallskip
\begin{tabular}{|c|}
\hline
Base $D({A})$\tabularnewline
\hline
\hline
ACD\tabularnewline
\hline
ABCE\tabularnewline
\hline
ABCE\tabularnewline
\hline
\end{tabular}\hfill{}\begin{tabular}{|c|}
\hline
Viable item\tabularnewline
\hline
\hline
B,E\tabularnewline
\hline
B,E\tabularnewline
\hline
\end{tabular}
\smallskip
\tabularnewline\hline
\end{tabular}
\caption{Trace d'exécution de l'algorithme \textsc{Minit}.}

\label{trace_minit}
\end{table}

\begin{exemple} \textbf{Déroulement de l'algorithme \textsc{Minit}}\\
Soit le contexte d'extraction cité par le tableau \ref{base1} avec les seuils $\textit{minsupp}=3$ et $\textit{maxc}=2$.
Au départ, les paramètres d'entrée de l'algorithme \textsc{Minit} sont les suivants : un contexte de données (\textit{cf.} Tableau \ref{base1}), \emph{minsupp} fixé à 3 et \emph{maxc} fixé à 2. L'algorithme commence par trier l'ensemble des motifs de taille
1 d'une manière décroissante selon le nombre d'apparition (support). Dans notre cas, $\textit{maxc}=2$. Alors l'algorithme extrait les motifs rares de cardinalité égale à 2. Ainsi, le motif $D$ sera supprimé de
la liste des motifs puisqu'il ne donnera pas naissance à un motif
infréquent minimal de taille 2. La deuxième étape consiste à repérer
l'item de la tête de liste qui est A et à générer le contexte associé
à ce dernier (\textit{cf.} Tableau \ref{trace_minit}), un appel récursif aura lieu à cette
étape avec les données suivantes : contexte $Base({A})$, \emph{minsupp}
égal à 3 et \textit{maxc} décrémenté de 1. Puisque la variable $\textit{maxc}$ vaut 1 alors l'algorithme s'arrête
et renvoie l'ensemble des motifs apparaissant avec $A$ tout en
supprimant les items déjà élagués (le cas de $D$) ainsi que les items qui sont apparus dans toutes les transactions (cas de l'item $C$). D'où, l'algorithme retourne seulement $\{B,E\}$. La troisième étape de l'algorithme assure la génération des motifs minimaux infréquents
de taille \emph{maxc}. Cette dernière étape se fait tout simplement en ajoutant
l'item $A$ à l'ensemble $\{B,E\}$ pour donner $\{AB,AE\}$ et nous pouvons
vérifier que ce sont les seuls motifs rares minimaux de taille 2.

\end{exemple}

\section{Étude comparative des algorithmes existants }
Nous avons présenté dans les sections précédentes les différentes
approches d'extraction des motifs rares à partir d'un
contexte formel, ainsi que les problèmes inhérents à chacune de ces
approches. Cependant, notre comparaison concernera
les trois premiers algorithmes cités dans l'état de
l'art. Ces trois algorithmes se ressemblent sur plusieurs
points, à savoir : l'inspiration de ces approches de l'algorithme  \textsc{
Apriori}, l'exploitation de l'espace de recherche {}``treillis'' d'une
manière horizontale {}``en largeur d'abord'' ainsi que l'utilisation des même paramètres d'entrée et la production des mêmes résultats.
Ces trois algorithmes se différent sur quelques points, à savoir :\\
Le sens du parcours, par exemple, les deux premiers suivent le sens classique (de bas en haut c'est-à-dire par taille croissante des motifs) et le troisième suit le sens inverse. En effet,
le choix de sens du parcours ainsi que la stratégie utilisée par les algorithmes d'extraction des motifs rares ont un effet sur les performances de ces derniers. Dans la prochaine section, nous montrons l'intérêt de ces critères par rapport aux benchmarks utilisés.\\
\section{Problèmes liés à l'extraction des motifs rares }
L'utilisation des bases benchmark qui sont à l'origine des contextes dédiés
aux algorithmes d'extraction des motifs fréquents est l'un des problèmes trouvés par les algorithmes d'extraction des motifs rares. En effet, l'ensemble des motifs de supports non nuls, dans de tels contextes, ne forment en général qu'un quart du treillis. De plus,
la partie des motifs nuls ou zéros, qui fait partie du résultat des algorithmes d'extraction des motifs rares, ne sert pratiquement pas à grand chose (puisque c'est la partie contenant les motifs de supports non nuls qui contient les motifs à la fois rares et fréquents) mais elle représente la grande partie du treillis. Nous avons deux façons de parcours, soit en suivant le sens ascendant ou bien le sens descendant. Si nous favorisons le parcours ascendant du treillis, nous sommes obligés de visiter des noeuds fréquents non intéressants.
L'exploitation inverse du treillis a été utilisée par l'algorithme de Adda \textit{ et al.} et les expérimentations
ont prouvé qu'un tel algorithme ne peut tourner que sur des super-machines à cause de la génération des trois quarts du treillis afin d'arriver à une partie exploitable.

La constatation des courbes \cite{AIS93,Haglin07,laszlo06} montre la relativité de chaque méthode par rapport aux contextes.
Aucun de ces algorithmes n'a prouvé ses performances
sur tous les benchmarks. L'explication est simple,
il existe une très forte liaison entre le sens d'exploitation
de treillis de Galois et la position de la bordure des motifs rares.
Nous constatons qu'\textsc{Apriori-rare} et \textsc{MRG-Exp}
donnent de bons résultats sur des contextes épars, par contre l'algorithme
\textsc{Afrim} est moins efficace dans ces contextes mais il prouve sa rapidité
dans les contextes denses. Nous ne pouvons pas comparer l'algorithme
\textsc{Minit} avec les trois autres algorithmes cités puisque ni le fonctionnement,
ni le raisonnement, ni le résultat sont les mêmes dans cet algorithme.
Le concepteur de l'algorithme \textsc{Minit}, n'a pas comparé son algorithme avec les algorithmes de la littérature et
il a mené ses expérimentations en modifiant les seuils \emph{maxsupp}
et \emph{maxc}. D'autre part, \textsc{Minit} ne fournit
en sortie qu'une partie de la liste des $\mathcal{MRM}$ dont la cardinalité
des motifs est égale à \emph{maxc}. Nous pouvons juste mentionner la complexité
exponentielle de \textsc{Minit}.\\
Pour avoir la totalité des motifs rares d'un contexte d'extraction à l'aide de l'algorithme \textsc{Minit}, il faut l'appliquer plusieurs fois sur le même contexte, tout en variant la valeur \emph{maxc}. Le tableau \ref{Tabcomp} représente une comparaison entre les différents algorithmes.

\begin{table}[!htbp]
\parbox{16.cm}{\hspace{-0.8cm}
\begin{tabular}{|l|l|l|}
\hline
\textbf{Méthode} & \textbf{Avantages} & \textbf{Inconvénients}\tabularnewline
\hline\hline
\textbullet{} \textsc{Apriori-rare} & \begin{tabular}{l}
- Simplicité du code\tabularnewline
- Réutilisation du code d'Apriori\tabularnewline
\end{tabular} & \begin{tabular}{l}
- Parcours inutiles \tabularnewline
des motifs fréquents\tabularnewline
- Pas d'information sur \tabularnewline
les supports des motifs \tabularnewline
rares.\tabularnewline
- La nécessité d'un deuxième \tabularnewline
algorithme pour fournir \tabularnewline
l'ensemble total \tabularnewline
des motifs rares\tabularnewline
\end{tabular}\tabularnewline
\hline
\textbullet{} \textsc{MRG-Exp} & \begin{tabular}{l}
- Réutilisation de code d'Apriori\tabularnewline
- Réduction du nombre de motifs \tabularnewline
visités a travers un \tabularnewline
nouveau critère d'élagage\tabularnewline
\end{tabular} & \begin{tabular}{l}
- Parcours inutile des \tabularnewline
motifs fréquents\tabularnewline
- L'algorithme fournit \tabularnewline
l'ensemble des motifs rares \tabularnewline
sans les valeurs des supports\tabularnewline
- La nécessité d'un deuxième \tabularnewline
algorithme pour fournir \tabularnewline
l'information sur les \tabularnewline
supports des motifs rares.\tabularnewline
\end{tabular}\tabularnewline
\hline
\textbullet{} \textsc{Minit} & \begin{tabular}{l}
- Utilisation des sous contextes \tabularnewline
pour réduire la taille des \tabularnewline
données dans la M.C.\tabularnewline
- Réduction de nombre \tabularnewline
d'itérations par rapport aux\tabularnewline
autres méthodes.\tabularnewline
\end{tabular} & \begin{tabular}{l}
- Le résultat de l'algorithme\tabularnewline
est très réduit puisqu'il \tabularnewline
faut donner la taille du motif\tabularnewline
qu'on souhaite extraire\tabularnewline
- Dans le cas pratique, \tabularnewline
il faut exécuter l'algorithme\tabularnewline
sur le même contexte d'extraction\tabularnewline
plusieurs fois \tabularnewline
(pour chaque \emph{maxc}) \tabularnewline
pour avoir la totalité de $\mathcal{MRM}$.\tabularnewline
\end{tabular}\tabularnewline
\hline
\end{tabular}}
\caption{Tableau comparatif des algorithmes d'extraction l'ensemble des motifs rares.} \label{Tabcomp}
\end{table}

Dans la littérature, aucun travail n'a était mené pour extraire une représentation concise exacte des motifs rares. Dans la section suivante, nous focalisons
notre étude sur les représentations concises exactes des motifs fréquents afin de pouvoir tirer quelques notions qui peuvent être utiles pour la conception
des nouvelles représentations concises des motifs rares.
\section{Représentations concises exactes des motifs fréquents}
Plusieurs travaux se sont intéressés à définir des représentations concises exactes. Dans cette section, nous présentons ses
représentations concises afin de voir si nous pouvons les adapter dans le domaine des motifs rares.\\
Les représentations concises des motifs fréquents sont :
\begin{itemize}
   \item La représentation basée sur les motifs fermés fréquents \cite{thesepasq},

    \item La représentation basée sur les motifs non dérivables fréquents \cite{cald02},

    \item La représentation basée sur motifs essentiels fréquents \cite{casali05}.

    \item La représentation basée sur les générateurs minimaux \cite{Liu08}.
\end{itemize}
Dans ce qui suit, nous détaillons ces représentations concises exactes et nous montrons lesquelles parmi celles qui peuvent être utilisées
dans le domaine des motifs rares.
\subsection{Représentation concise basée sur les fermés fréquents}
Dans plusieurs bases réelles, un ensemble d'itemsets
caractérise le même ensemble de transactions et
possède le  même support, ce qui constitue une forme de
redondance. Afin de l'éliminer, Pasquier \textit{et al.} ont
pensé à appliquer l'opérateur de fermeture
$h$ sur les itemsets fréquents afin de regrouper l'ensemble des itemsets fréquents caractérisant le même ensemble de transactions au sein du même
itemset. Ainsi, la représentation par les fermés
fréquents, notée dans la suite de ce mémoire par
 $\mathcal{MFF}$, est apparue.
\begin{exemple}
Soit le contexte d'extraction énoncé dans la figure \ref{DB1} (page \pageref{DB1}). Pour $\textit{minsupp}=2$, l'ensemble des motifs fermés fréquents
est illustré dans le tableau \ref{iff}.
\end{exemple}
\begin{table}
\begin{center}
\begin{tabular}{|c|c|}
  \hline
  Motif fermé fréquent & Support\\
  \hline
  \textsc{BCE} & 3 \\
  \textsc{AC} & 3 \\
  \textsc{BE} & 4 \\
  \textsc{C} & 4 \\
  \hline
\end{tabular}
\end{center}
\caption{La représentation basée sur les $\mathcal{MFF}$ associés au contexte $\mathcal{K}$ pour $\textit{minsupp}$ $=$ 2.}\label{iff}
\end{table}
\begin{theorem}\cite{thesepasq} \textbf{Représentation basée sur les motifs fermés fréquents}

\medskip
L'ensemble des motifs fermés fréquents munis de leurs
supports respectifs forme une représentation concise exacte de
l'ensemble des motifs fréquents.
\end{theorem}
\subsection{Représentation concise basée sur les non-dérivables fréquents}
Dans cette section, nous présentons les motifs non-dérivables fréquents. Il faut signaler que les motifs non-dérivables vont être notés par $\mathcal{MNDF}$.
Cette représentation se base sur des fondements mathématiques \textit{e.g.}, les règles de déduction. Dans ce mémoire nous ne présentons pas ces
formules mathématiques puisqu'ils sont présentées et démontrées par Calders \emph{et al.} \cite{cald02,calders03}.
L'originalité de cette représentation réside dans le fait qu'elle permet de déduire le support d'un itemset $I$ à partir de ses sous-ensembles stricts.
Le motif que son support peut être déduit, est appelé \emph{motif dérivable} ou bien itemset dérivable. Ce motif ne figure pas dans l'ensemble résultat
puisqu'il peut être trouvé grâce à ses sous-ensembles stricts.
\begin{exemple}
Considérons le contexte d'extraction énoncé dans la figure \ref{DB1} (page \pageref{DB1}).
Pour $\textit{minsupp}=2$, l'ensemble des motifs dérivables fréquents est illustré dans le tableau \ref{indf}.
Afin de simplifier l'écriture, nous avons mis les détails de calcul des supports des motifs dérivables à l'aide des règles de déduction.
Après une évaluation de systèmes des règles de déduction, nous avons pu trouver le support du motif $ABC$ à l'aide de deux règles suivantes :\\
La règle associée à $AC$, noté $\mathcal{R}(AC)$, donne : $Supp(ABC) \leq 2$\\
La règle associée à $AB$, noté $\mathcal{R}(AB)$, donne : $Supp(ABC)\geq 2$\\
Ceci signifie que $ABC$ est un motif dérivable de support 2.
\end{exemple}
\begin{table}
\begin{center}
\begin{tabular}{|c|c||c|c|}
  \hline
  Motif non dérivable & Support& Motif non dérivable & Support\\
  \hline
  \textsc{A}&3&\textsc{AB}&2\\
  \textsc{B}&4&\textsc{AC}&3\\
  \textsc{C}&4&\textsc{AE}&2\\
  \textsc{E}&4&\textsc{BC}&3\\
  -&-&\textsc{BE}&4\\
  -&-&\textsc{CE}&3\\
  \hline
\end{tabular}
\end{center}
\caption{La représentation basée sur les $\mathcal{MNDF}$ associés au contexte $\mathcal{K}$ pour \textit{minsupp} = 2.}\label{indf}
\end{table}
\begin{theorem} \cite{cald02} \textbf{Représentation par les itemsets non dérivables fréquents }\\
Soient $\mathcal{K}$ un contexte d'extraction, \textit{minsupp} un
seuil minimal du support  défini par l'utilisateur et
\textsc{MND}$($$\mathcal{K}$,\textit{minsupp}$)$ l'ensemble
défini comme suit:
\begin{displaymath}
\textsc{MND}(\mathcal{K},\textit{minsupp}) := \{(I,Supp(I))\mid l_I
\neq u_I\}
\end{displaymath}
\textsc{MND}$($$\mathcal{K}$,\textit{minsupp}$)$ est une
représentation concise exacte de l'ensemble des itemsets
fréquents. $l_I$ et $u_I$ sont les limites de support de $I$.
\end{theorem}
Dans ce qui suit, nous présentons la dernière représentation concise exacte à savoir celle basée sur les motifs essentiels fréquents.
\subsection{Représentation concise basée sur les  essentiels fréquents}
La notion des motifs essentiels fréquent a été introduite par Casali \emph{et al.} dans \cite{casali05}. Dans la suite du mémoire, nous adoptons
les notations suivantes :
\begin{itemize}
  \item $\mathcal{MEF}$ : Motifs essentiels fréquents.
  \item $\mathcal{RCEMF}$ : Représentation concise exacte des motifs essentiels fréquents.
\end{itemize}

Avant d'entamer la partie descriptive des motifs essentiels nous définissons la notion de motif essentiel.\\

\begin{definition}\label{définitionmotifessentiel} \textbf{Motif essentiel}

\medskip
Soit $\mathcal{K}=(\mathcal{O},\mathcal{I},\mathcal{R})$ un contexte
d'extraction, et $I \in \mathcal{I}$. $I$ est un itemset essentiel
si et seulement si :
\begin{displaymath}
Supp(\vee I)\mbox{ } > \mbox{ }\max\{Supp(\vee I\backslash i)\mid
i\in I\}
\end{displaymath}
\end{definition}
L'originalité de cette représentation découle de l'utilisation de support disjonctif. Cette représentation se base sur les motif essentiels pour réduire l'ensemble des motifs fréquents. Un motif $I$ est dit essentiel si son support conjonctif est différent des
supports disjonctifs de ses sous-ensembles stricts.
\begin{exemple}
Soit la base de transaction présentée par le tableau \ref{DB1} (page \pageref{DB1}). Pour $\textit{minsupp}=2$, la représentation basée sur
les essentiels fréquents est donnée par le tableau \ref{ief}. Nous remarquons que le motif $AC$ n'est pas un essentiel puisque $Supp(AC) = Supp(A)=3$.
Cependant, $AB$ est considéré comme un motif essentiel puisque $Supp(AB) \neq Supp(A)$ et $Supp(AB) \neq Supp(B)$.
\end{exemple}
\begin{table}
\begin{center}
\begin{tabular}{|c|c|}
  \hline
  Motif essentiel fréquent & Support disjonctif\\
  \hline
  \textsc{A} & 3 \\
  \textsc{B} & 4 \\
  \textsc{C} & 4 \\
  \textsc{E} & 4 \\
  \textsc{AB} & 5 \\
  \textsc{AE} & 5 \\
  \textsc{BC} & 5 \\
  \textsc{DE} & 5 \\
  \hline
  \hline
  Motif maximal fréquent & Support conjonctif\\
  \hline
   \textsc{ABCE} & 2 \\
  \hline
\end{tabular}
\end{center}
\caption{La représentation basée sur les $\mathcal{MEF}$ associés au contexte $\mathcal{K}$ pour \textit{minsupp} = 2.}\label{ief}
\end{table}

\begin{theorem}
\textbf{Représentation basée sur les essentiels fréquents}

\medskip
Soit $\mathcal{MFM}$ l'ensemble des motifs
maximaux fréquents. Alors, les deux ensembles  $\mathcal{MEF}$ et $\mathcal{MFM}$ forment une représentation concise exacte des motifs fréquents.\\
\begin{center}
$\texttt{RCEMF} = \mathcal{MEF}\mbox { }\cup\mbox { } \mathcal{MFM}$
\end{center}
\end{theorem}
\subsection{Représentation concise basée sur les générateurs minimaux}
La notion des générateurs minimaux fréquents a été introduite par  Liu \emph{et al.} dans \cite{Liu08}. Dans la suite du mémoire, nous adoptons la notation $\mathcal{GMF}$ pour designer l'ensemble des générateurs minimaux fréquents. Dans les bases réelles, plusieurs itemsets caractérisent le même ensemble de transaction et possèdent le même support. Afin de bénéficier de cette notion de classe d'équivalence, Liu \emph{et al.} ont pensé à conserver les motifs minimaux de chaque classe d'équivalence \textsc{GMF}s. Cependant, cet ensemble ne représente pas une représentation concise exacte des motifs fréquents. Ainsi, l'augmentation de cet ensemble par celui des motifs de la bordure positive constitue une représentation concise exacte des motifs fréquents.

\begin{exemple}
Soit le contexte d'extraction énoncé dans la figure \ref{base1} (page \pageref{base1}). Pour $\textit{minsupp}=2$, l'ensemble des générateurs minimaux ainsi que l'ensemble des motifs de la bordure positive sont illustrés dans le tableau \ref{igf}.
\begin{table}
\begin{center}
\begin{tabular}{|c|c|}
  \hline
  Générateur minimal fréquent & Support\\
  \hline
  \textsc{A} & 3 \\
  \textsc{B} & 4 \\
  \textsc{C} & 4 \\
  \textsc{BC} & 3 \\
  \textsc{BE} & 3 \\
  \textsc{CE} & 4 \\
  \hline
  \hline
  \multicolumn{2}{|c|}{Motif de la bordure positive}\\
  \hline
  \multicolumn{2}{|c|}{\textsc{AC}; \textsc{BCE}}\\
  \hline
\end{tabular}
\end{center}
\caption{La représentation basée sur les $\mathcal{GMF}$ associés au contexte $\mathcal{K}$ pour \textit{minsupp} = 2.}\label{igf}
\end{table}
\end{exemple}

\begin{theorem}
L'ensemble des générateurs minimaux fréquents munis de leur supports, augmenté par l'ensemble des motifs de la bordure positive, forme une représentation concise exacte des motifs fréquents.
\[\texttt{RCEMF}\equiv \mathcal{GMF} \cup \mathcal{MFM}\]
\end{theorem}

\subsection{Discussion}
Après cette étude des représentations concises exactes, nous déduisons que seules les deux représentations basées respectivement sur les motifs fermés et sur les générateurs minimaux peuvent être intéressantes dans l'exploitation des motifs rares puisqu'elles utilisent la notion de classe d'équivalence. En effet, la localisation détermination du statut d'un motif --fermé ou générateur minimal-- nécessite simplement de comparer son support à celui de ses sur-ensembles immédiats ou ses sous-ensemble immédiats respectivement.\\
Par ailleurs, si nous optons pour une représentation basée sur les motifs essentiels, la dérivation du support conjonctif d'un motif rare nécessite la connaissance du support disjonctif de tous ses sous-ensembles. Toutefois, une partie de ces derniers est fréquente et par conséquent elle ne peut pas être régénérable à partir d'une représentation concise dédiée aux motifs rares. Le même problème sera rencontré dans le cas des motifs non-dérivables où la connaissance du support de tous les sous-ensembles est nécessaire pour :
\begin{itemize}
  \item Dériver le statut d'un motif (dérivable ou non).
  \item Régénérer le support d'un motif dérivable rare à partir des motifs qui seront retenus.
\end{itemize}

Il en résulte que les représentations basées sur les motifs fermés et les générateurs minimaux sont les mieux adaptées à être étendues au cas des motifs rares.
\section{Conclusion}
Dans ce chapitre, nous avons présenté les méthodes proposées dans la littérature. nous avons également mené une étude critique sur ces méthodes en indiquant leur principaux avantages et inconvénients. En effet, nous avons remarqué que chacune des approches considérées isolément reste assez limitée et impose toujours des restrictions au niveau
de l'apprentissage. Dans ce chapitre, nous avons présenté
différentes méthodes de l'extraction des motifs rares
à partir d'un contexte d'extraction. Nous avons discuté
leur particularités, leurs propriétés et leurs limites. Nous avons
abouti au fait que ces méthodes présentent des limites qui les rendent
inadaptées pour certains cas particuliers ou dans des cas réels. Ceci nous a motivé
à proposer de nouvelles approches permettant d'extraire des ensembles réduits sans perte d'information des motifs rares.
\chapter{Représentations concises exactes des motifs rares}
\minitoc

\section{Introduction}
Dans le chapitre précédent, nous avons présenté les travaux de la
littérature qui mènent à extraire les motifs rares. Après une étude
comparative, nous avons dégagé les avantages et les inconvénients de
chaque méthode. Tout de même, nous avons constaté deux remarques liées à l'extraction des motifs rares. Premièrement, aucune
des méthodes ne permet d'extraire une représentation concise exacte
des motifs rares. Deuxièmement, une combinaison entre les méthodes existantes peut
donner naissance à une approche robuste et efficace puisque n'importe
quelle approche de l'état de l'art n'a prouvé sa performance.\\
Dans ce chapitre, nous présentons deux nouvelles représentations concises exactes
des motifs rares. Ainsi, chaque approche consiste à extraire un sous-ensemble de l'ensemble total des motifs rares
à partir duquel nous pouvons générer avec exactitude la totalité de l'ensemble $\mathcal{MR}$.\\
Tout de même, et de point de vue implémentation, les deux méthodes ont utilisé
un parcours du treillis à double sens pour minimiser les motifs candidats visités
tout en se basant sur deux propositions :
\begin{itemize}
\item Pas de génération de sous-ensembles d'un motif fréquent : Si un
motif $I$ est fréquent alors pas de génération des sous motifs de
$I$ dans les nouvelles regénérations.
\item Pas de génération de sur-ensembles d'un motif zéro : Si le support
d'un motif $I$ est égal à zéro alors pas de génération des sur-ensembles
de $I$ dans les nouvelles regénérations.
\end{itemize}

\section{Représentation concise exacte des motifs rares}
Après une étude de la littérature, nous avons remarqué qu\textquoteright{}aucune
étude n'a été menée pour extraire une représentation concise exacte des motifs rares, En effet, la plupart des travaux se sont focalisés sur l'extraction d'un ensemble séparateur entre les motifs rares et ceux fréquents. D'autres travaux se sont focalisés à la génération des motifs rares à partir d'un ensemble séparateur. Notre objectif est de concevoir un algorithme qui permet d'extraire à partir d'un contexte de données un sous-ensemble des motifs rares qui servira par la suite à trouver tous les motifs rares munis de leurs supports. Dans ce qui suit, nous présentons les nouvelles représentations concises exactes des motifs
rares l'une basée sur les générateurs minimaux et l'autre sur les
fermés rares. En ce qui concerne la première représentation, l\textquoteright{}idée
est de conserver les générateurs minimaux
de chaque classe d\textquoteright{}équivalence dont le support est
non nul. La deuxième représentation consiste à conserver les
fermés rares des classes d'équivalence ainsi que l'ensemble de motifs formant la bordure séparatrice entre les itemsets fréquents et ceux rares.\\
Dans ce qui suit, nous présentons une étude de nos représentations concises exactes afin d'arriver à dégager les avantages apportés par ces méthodes par rapport aux autres.\\
%
\subsubsection*{Notation }
Dans ce qui suit, nous notons l'ensemble des générateurs minimaux rares par $\mathcal{GMR}$, l'ensemble des motifs fermés rares par $\mathcal{MFR}$, la représentation concise exacte des motifs rares basée sur les générateurs minimaux par $\texttt{RCEGM}$ et la représentation concise exacte basée sur les fermés rares par $\texttt{RCEFR}$
\subsection{Représentation concise exacte basée sur les générateurs minimaux}

Dans ce qui suit, nous présentons notre première représentation concise exacte à savoir celle basée sur les générateurs minimaux rares. Cette méthode repose sur les motifs clés \cite{PBT99}. Un motif clé est un motif minimal d'une classe d'équivalence regroupant tous les motifs contenus dans les mêmes objets du contexte d'extraction. Tous les motifs d'une classe d'équivalence possèdent le même support, et le support de motif non-clé peut être retrouvé à partir des motifs clés d'une même classe d'équivalence.

\subsubsection{Description de la représentation}
L'idée de l'approche est d'utiliser les générateurs minimaux rares pour déduire la totalité des motifs rares munis de leur supports. Les générateurs minimaux représentent les minimaux d'une classe d'équivalence. Ainsi, ces motifs admettent deux propriétés importantes : (i) ils sont minimaux au sens de l'inclusion; et (ii) les motifs rares représentent un filtre d'ordre dans le treillis de Galois. En se basant sur ces deux propriétés, nous pouvons affirmer que la délimitation entre les motifs rares et ceux fréquents est assurée. Par ailleurs, les supports sont déduits grâce à la définition même d'un générateur minimal.

\begin{theorem}
L'ensemble des générateurs minimaux rares forme un idéal d'ordre dans le treillis.
\end{theorem}
\textbf{Preuve.}\\
Nous allons démontrer que l'ensemble des motifs générateurs minimaux (fréquents ou infréquents) forme un idéal d'ordre. Soient $ P,Q \in \mathcal{I}$ et la fonction $f$ définie la correspondance de Galois d'une relation binaire (\textit{cf.} Définition \ref{cx_galois} page \pageref{cx_galois}). Nous supposons que $P$ n'est pas un $\mathcal{GM}$ et si $P\subseteq Q$, alors $Q$ n'est pas un $\mathcal{GM}$.
$P$ n'étant pas $\mathcal{GM}$ par hypothèse, il existe $P'\in \mathcal{GM}$ avec $P'\subset P$. Sachant que
$f(P)=f(P')$,
montrons que $f(Q)= f(Q\setminus (P\setminus P'))$. Nous savons que $f(P')= f(P)$.
$f(P\setminus P')$ (antimonotonie de $f$ ). $f(Q)=f((Q\setminus (P\setminus P'))\cup (P\setminus P'))=f(Q\setminus (P\setminus P')) \cap
f(P\setminus P')$. Donc, $f(Q) \supseteq f(Q\setminus (P\setminus P')) \cap f(P')$. Cette expression se récrit $f(Q) \supseteq
f((Q\setminus (P\setminus P'))\cup P')$. Puisque $P' \subseteq Q\setminus (P\setminus P')$, nous trouvons $f(Q) \supseteq f(Q\setminus (P\setminus P')).$
L'inclusion opposée est immédiate, et notre égalité est vérifiée : $Q$ n'est donc pas minimal dans sa classe d'équivalence.\\

Le théorème déjà cité permet de réduire le nombre de motifs à visiter. Ainsi, un motif qui n'est pas un $\mathcal{GM}$ (générateur minimal) ne peut pas donner naissance à un $\mathcal{GM}$ et donc il serait à élaguer.

Si $P$ et $Q$ sont des motifs appartenant à la même classe d'équivalence, on voit
aisément qu'ils ont le même support. De même, deux motifs de support égal et dont
l'un est inclus dans l'autre font partie de la même classe d'équivalence :
\begin{proposition} \label{pro_ferm} Soient deux motifs $P$ et $Q$ et $\theta$ la relation d'équivalence induite par l'opérateur de fermeture de Galois. On a :\\
\begin{enumerate}
  \item $P \theta Q \Longrightarrow Supp(P) = Supp(Q)$
  \item $P\subseteq Q$ et Supp(P) = Supp(Q) $\Longrightarrow P \theta Q$
\end{enumerate}

\end{proposition}
\textbf{Preuve.}\\La première partie de cette proposition découle directement de la définition de la relation d'équivalence $\theta$ (\textit{cf.} Définition \ref{def_ce}, page \pageref{def_ce}). Pour démontrer la seconde, on remarque que $f(P)\supseteq f(Q)$. Puisque $Supp(P) = Supp(Q)$ est équivalent à $card(f(P))=card(f(Q))$, on a $f(P)=f(Q)$.
Si la relation $\theta$ était connue d'avance, on pourrait compter le support d'un seul
motif pour chaque classe d'équivalence. Cependant, On ne connaît évidemment pas cette relation; mais elle peut être construite au fur et à mesure. D'une manière générale, on calculera le support d'au moins un motif d'une classe d'équivalence. Si on a déterminé
le support d'un motif P et qu'on trouve plus tard un motif $Q\in[P]$, il est inutile
d'accéder au contexte d'extraction pour calculer le support de $Q$.

\begin{theorem}
\[GMR \equiv ~Repres\acute{e}ntation~concise~exacte~des~motifs~rares~~(\texttt{RCEMR})\]
L'ensemble des générateurs minimaux rares (les motifs clés) forme une représentation concise exacte des motifs rares.
\end{theorem}
\textbf{Preuve.}\\ L'ensemble des générateurs minimaux rares,  peut former tout seul une représentation concise exacte des motifs rares. Les motifs rares, comme nous l'avons montré, forment un filtre d'ordre et l'ensemble des générateurs minimaux représentent les motifs minimaux de chaque classe d'équivalence (motifs clés). La frontière entre les motifs fréquents et ceux rares, nommée bordure négative des motifs fréquents ou couramment appelée ensemble des motifs rares minimaux, est incluse dans l'ensemble $\mathcal{GMR}$ (\textit{cf.} Chapitre 1). Cette partie commune nous garantit la distinction entre les deux types de motifs (rare ou fréquent), le support de chaque motif rare est assuré par les générateurs minimaux de chaque classe d'équivalence. En effet, l'ensemble des générateurs minimaux forme une représentation concise exacte des motifs rares.\\
Pour chaque motif $P \subseteq \mathcal{I}$, nous pouvons déduire sa nature (rare ou fréquent) et son support exact s'il s'agit d'un motif rare. D'une manière formelle, soit $\mathcal{GMR}$ l'ensemble des générateurs minimaux rares, alors $\forall$  $P$ $\subseteq$ $\mathcal{I}$, deux cas peuvent se présenter :
\begin{enumerate}
  \item   $\{\forall v \in \mathcal{GMR} ~~tel que ~~P \nsupseteq v \Longrightarrow Supp(P) > \emph{maxsupp}\}$
  \item $\{ \exists v \in \mathcal{GMR}~~tel que ~~P\supseteq v \Longrightarrow Supp(P) \leq \emph{maxsupp}\}$ \\
  \textbf{et} $ \{\forall v' \in \mathcal{GMR}~~tel que~~P \supseteq v' ~~\Rightarrow Supp(P) = Min(Supp(v'))\}$
\end{enumerate}
Il est important de noter que l'ensemble des générateurs minimaux forme un idéal d'ordre dans le treillis booléen. Ceci constitue un critère d'élagage des motifs qui ne sont pas des générateurs minimaux puisque au moins un de leurs sous-ensembles n'est pas un générateur minimal.
\subsubsection{Description de l'algorithme}
Dans cette section, nous expliquons le déroulement de l'algorithme
permettant d'extraire un sous-ensemble (ensemble des générateurs minimaux) à partir
duquel nous déduisons les supports de tous les motifs rares.
L'algorithme prend en entrée un contexte d'extraction ainsi qu'un seuil maximal \emph{maxsupp}. Il fournit en sortie
: l'ensemble des générateurs minimaux notés $\mathcal{GMR}$ ainsi que l'ensemble des motifs zéros notés $\mathcal{GMZ}$.\\
En adoptant la stratégie ``Générer et Tester'', notre algorithme parcourt l'espace de recherche par niveau pour déterminer l'ensemble des générateurs minimaux rares, noté $\mathcal{GMR}$, ainsi que l'ensemble des générateurs zéros, noté $\mathcal{GMZ}$.
En effet, l'algorithme commence par calculer les supports des motifs de taille 1, élague ceux qui sont des motifs zéros et insère ceux qui sont rares dans l'ensemble $\mathcal{GMR}$. Au fur et à mesure que l'algorithme génère les motifs candidats de chaque niveau, il conserve ceux qui vérifient la propriété d'un \textsc{GM}. L'algorithme \textsc{GMRare} s'arrête lorsqu'il ne reste aucun motif à générer, en s'arrêtant l'algorithme fournit comme résultat l'ensemble de tous les \textsc{GM}s rares.\\
L'algorithme GMRare utilise deux critères d'élagage pour les motifs candidats :
\begin{itemize}
  \item (1) Pas de génération des sur-ensembles d'un motif zéro.
  \item (2) Pas de génération d'un motif non générateur minimal.
\end{itemize}
Ces deux critères d'élagage sont basés sur deux propriétés :
\begin{itemize}
  \item (1) Tous les sur-ensembles d'un motif zéro sont des motifs zéros.
  \item (2) Si un motif $x \notin GM$, alors pour chaque $y$ sur-ensemble de $x$, $y\notin GM$.
  \end{itemize}
Il est à noter que ces propriétés sont détaillées dans le chapitre 1 (\textit{cf.}, Propriété \ref{pro_gm}, page \pageref{pro_gm}).\\

Le pseudo-code de notre algorithme est donné par l'algorithme \ref{algoRepRare}. Les notations utilisées sont résumées dans le tableau \ref{notationsGMR}.

\begin{table}[bt]
\begin{center}
\begin{tabular}{|p{370pt}|}
 \hline
\begin{description}
\item[$\mathcal{C}_{i}$]: l'ensemble des motifs candidats à générer.
\item[$\mathcal{GMZ}$]: l'ensemble des motifs zéros minimaux sert à faire l'élagage.
\item[$\mathcal{GMR}$]: l'ensemble des générateurs minimaux.
\item[\textit{maxsupp}]: c'est un seuil'utilisateur, $\emph{maxsupp=minsupp}-1$.
\end{description}
\\
\hline
\end{tabular}
\end{center} \caption{Notations utilisées dans
l'algorithme \textsc{GMRare}.}\label{notationsGMR}
\end{table}

\linesnumbered
\begin{algorithm}[tb]
{
    \SetVline
    \setnlskip{-3pt}


\Donnees {\begin{enumerate}
    \item Contexte d'extraction
    \item Seuil \emph{maxsupp}.
\end{enumerate} }

\Res{\begin{enumerate}
    \item Générateurs minimaux des motifs rares $\mathcal{GMR}$.
\end{enumerate}}

    \Deb
{
    $\mathcal{C}_{1}={1\_motif}$\\
    $i \leftarrow 1$\\
    \Tq {$\mathcal{C}_{i}\neq \emptyset$}
    {
    /* Calcul de support des motifs de l'ensembles $\mathcal{C}_{i}$ */
    \\
    \PourCh {$e_{i} \in \mathcal{C}_{i}$}
    {
    $\emph{CountSupport}~(e_{i})$\\
     \eSi{$Supp(e_{i}) = 0$}
     {
      $\mathcal{GMZ}\leftarrow \mathcal{GMZ} \cup \{e_{i}\}$;\\
      $\mathcal{C}_i \leftarrow \mathcal{C}_i \setminus \{e_i\}$;// Pas de génération des sur-ensembles d'un motif zéro.\\

     }
     {

       \Si{$Supp(e_{i}) \leq maxsupp$ ET $(\forall~ e'_i \subset e_i$, $Supp(e'_i) \neq Supp(e_i))$}
            {
            $\mathcal{GMR} \leftarrow \mathcal{GMR} \cup \{e_{i}\}$;\\
            }
     }
    }
    $\mathcal{C}_{i+1}\leftarrow \emph{Gen\_Candidats}~(\mathcal{C}_{i},\mathcal{GMZ})$;\\

    }
\textbf{retourner} $\mathcal{GMR}$
}
}
  \caption{\textsc{GMRare}}
  \label{algoRepRare}
\end{algorithm}

\begin{table}[bt]
\begin{center}
\begin{tabular}{|p{370pt}|}
 \hline
\begin{description}
\item[$\mathcal{C}_i$]: Ensembles d'itemsets candidats de cardinalité $i$.
\item[$\textit{k}$]: Une variable de type itemset sert à récupérer le motif résultat.
\item[$\textit{i}$]: entier.
\end{description}\\
\\
\hline
\end{tabular}
\end{center}
\caption{Notations utilisées dans la procédure \textsc{Gen\_Candidat}.}
\label{notationsGenCand}
\end{table}

\medskip{}
\begin{algorithm}[tb]
{

    \setnlskip{-3pt}


\Donnees {\begin{enumerate}
    \item Liste des motifs de taille $i$ noté $\mathcal{C}_i$.
\end{enumerate} }

\Res{\begin{enumerate}
    \item liste des motifs de taille $i+1$ noté $\mathcal{C}_{i+1}$.
\end{enumerate}}

    \Deb
{

    \eSi {$i=1$}
      {
       \PourCh{$x\in \mathcal{C}_i $}
        {
         \PourCh{$y\in \mathcal{C}_i$}
         {
         $k=x \cup y$;\\
         $\mathcal{C}_{i+1}=\mathcal{C}_i \cup {k};$ // génération des motifs de taille 2.\\
         }
        }
      }
      {
      \PourCh {$x\in \mathcal{C}_i~ET~y\in \mathcal{C}_i $}
         {
         $k=\{k_{xy}~|~ x=x_{1},x_{2},...,x_{p}~et~y=y_{1},y_{2},...,y_{p}~et~x_{q}=y_{q}~ \forall q ~\in [1..p-1]~alors~ k_{xy}=x_{1},x_{2},...,x_{p},y_{p}\};$\\
         $\mathcal{C}_{i+1}=\mathcal{C}_i \cup {k};$
         }
      }

}
}
  \caption{\textsc{Gen\_Candidats}}
  \label{algGencand}
\end{algorithm}

L'élagage à travers ces deux critères est intéressant. En effet, ils permettent d'éviter l'extraction des motifs zéros qui représente une large partie des motifs dérivés à partir d'un contexte. Par ailleurs, ces deux critères permettent aussi d'éviter le traitement des motifs qui ne donneront pas lieu à des \textsc{GM}s.

Dans les paragraphes suivants, nous détaillons le rôle de chacune des fonctions et des procédures utilisées par l'algorithme \textsc{GMRare}.

\subsubsection{Fonctions et procédures utilisées par l'algorithme \textsc{GMRare}}

\textbf{$\bullet$ La fonction \emph{CountSupport} :}
La fonction \emph{CountSupport} prend en entrée un seul paramètre de type \emph{itemset} et retourne un résultat de type \emph{entier} (\textit{cf.} Algorithme \ref{algoCountSupp}). L'unique paramètre passé en argument est l'itemset, dont nous voulons calculer le support. Le résultat retourné est le support de l'itemset dans le contexte d'extraction. La fonction \emph{CountSupport} calcule le support d'un motif donné à travers une navigation intelligente dans l'arbre représentant le contexte. Le cas le plus simple est de calculer le support d'un motif de cardinalité égale à 1, puisque une recherche dans l'arbre permet de trouver directement son support. Le cas le plus général revient à calculer le support d'un motif $I$ composé de $n$ items $I=i_{1},i_{2},...,i_{n}$, la fonction commence par chercher le premier item à savoir $i_{1}$ si le processus de recherche a réussi à trouver l'élément, alors un appel récursif de la fonction aura lieu pour le reste des items $i_{2},i_{3},...,i_{n}$. En effet, quand tous les items sont trouvés, alors la fonction retourne le support du dernier item, tandis que la valeur zéro est retournée dans le cas contraire.\\
\begin{table}[tb]
\begin{center}
\begin{tabular}{|p{370pt}|}
 \hline
\begin{description}
\item[$v$]: Un itemset de taille $k$ dont nous voulons calculer leur support.
\item[$p$]: Représente l'adresse de l'item retourné par la fonction \texttt{SeekItem()}.
\item[$racine$]: L'adresse de la racine de l'arbre contenant le contexte de données.
\item[\texttt{SeekItem()}]: Une fonction permet de chercher un item dans un arbre, l'item et la racine de l'arbre sont passés en paramètre.
\end{description}
\\
\hline
\end{tabular}
\end{center} \caption{Notations utilisées de la fonction \textsc{CountSupport}.}\label{notationsCountsupp}
\end{table}

\begin{algorithm}[tb]
{
    \SetVline
    \setnlskip{-3pt}


\Donnees {\begin{enumerate}
    \item Un itemset $v$ de taille $k$.
    \item La racine de l'arbre contenant le contexte de données noté \emph{racine}.
\end{enumerate} }

\Res{\begin{enumerate}
    \item Le support d'itemset passé en paramètre.
\end{enumerate}}

    \Deb
{
   $v$=$\{v_{1},...,v_{k-1},v_{k}\}$\\
    $p$=\texttt{SeekItem}($v_{1}$,$racine$)\\
    \eSi {($p$ existe ) }
            {
                 \eSi{$v$ contient un seul item $v_k$}
                 {
                  retrun ($Supp(v_k)$)//Support de l'unique item $v_k$
                 }
                 {
                 $v'$=$\{v_{2},...,v_{k-1},v_{k}\}$\\
                 \textbf{retourner} (CountSupport($v'$,$p$))\\
                 }

            }
        { \textbf{retourner} (0);}

}}
  \caption{\textsc{CountSupport}}
  \label{algoCountSupp}
\end{algorithm}

\textbf{$\bullet$ La fonction \emph{intersectItemset} :}
Cette fonction prend en paramètre deux motifs et retourne une valeur booléenne. La valeur ``\emph{vrai}'' est retournée si une inclusion existe entre les deux ensembles passés en paramètre. Dans le cas contraire, la valeur \emph{faux} est retournée. L'inclusion signifie que l'un des motifs est un sous-ensemble de l'autre.

\textbf{$\bullet$ La procédure \emph{Gen\_Candidat} :}
Cette procédure sert à générer à partir d'un ensemble des motifs de taille $m$, passé en paramètre, un ensemble de motif de taille $m+1$. La procédure \emph{Gen\_Candidat} sépare la génération des motifs de taille 1 à celui de taille supérieure à 1 (\textit{\textit{cf.}} Algorithme \ref{algGencand}, ligne 8). Les notations utilisées sont résumées dans le tableau \ref{notationsGenCand}.

\textbf{$\bullet$ La fonction \emph{intersect} :}
intersect est une fonction appelée à partir des fonctions de génération des candidats, elle retourne une valeur booléenne et prend en paramètre d'entrée deux pointeurs
de type \emph{itemset }et un \emph{entier}. La fonction admet un rôle au
niveau de la génération des candidats puisque c'est grâce à sa valeur
de retour que la procédure \emph{Gen\_Candidat } génère un motif candidat. La valeur \emph{vrai} retournée par \emph{intersect }affirme
que les deux motifs peuvent être utilisés pour donner un nouveau motif. Dans le cas contraire, la valeur \emph{faux} est retournée.\\
Après une explication générale de cette fonction, nous présentons le rôle de chaque variable d'entrée et le déroulement de la fonction pour aboutir au résultat désiré. En effet, la génération d'un motif se fait suite à une vérification que deux motifs aient en commun les $m$ premiers items. Par exemple\emph{ ABC} et \emph{ABD} admettent les deux premiers items en commun. Si deux motifs de taille \emph{m+1} admettent les \emph{m} premiers items en commun alors un candidat de taille \emph{m+2} peut être généré à partir de ces deux motifs. En considérant notre exemple, les deux motifs \emph{ABC}
et \emph{ABD} de taille trois admettent deux items en commun et permettent ainsi de générer un candidat de taille quatre à savoir $ABCD$.

\textbf{$\bullet$ La fonction \emph{existInGMZ }:}
La fonction  \emph{existInGMZ} vérifie l'existence d'un motif dans l'ensemble
des motifs $\mathcal{GMZ}$ (ensemble des générateurs zéros minimaux). En effet, une recherche séquentielle dans l'ensemble $\mathcal{GMZ}$ permet de retrouver le motif passé en paramètre. Ainsi, cette fonction fournit comme résultat une valeur booléenne et prend en paramètre deux ensembles de type \emph{itemset} : le premier représente le motif candidat qu'on souhaite
insérer dans la liste $\mathcal{GMZ}$; et le deuxième représente la liste $\mathcal{GMZ}$. La fonction
vérifie l'existence des sous-ensembles d'un motif y compris lui même
dans la liste des $\mathcal{GMZ}$. Si l'un de ces sous-ensembles a été trouvé, alors
la fonction retourne la valeur \emph{vrai} et le motif ne s'ajoute pas à
la liste des $\mathcal{GMZ}$ (puisque son support peut être déduit à partir de
l'un de ces sous-ensembles). Dans le cas contraire, la valeur \emph{faux} est retournée et nous insérons le motif dans l'ensemble des $\mathcal{GMZ}$ à l'aide de la fonction \emph{insertItemToGMZ} qui fait l'objet du prochain paragraphe.

\textbf{$\bullet$ La procédure \emph{insertItemToGMZ} :}
La procédure \emph{insertItemToGMZ }prend en entrée un ensemble de
type \emph{itemset }qui représente le motif que nous voulons insérer
dans la liste $\mathcal{GMZ}$. En effet, le rôle de cette
procédure se restreint à l'insertion du motif passé en paramètre dans la liste tout en respectant
l'ordre des motifs. L'ordre des motifs de la liste $\mathcal{GMZ}$ dépend de leur cardinalités. Les motifs admettant des cardinalités
identiques sont classés selon l'ordre lexicographique. Dans cette fonction, nous utilisons un tel ordre afin de faciliter la recherche des motifs dans la liste. Ainsi, la recherche d'un motif de taille $n$ échoue à la rencontre
du premier motif de taille $n+1$. Au pire des cas, la recherche nécessite
un parcours total de la liste des $\mathcal{GMZ}$. Au meilleur des cas, la recherche
nécessite seulement une seule vérification.


\textbf{$\bullet$ La procédure \emph{InsertItemToGm} :}
La procédure \emph{InsertItemToGm }prend en paramètre un ensemble de type
\emph{itemset}, qui représente le motif à insérer
dans la liste $\mathcal{GMR}$. En premier lieu, la procédure vérifie l'absence du motif dans la liste ainsi que l'absence
de tous ces sous-ensembles. En deuxième lieu, la procédure se restreint
à l'insertion du motif dans la liste tout en respectant l'ordre des motifs. L'ordre des motifs dans la liste dépend de leurs cardinalités,
les motifs admettant des cardinalités identiques sont classés selon
l'ordre lexicographique.

\textbf{$\bullet$ La procédure \emph{elagage } :}
Afin de mieux présenter cette procédure, nous commençons par décrire son rôle principal. Cette procédure permet d'élaguer l'ensemble des motifs en se basant sur deux propriétés :
\begin{itemize}
  \item Pas de génération des sur-ensembles d'un motif zéro (i);
  \item Pas de génération d'un motif non générateur minimal puisqu'il ne donne pas naissance à un générateur minimal (ii).
\end{itemize}
En effet, la procédure \emph{elagage} sera appelée à chaque génération de nouveaux motifs candidats, à partir du support de motif $I$ et les supports de ces sous-ensembles, la procédure \emph{elagage}, décide l'élagage du motif $I$ tout en se basant sur les deux propriétés (i) et (ii).

\begin{exemple} \label{expgmrare} \textbf{Exemple illustratif}\\
Rappelons le contexte d'extraction donné dans le premier chapitre table \ref{DB1} (page \pageref{DB1}). Pour $\textit{maxsupp}= 3$,
nous appliquons l'algorithme \textsc{GMRare}.

\begin{small}
\begin{center}\textit{\textbf{\underline{Étapes de déroulement de l'algorithme}}}\end{center}
\begin{minipage}[b]{1 \linewidth}
\noindent\fbox{\parbox{\linewidth}{
\centering \textbf{Initialisation}
}}
\noindent\fbox{\parbox{\linewidth}{
$\mathcal{C}_{i}$ $\longleftarrow$ $\{A,B,C,D,E\}$ // Les candidats générés.\\
$\mathcal{GMR}$ $\longleftarrow$ $\textrm{\O}$ // Ensemble des générateurs minimaux.\\
$\mathcal{GMZ}$ $\longleftarrow$ $\textrm{\O}$ // Ensemble des motifs zéros.\\
}}
\end{minipage}

\begin{minipage}[b]{0.25 \linewidth}
\noindent\fbox{\parbox{\linewidth}{
\textbf{Première itération}
}}
\noindent\fbox{\parbox{\linewidth}{
$Supp(A)=3$\\
$Supp(B)=4$\\
$Supp(C)=4$\\
$Supp(D)=1$\\
$Supp(E)=4$
}}
\end{minipage}
$~~~~\rightarrow~~$
\begin{minipage}[b]{0.55 \linewidth}
\noindent\fbox{\parbox{\linewidth}{
$\mathcal{GMR}$ $\longleftarrow$ $\{(A,3);(D,1)\}$\\
$\mathcal{GMZ}$ $\longleftarrow$ $\{ABCD,ABDE,ACDE,BCDE\}$
}}
\end{minipage}
\\

\begin{minipage}[b]{0.25 \linewidth}
\noindent\fbox{\parbox{\linewidth}{
\textbf{Deuxième itération}
}}
\noindent\fbox{\parbox{\linewidth}{
$Supp(AB)=2$\\
$Supp(AC)=3$\\
$Supp(AD)=1$\\
$Supp(AE)=2$\\
$Supp(BC)=3$\\
$Supp(BD)=0$\\
$Supp(BE)=0$\\
$Supp(CD)=4$\\
$Supp(CE)=2$\\
$Supp(DE)=1$
}}
\end{minipage}
$~~\rightarrow$
\begin{minipage}[b]{0.70 \linewidth}
\noindent\fbox{\parbox{\linewidth}{
$\mathcal{GMR}$ $\longleftarrow$ $\{(A,3);(D,1);(AB,2);(AE,2);(BC,3);(CE,3)\}$\\
$\mathcal{GMZ}$ $\longleftarrow$ $\{BD,DE\}$
}}
\end{minipage}
\end{small}
\\

Au départ, l'algorithme \textsc{GMRare} génère les motifs de taille $1$. Après une évaluation de leur supports l'algorithme retient les motifs rares (motif $A$ et $D$ dans l'exemple), insère ces motifs dans l'ensemble $\mathcal{GMR}$ et génère les motifs de taille 2.
Lors de la deuxième itération, l'ensemble $\mathcal{GMR}$ contient les motifs $\{(A,3)~~et~~(D,1)\}$. Pour chaque motif $v$ de taille 2, dont le support est inférieur à $maxsupp$, l'algorithme teste la possibilité d'insérer $v$ à l'ensemble $\mathcal{GMR}$.
Dans l'exemple le motif ($AB$,2) est inséré puisque son support est différent de celui de $A$. Le motif ($AC$,3) n'est pas inséré puisque son support peut être déduit à partir de celui de ($A$,3).

\end{exemple}

\subsection{Représentation concise exacte basée sur les fermés rares}
Dans ce qui suit, nous présentons notre deuxième représentation concise exacte à savoir celle basée sur les fermés rares. Un motif fermé d'une classe d'équivalence  est un motif unique dans cette classe. Une classe d'équivalence regroupe tous les motifs qui possèdent le même support. Les supports des motifs non fermés peuvent être trouvés à partir des motifs fermés de la même classe d'équivalence.

\subsubsection{Description de la représentation}
Nous proposons une nouvelle représentation concise  des motifs rares basée sur les motifs fermés d'une classe d'équivalence. Dans ce qui suit, nous rappelons la sémantique d'une classe d'équivalence : l'idée de base est de regrouper les motifs considérés comme équivalents, il s'agit d'un principe puissant, qui permet de regrouper dans des classes d'équivalence les motifs correspondants aux mêmes objets (\textit{cf.} Définition \ref{def_ce}, page \pageref{def_ce}).

Les motifs fermés d'une manière générale (fermés rares ou bien fréquents) sont les maximaux d'une classe d'équivalence. Ils représentent les sur-ensembles d'une classe d'équivalence dont le support est inférieur à un seuil dit \emph{maxsupp} (fermé rare), et les motifs rares forment un filtre d'ordre. Ainsi, la connaissance des fermés n'entraîne pas la connaissance de tous les motifs rares. L'ensemble des fermés rares ne forme pas une représentation concise exacte des motifs rares. Cependant, la solution est de trouver un ensemble à partir duquel nous pouvons utiliser la notion de filtre d'ordre des motifs rares, pour délimiter les motifs fréquents de ceux rares et de bénéficier de la définition des motifs fermés pour déterminer le support de chaque motif rare (\textit{cf.} Proposition \ref{pro_ferm}). En effet, l'ensemble séparateur ne peut être que l'ensemble des motifs minimaux au sens de l'inclusion pour les motifs rares (bordure négative) ou bien l'ensemble des motifs maximaux au sens de l'inclusion pour les motifs fréquents (bordure positive).\\
Si $P$ et $Q$ deux motifs appartenant à la même classe d'équivalence, on voit
aisément qu'ils ont le même support (\textit{cf.} Proposition \ref{pro_ferm}).
\begin{proposition}
L'ensemble des motifs rares forme un filtre d'ordre dans $(2^{n}$, $\subseteq)$ où $n$ est le nombre d'items (par rapport à la contrainte de fréquence):
\begin{itemize}
  \item Tout sur-ensemble d'un motif rare est aussi rare.
  \item Tout sous-ensemble d'un motif non rare est aussi non rare.
  \end{itemize}
\end{proposition}
\mbox{}\\\textbf{Notations.} Rappelons les notations du premier chapitre, l'ensemble des motifs rares minimaux et la bordure positive des motifs rares est noté par $\mathcal{MRM}$. De même pour l'ensemble des motifs fréquents maximaux et la bordure négative, ils sont notés par $\mathcal{MFM}$.
\begin{theorem}
\[\mathcal{MFR} \cup \mathcal{MRM} \equiv Repr\acute{e}sentation~concise~exacte~des~motifs~rares~~ \texttt{R C E M R}\]
Les deux ensembles $\mathcal{MFR}$ et $\mathcal{MRM}$ forment une représentation concise exacte des motifs rares.
\end{theorem}
\textbf{Preuve.}  L'ensemble des fermés rares et l'ensemble des minimaux rares (bordure positive des motifs rare) forment une représentation concise exacte des motifs rares si et seulement si  nous pouvons déduire pour chaque motif $P \subseteq \mathcal{I}$  sa nature (rare ou fréquent) et son support exact s'il s'agit d'un motif rare.\\
Soient $\mathcal{MFR}$ l'ensemble des fermés rares, $\forall$ $P$ $\subseteq$ $\mathcal{I}$ deux cas se présentent :
\begin{enumerate}
  \item   $\{\forall v \in \mathcal{MRM} ~~tel que ~~P \not\subset v \Longrightarrow Supp(P) > \emph{maxsupp}\}$ ($P$ motif fréquent)
  \item $\{ \exists v \in \mathcal{MRM} ~~tel que ~~P\supseteq v \Longrightarrow Supp(P)\leq \emph{maxsupp}\}$ \\
  \textbf{et}$ \{\exists v' \in \mathcal{MFR}~~tel que~~P\subseteq v'~~et~~h(v')=h(P)\Rightarrow Supp(v')=Supp(P)\}$
\end{enumerate}

\subsubsection{Description de l'algorithme}
L'idée est semblable à la méthode décrite précédemment. Dans cette section, nous expliquons
le déroulement de l'algorithme permettant d'extraire un sous-ensemble
(fermés rares union la bordure positive) à partir duquel nous déduisons
les supports de tous les motifs rares. L'algorithme prend en entrée
un contexte d'extraction ainsi qu'un seuil maximal \emph{maxsupp}. Il fournit en résultat la liste des fermés rares enrichie
par les motifs de la bordure positive ainsi que la liste des générateurs
minimaux des motifs zéro $\mathcal{GMZ}$.\\
En adoptant la stratégie ``Générer et Tester'', notre algorithme parcourt l'espace de recherche par niveau pour déterminer l'ensemble des fermés rares, noté $\mathcal{MFR}$, ainsi que l'ensemble des motifs rares minimaux et l'ensemble des générateurs zéros, noté $\mathcal{GMZ}$.\\
Nous signalons aussi, l'utilisation de plusieurs fonctions et procédures qui font partie de l'algorithme précédent à l'exemple de \emph{loadToTree, CountSupp, intersectItemset, intersect, existInGMZ} et \emph{insertItemToGMZ}. Ainsi, et pour ne pas refaire les même descriptions de ces modules nous présentons seulement les modules non décrits précédemment.\\

Le pseudo-code de notre algorithme est donné par l'algorithme \ref{algomfrare}. Les notations utilisées sont résumées dans le tableau \ref{notationsMFR}.

\begin{table}[tb]
\begin{center}
\begin{tabular}{|p{370pt}|}
 \hline
\begin{description}
\item[$\mathcal{C}_{i}$]: l'ensemble des motifs candidats à générer.
\item[$\mathcal{GMZ}$]: l'ensemble des motifs zéros minimaux.
\item[$\mathcal{MFR}$]: l'ensemble des fermés rares.
\item[$\mathcal{MRM}$]: l'ensemble des motifs rares minimaux.
\item[\emph{maxsupp}]: c'est un seuil utilisateur, $\emph{maxsupp=minsupp}-1$.
\end{description}
\\
\hline
\end{tabular}
\end{center} \caption{Notations utilisées dans
l'algorithme \textsc{MFRare}.}\label{notationsMFR}
\end{table}

\begin{algorithm}[tb]
{
    \SetVline
    \setnlskip{-3pt}


\Donnees {\begin{enumerate}
    \item Contexte de données et seuil \emph{maxsupp}.
\end{enumerate} }

\Res{\begin{enumerate}
    \item L'ensemble des fermés rares et l'ensemble des motifs \\rares minimaux $\mathcal{MRM}$.
\end{enumerate}}

    \Deb
{
    $\mathcal{C}_{1}={1\_motif}$\\
    $i\longleftarrow 1$\\
    \Tq {$\mathcal{C}_{i}\neq \emptyset $}
    {
    /* Calcul de supports des motifs de l'ensembles $\mathcal{C}_{i}$   */

    \PourCh {$e_{i} \in \mathcal{\mathcal{C}}_{i}$}
    {
    $\emph{countSupport}~(e_{i})$;\\
     \eSi {$Supp~(e_{i}) = 0$}
     {
      $\mathcal{GMZ}\leftarrow \mathcal{GMZ} \cup \{e_{i}\}$\\
 $\mathcal{C}_i \leftarrow \mathcal{C}_i \setminus \{e_i\}$;// Pas de génération des sur-ensembles d'un motif zéro.\\
     }
     {
     /* Si $e_i$ est rare ainsi que tous ses sous-ensembles sont fréquents alors $e_i$ appartient à l'ensemble des motifs rares minimaux */ \\
       \eSi {$Supp~(e_{i}) \leq maxsupp$ ET $(\forall~ e'_i \subset e_i$, $e'_i\notin \mathcal{MRM})$}
            {
            $\mathcal{MRM}\leftarrow \mathcal{MRM} \cup \{e_{i}\}$\\
            }
            {
            \Si {$Supp~(e_{i}) \leq maxsupp$ ET $(\forall~ e'_i \supset e_i$, $Supp(e_i)\neq Supp(e'_i))$}
                {
                $\mathcal{MFR}\leftarrow \mathcal{MFR} \cup \{e_{i}\}$\\
                }

            }
     }
    }
    $\mathcal{C}_{i+1}\leftarrow \emph{Gen\_Candidats }(\mathcal{C}_{i},\mathcal{GMZ})$\\
    $i\leftarrow i + 1 $\\

    }
\textbf{retourner} $\mathcal{MFR} \cup \mathcal{MRM}$
}
}
  \caption{\textsc{MFRare}}
  \label{algomfrare}
\end{algorithm}

En ce qui concerne le résultat de l'algorithme, et contrairement à l'algorithme des générateurs minimaux, nous ne pouvons pas se restreindre à l'unique ensemble des fermés rares puisque les fermés d'une manière générale
(fermé rare ou bien fréquent) sont les maximaux d'une classe d'équivalence. Ainsi,  l'utilisation
des motifs fermés rares oblige l'ajout d'un deuxième ensemble qui forme une bordure séparatrice entre les motifs fréquents et ceux rares afin de pouvoir former une représentation concise exacte. Nous prendrons un exemple illustratif pour montrer le rôle de l'ensemble séparateur. Soit l'ensemble des fermés rares contenant le motif $ABC$ du support 2, plusieurs
problèmes vont être posés. L'application de la propriété de la fermeture comme c'était
dans le cas des motifs fréquents implique le résultat suivant :
chaque sous-ensemble inclus dans $ABC$ à le même support que $ABC$
(bien sur s'il n'existe pas un autre fermé plus réduit que $ABC$ en
terme d'inclusion). Ainsi, les motifs $AB,AC,BC,A,B,C$ et $\mbox{\textrm{\O}}$ ont
le même support que $ABC$ qui est égaux à 2, ce qui est absurde. À cause de
ce problème, il faut avoir une bordure à partir de laquelle nous nous limitons aux motifs rares. La bordure peut contenir les motifs fréquents
maximaux, et par la suite les motifs qui se trouvent au-dessous de cette bordure ne nous intéressent pas, ou bien nous pouvons utiliser la bordure contenant les motifs rares minimaux, et dans ce cas, les motifs qui se trouvent au-dessus de cette bordure forment les motifs rares dans le treillis. Revenant à notre exemple, si nous optons à l'utilisation de la bordure
positive contenant les motifs $A, B $ et $C$ de plus l'ensemble des fermés rares contenant l'itemset $ABC$ de support 2,  nous sommes sûrs que les sur-ensembles de la bordure sont des motifs rares et que les supports de $AB, AC, BC$ sont égal à 2 puisqu'ils appartiennent à la même classe d'équivalence dont le fermé est $ABC$.\\
En premier lieu, nous présentons une description générale de l'algorithme. L'algorithme \textsc{MFRare} commence par calculer les supports des motifs de taille 1, ces motifs forment initialement les éléments de l'ensemble $\mathcal{MFR}$. En deuxième lieu, l'algorithme génère, pour chaque niveau, des motifs candidats afin de fournir l'ensemble final $\mathcal{MFR}$. Un motif candidat est inséré dans la liste des $\mathcal{MFR}$ si :
\begin{itemize}
  \item Aucun de ses sous-ensembles n'existe dans l'ensemble $\mathcal{MFR}$.
  \item L'un de ses sous-ensembles existe et admet le même support que lui.
\end{itemize}
L'insertion d'un motif candidat engendre la suppression de tous ces sous-ensembles à partir de l'ensemble $\mathcal{MFR}$. Nous prenons un exemple, soit $\mathcal{I}=\{A,B,C,D\}$ la liste des items d'un contexte de données. Initialement, et après calcul des supports des motifs de la liste $\mathcal{I}$, l'algorithme insère les motifs rares de $\mathcal{I}$ dans $\mathcal{MFR}$. À partir de la deuxième phase, l'algorithme aura deux ensembles $\mathcal{MFR}=\{(A,2),(B,2),(C,1),(D,2)\}$ et $\mathcal{C}_{i}=\{(AB,2),(AC,0),(AD,1),(BC,1),(BD,2),(CD,1)\}$ où $\mathcal{C}_{i}$ est l'ensemble des motifs candidats munis de leurs supports. La dernière étape consiste à remplacer chaque motif de $\mathcal{MFR}$ par son sur-ensemble s'ils ont le même support. D'où, l'ensemble des fermés rares sera $\mathcal{MFR}=\{(AB,2),(AD,2),(BC,1),(BD,2),(CD,1)\}$.\\

Dans ce qui suit, nous présentons les procédures et les fonctions ainsi que les pseudo-codes relatifs pour chaque module.

\subsubsection{Fonctions et procédures utilisées par l'algorithme \textsc{MFRare}}

\textbf{$\bullet$ La procédure insertItemToMFR :} cette procédure prend en entrée l'itemset que l'on veut insérer dans la liste des \textsc{MFR}s. Cette procédure, et selon la valeur de retour de la fonction  \emph{existInMFR}, insère le motif dans la liste des \textsc{MFR}s. La procédure  \emph{insertItemToMFR} n'assure que l'insertion du motif dans la bonne position de l'ensemble $\mathcal{MFR}$. L'ensemble $\mathcal{MFR}$ est organisé selon deux critères à savoir la cardinalité  et l'ordre lexicographie des motifs.

\textbf{$\bullet$ La fonction existInMFR : }la fonction  \emph{existInMFR} prend en entrée l'itemset candidat et retourne comme résultat une valeur booléenne. La valeur de retour \emph{vrai}, signifie l'existence d'un sous-ensemble de l'itemset passé en paramètre dans l'ensemble des fermés $\mathcal{MFR}$. La valeur \emph{faux}, sera interprétée par l'absence des sous-ensembles du motif passé en paramètre.

\textbf{$\bullet$ La procédure elagage : }
Cette procédure élague les motifs qui ont des supports égal à zéro, tout en se basant sur la propriété du filtre d'ordre ``\emph{tous les sur-ensembles  d'un motif zéro sont des motifs zéros}''. Ainsi, cette procédure, et contrairement à la procédure d'élagage de l'algorithme \textsc{GMRare}, n'utilise que cet unique critère d'élagage.
\section{Structures de données utilisées}
Afin d'implementer les deux algorithmes à savoir \textsc{MFRare} et \textsc{GMRare}, nous avons utilisé principalement les arbres préfixés nommés aussi \emph{Trie}.
\subsubsection{Les arbres \emph{Trie}}
Pour stocker notre contexte d'extraction en mémoire centrale, nous avons utilisé un arbre préfixé \emph{trie}. L'avantage d'une telle structure c'est qu'elle réduit la redondance dûe au fait que plusieurs itemsets peuvent avoir un ensemble d'items en commun. Ainsi, elle réduit considérablement l'espace mémoire consommé lors du stockage des itemsets candidats.

\begin{figure}
\begin{center}
\includegraphics[scale=0.3]{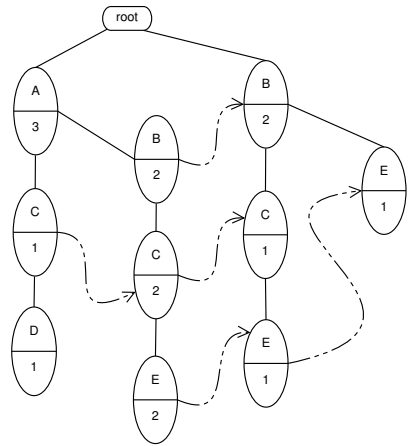}
 \end{center}
\caption{Exemple d'un \textit{trie}.}\label{imagetrie}
\end{figure}

Un \textit{trie} est un arbre de recherche (\textit{cf.} Figure \ref{imagetrie}), dont les données
sont stockées d'une fa\c{c}on condensée
\cite{bodon03fimi}. La structure de données \textit{trie}
était à l'origine introduite pour stocker et pour
récupérer les mots d'un dictionnaire \cite{knuth68}. Un
\textit{trie} est un arbre dirigé du haut vers le bas comme
dans le cas d'un arbre de hachage. Néanmoins, dans le cas d'un
\textit{trie}, nous ne distinguons pas entre un n\oe ud interne et
un n\oe ud feuille contrairement au cas d'un arbre de hachage. En
effet, dans ce dernier, le n\oe ud interne est caractérisé par une
table de hachage et un n\oe ud feuille contenant un ensemble
d'itemsets \cite{bodon03trie}. Dans un \textit{trie}, la racine
est considérée à une profondeur 0, et un n{\oe}ud à une
profondeur $d$ ne peut pointer qu'aux n{\oe}uds de profondeur
($d+1$). Un pointeur est appelé aussi \textit{branche} ou
\textit{lien}, et est étiqueté par~une lettre. Si un
n{\oe}ud \texttt{u} pointe sur un n{\oe}ud \texttt{v}, \texttt{u}
est appelé le \textit{parent} de \texttt{v} et \texttt{v} un
\textit{enfant} de \texttt{u}.

Chaque feuille \textit{l} représente un mot qui est la
concaténation des lettres qui se trouvent sur le chemin de la
racine à \textit{l}. Notons que si les premières $k$ lettres
sont les mêmes pour deux mots, alors les premières $k$
branches sur leur chemins sont aussi les mêmes
\cite{bodon03fimi}.

La structure de données \textit{trie} permet de stocker et de
récupérer non seulement les mots mais n'importe quel
ensemble $E$ fini et ordonné. Dans ce cas, chaque lien est
étiqueté par un élément de $E$, et le
\textit{trie} contient un sous-ensemble $F$ de $E$ s'il y a un
chemin où les liens sont étiquetés par les
éléments de $F$ dans l'ordre choisi.

\bigskip
Dans le contexte de la fouille de données, l'alphabet est
l'ensemble (ordonné) des items $\mathcal{I}$. Un $k$-itemset
$c$ = ${\{}i_{_{1}} < i_{_{2}} < i_{_{3}}< \ldots <i_{_{k}}{\}}$
peut être vu comme étant le mot $i_{_{1}}i_{_{2}}i_{_{3}}
\ldots i_{_{k}}$ composé des lettres de $\mathcal{I}$. Dans
notre cas, le chemin menant de la racine à chaque n\oe ud du
\textit{trie} représente un (candidat) générateur minimal.\\
Dans \cite{bodon03trie}, les auteurs montrent l'intérêt de
l'utilisation de la structure de données \textit{trie} comparée à
la structure de donnée d'arbre de hachage en considérant
différents critères, tels que la simplicité d'utilisation,
l'extraction des informations, etc. Dans \cite{thesebastide},
l'auteur compare différentes implémentations possibles de la
structure de données \textit{trie}. Dans
\cite{bodon03fimi,bodon04fimi,mueller951}, les auteurs montrent
l'efficacité de la structure \textit{trie}, appliquée à
l'algorithme \textsc{Apriori} \cite{Agra94}, le premier à avoir
introduit l'utilisation de l'arbre de hachage, durant l'étape de
calcul des supports des itemsets.
\begin{exemple} Considérons l'ensemble des motifs $\{ABC,ABD,ACD,BCD\}$. Le trie stockant cet ensemble est donné par  la figure \ref{imagetrie}.
\end{exemple}
Le trie possède une caractéristique intéressante favorisant son utilisation dans les algorithmes de fouille de données du type ``Générer et Tester'', à savoir le coût de l'étape de génération des candidats des itemsets candidats. En effet, deux noeuds de profondeur $j$ qui possèdent le même parent, ont nécessairement ($j-1$) items en commun. Par conséquent, un itemset candidat $\{i_{1},...,i_{j-1},i_{j}\}$ est construit en insérant directement $i_{j}$  dans la liste des fils du noeud $i_{j-1}$ de son parent $\{i_{1},...,i_{j-1}\}$.\\
Ainsi, la complexité de génération d'un itemset candidat à partir d'itemset stockés dans un trie est de l'ordre de $\mathcal{O}(1)$.
\section{Preuves théoriques}
Dans cette section, nous montrons la validité des algorithmes \textsc{GMRare} et \textsc{MFRare}. Ensuite, nous prouvons la terminaison et la complétude des deux algorithmes.
\begin{proposition} L'algorithme \textsc{GMRare} est valide. En effet, il permet de déterminer tous les générateurs minimaux rares ainsi que les générateurs zéros minimaux.
\end{proposition}
\textbf{Preuve.}\\
\textsc{GMRare} détermine tous les générateurs minimaux. En effet, \textsc{GMRare} parcourt l'espace de recherche niveau par niveau. Tout au long du parcours, notre algorithme élimine un candidat $v$ de taille $k$ ayant un de ses sous-ensembles de taille $k-1$ noté $v_{1}$, si l'une de deux conditions suivantes sont satisfaites :
\begin{enumerate}
  \item  Le Support de $v$ est nul, $v$ sera élagué et inséré dans l'ensemble des $\mathcal{GMZ}$.
  \item  Le support de $v_{1}$ est égal au support de $v$. Ce dernier ne pourra pas être un générateur minimal car il existe un sous-ensemble de $v$ (en l'occurrence $v_{1}$ ) ayant la même fermeture que $v$
\end{enumerate}
Si aucune de ces deux conditions n'est vérifiée et si $v$ est rare, alors il sera inséré dans l'ensemble des générateurs minimaux rares.
Si $v$ n'est pas un générateur minimal alors l'algorithme élague les sur-ensembles de $v$ puisqu'aucun d'entre eux ne peut être un générateur minimal.\\

\textsc{GMRare} détermine tous les générateurs zéros minimaux. En effet, \textsc{GMRare} parcourt l'espace de recherche par niveau. À chaque rencontre d'un motif zéro $v$ d'un niveau $k$, l'algorithme insère ce dernier dans l'ensemble des $\mathcal{GMZ}$ avant de l'élaguer. Puisque l'algorithme a éliminé $v$, alors pas de génération des sur-ensembles de $v$ qui sont tous des motifs zéros. Ainsi, il n'existe aucun motif $w$ tel que support de $w$ est égal à zéro et $w$ est inclus dans $v$. ceci signifie que $v$ est un générateur zéro minimal.

\begin{proposition} L'algorithme \textsc{GMRare} termine.
\end{proposition}
\textbf{Preuve.}\\
Le nombre de motifs généré par l'algorithme est fini. En effet, le nombre maximal des motifs candidats généré par l'algorithme pour un contexte de $n$ items est au plus égal à $2^{n}$.\\

Nous concluons que \textsc{GMRare} permet d'extraire avec exactitude tous les éléments de l'ensemble $\mathcal{GMR}$ ainsi que les éléments de l'ensemble $\mathcal{GMZ}$ muni de leurs supports corrects. Il est donc correct et complet.
\begin{proposition} L'algorithme \textsc{MFRare} est valide. En effet, il permet de déterminer tous les fermés rares ainsi que la bordure positive.
\end{proposition}
\textbf{Preuve.}\\
\textsc{MFRare} détermine tous les fermés rares. En effet, \textsc{MFRare} parcourt l'espace de recherche niveau par niveau. Tout au long du parcours, notre algorithme construit l'ensemble des fermés de manière progressive en se basant sur l'élimination. Au début, l'ensemble des fermés rares est initialisé  à l'ensemble total des items munis de leurs supports. Au fur et à mesure, l'algorithme élimine un  motif $v$ de taille $k-1$ de l'ensemble des fermés rares si l'un de ces sur-ensembles de taille $k$ noté $v_{1}$ ayant le même support que lui. Dans ce cas, le motif $v_{1}$ sera inséré à la place de $v$.
Tout au long du parcours, notre algorithme élimine un candidat $w$ de taille $k$ de support zéro puisque tous ces sur-ensembles vont être des motifs zéros.\\
\textsc{MFRare} détermine la bordure positive. Cette bordure contient les motifs qui sont rares ainsi que ses sous-ensembles qui sont fréquents. En effet, \textsc{MFRare} parcourt l'espace de recherche par niveau. À chaque rencontre d'un motif rare $v$ d'un niveau $k$ l'algorithme insère ce dernier dans l'ensemble des motifs maximaux fréquents noté $\mathcal{MFM}$, dans le cas où aucun de ses sous-ensembles n'appartient à cet ensemble.

\begin{proposition} L'algorithme \textsc{MFRare} termine.
\end{proposition}
\textbf{Preuve.}\\
Le nombre de motifs généré par l'algorithme est fini. En effet, le nombre maximal des motifs candidats généré par l'algorithme pour un contexte de $n$ items est au plus égal à $2^{n}$.\\

Nous concluons que \textsc{MFRare} permet d'extraire avec exactitude tous les éléments de l'ensemble $\mathcal{MFR}$ ainsi que l'ensemble $\mathcal{MRM}$ muni de leurs supports corrects. Il est donc correct et complet.

\section{Complexité}
Dans cette section, nous nous proposons de déterminer la complexité au pire de cas de l'algorithme \textsc{GMRare}. Il est à noter que le contexte pire des cas contient un nombre des classes d'équivalences égal au nombre des générateurs minimaux qui sont de même égaux au nombre des motifs fermés, \textit{i.e.}, chaque classe d'équivalence contient un unique générateur minimal qui est de même le fermé de cette classe. Par ailleurs, l'algorithme $\textsc{GMRare}$ et l'algorithme $\textsc{MFRare}$, dans un tel contexte, ont la même complexité théorique. Dans ce qui suit, nous détaillons le calcul de la complexité du premier algorithme.

La complexité de l'algorithme $\textsc{GMRare}$ est la somme des complexités de trois parties qui le constituent. La première partie est consacrée à la génération des candidats afin de conserver les \textsc{GMR}s. La deuxième partie, nommée \textit{Calcul support}, consiste à calculer les supports des motifs candidats afin de ne retenir que ceux qui sont des motifs non-zéros. La troisième partie, nommée \textit{élagage et génération du résultat}, consiste à conserver les motifs rares, qui vérifient la définition du générateur minimal. Étant donné que nous n'avons pas trouvé un contexte qui maximise la somme des complexités des trois étapes, nous allons calculer la complexité de notre algorithme \textsc{GMRare} comme étant la somme des complexités des trois parties dans le pire des cas.

Soit $\mathcal{K}=(\mathcal{O},\mathcal{I},\mathcal{R})$ un contexte d'extraction avec $|\mathcal{O}|=m$ et $|\mathcal{I}|=n$. Dans ce qui suit, nous supposons que toutes les transactions ont pour longueur $n$. D'où, pour chaque transaction il existe $2^n$ sous-ensembles à générer afin de calculer leurs supports. En effet, le coût des tests d'inclusion et de l'opération d'intersection entre deux motifs de taille $n$ est de l'ordre de $\mathcal{O}(n)$.\\
Dans la première partie, nommée génération des candidats, le nombre des candidats à générer est égal à ($2^{n}-1$), d'où, le coût de cette génération est de l'ordre de $\mathcal{O}(2^{n})$.\\
Dans la deuxième partie, le coût de calcul des supports est la multiplication du coût de calcul du support d'un motif de taille $n$ dans une transaction par le nombre de transaction. Ainsi, le coût de l'opération de calcul du support d'un motif est de l'ordre de $\mathcal{O}(n)$, le coût de calcul du support d'un motif dans un contexte de données est de l'ordre de $\mathcal{O}(n \times m)$ (nous avons supposé l'existence de $m$ transactions). D'où, le coût de la phase de calcul des support d'un contexte d'extraction est de l'ordre de $(2^{n}-1) \times (m \times n)$. cette quantité peut être remplacée par $\mathcal{O}((m \times n) \times 2^{n}$).\\
La dernière étape, nommée élagage et génération du résultat, comporte deux manières d'élagage à savoir l'élagage à travers le support du motif qui est de l'ordre de $\mathcal{O}(2^n-n)$ et l'élagage selon la vérification de l'idéal d'ordre pour les générateurs minimaux est de l'ordre de $\mathcal{O}(n^2 \times (2^n-n))$.\\
De plus, le coût des stockage des liens et le coût d'initialisation des ensembles est de l'ordre de $\mathcal{O}(1)$.

Ainsi, La complexité de l'algorithme \textsc{GMRare}, au pire des cas, est de l'ordre de :\\
$\mathcal{C}_{pire} \equiv \mathcal{O}(2^{n}) + \mathcal{O}(2^{n} \times (m \times n)  + \mathcal{O}(2^n-n)+ \mathcal{O}(n^2 \times (2^n-n))+\mathcal{O}(1) \\ \equiv \mathcal{O}(2^{n} + 2^{n} \times (m \times n)+
(2^n-n)+ n^2 \times (2^n-n)+1) \equiv \mathcal{O}(2^{n} \times (n^2+m \times n+2)) \equiv \mathcal{O}((n^2+m \times n) \times 2^n )$.\\
D'où, $\mathcal{C}_{pire} \equiv \mathcal{O}((n^2+m \times n) \times 2^n )$.\\

Il est important de noter que la complexité des algorithmes est exponentielle en fonction du nombre d'items noté $n$ et polynomiale en fonction du nombre de motifs noté $2^n$.

\section{Conclusion}
Dans ce chapitre, nous avons introduit deux nouveaux algorithmes, nommés respectivement \textsc{GMRare} et \textsc{MFRare}, permettant l'extraction des deux représentations concises exactes des motifs rares basés respectivement sur les générateurs minimaux rares noté $\mathcal{GMR}$ et sur les fermés rares noté $\mathcal{MFR}$. Nous avons également présenté les principales structures de données mises en oeuvre afin d'implémenter ces algorithmes et nous avons prouvé la complétude de ces deux algorithmes.\\

Dans le chapitre suivant, nous nous proposons de mener une étude expérimentale grâce à laquelle nous évaluons la taille de nos représentations concises exactes par rapport à celle des représentations concises de la littérature. À cet effet, nous quantifions le temps consommé par nos deux algorithmes à savoir \textsc{GMRare} et \textsc{MFRare} en les comparant aux temps des algorithmes d'extraction des motifs rares.
\chapter{Étude expérimentale}
\minitoc

\section{Introduction}

Dans le chapitre précédent, nous avons introduit deux algorithmes dédiés à extraire deux représentations concises exactes de motifs rares.
À cet effet, ces algorithmes utilisent la structure de données trie pour stocker les motifs et un parcours  par niveau du treillis, afin d'atteindre les objectifs suivants :
 \begin{itemize}
   \item Minimiser le temps d'extraction des représentations concises exactes à savoir celle basée sur les générateurs minimaux et celle basée sur les fermés rares.
   \item Réduire la cardinalité de l'ensemble des motifs à extraire.
 \end{itemize}

Dans ce chapitre, nous nous proposons de mener notre étude expérimentale
qui permet : d'une part, d'étudier l'aspect cardinalité des ensembles
des motifs rares extraits suivant l'approche associée, et d'autre part, le temps d'extraction des algorithmes proposés en comparaison avec ceux de la littérature. Les expérimentations ont été réalisées sur des bases benchmark denses et éparses.

\section{Environnement d'expérimentation}

\subsection{Environnement matériel et logiciel}
Toutes les expérimentations ont été réalisées sur un PC muni d'un processeur
Intel Pentium IV ayant une fréquence d'horloge 3.0Ghz et 755Mo de
mémoire vive (avec 1G d'espace d'échange swap) tournant sous la plate-forme
Ubuntu dans sa version 8.10 (Gnome 2.24.1). Il faut mentionner que nous avons utilisé trois algorithmes de la littérature dont nous avons les codes sources à l'instar de \textsc{Minit} est implémenté en C++, \textsc{Apriori-rare} et \textsc{Arima}  sont implémentés en Java. De plus, nos algorithmes \textsc{MFRare} et \textsc{GMRare} sont implémentés en C++.

\subsection{Description des exécutables}
Le tableau \ref{tab_algo} présente la description détaillée des exécutables de chaque algorithme présenté dans ce chapitre.

\newcolumntype{M}[1]{>{\raggedright}p{#1}}
\begin{table}[!htbp]
\begin{tabular}{|c||M{4.5cm}|M{8cm}|}
\hline
\textbf{Algorithme} & \textbf{Entrée} & \textbf{Sortie}\tabularnewline
\hline
\hline
\textsc{Minit} & Fichier ascii (.dat), \textit{maxsupp}, \textit{maxc}&
Ensemble des motifs rares minimaux, dont la cardinalité de ces motifs est  égale à \textit{maxc}.%
\tabularnewline
\hline
GMRare & Fichier ascii (.dat), \textit{maxsupp}&
Ensemble des générateurs minimaux des motifs rares munis de leur supports et l'ensemble des GMZ.%
\tabularnewline
\hline
MFRare & Fichier ascii (.dat), \textit{maxsupp}&
Ensemble des fermés rares, la bordure négative des motifs rares et l'ensemble des GMZ.%
\tabularnewline
\hline
Apriori-rare&Fichier ascii (.dat), \textit{maxsupp}&
Ensemble des motifs rares minimaux, ensemble des générateurs zéros GMZ.%
\tabularnewline
\hline
ARIMA&Ensemble des motifs rares minimaux et l'ensemble des générateurs zéros&
Ensemble total des motifs rares munis de leurs supports.%
\tabularnewline
\hline
\end{tabular}
\caption{Description des exécutables utilisés.}
\label{tab_algo}
\end{table}

\subsection{Description des bases de test}
Dans le cadre de nos expérimentations, nous avons comptabilisé les cardinalités des sorties des algorithmes déjà cités dans les bases benchmarks usuelles.
Les bases benchmark sont divisées en deux classes :
\begin{itemize}
           \item les bases denses.
           \item les bases éparses.
         \end{itemize}

L'ensemble de ces bases est disponible à l'adresse suivante: \textsl{http://fimi.cs.helsinki.fi/data}.
Les caractéristiques de ces bases sont résumées par le tableau \ref{Carac_BDTs}. Ce tableau définit pour chaque base, son type (dense ou épars), le nombre de transactions, le nombre d'items et la taille moyenne des transactions. Les bases \textsc{Chess} et \textsc{Connect} sont dérivées à partir des étapes de jeux respectifs qu'elles représentent. Ces bases sont fournies par \textsc{UC} Irvine Machine Learning Database Repository $^{(}$\footnote{Accessible à l'adresse : \textsl{http://www.ics.uci.edu/~mlearn/MLRepository.html}}$^{)}$. Les deux bases synthétiques \textsc{T10I4D100K} et
\textsc{T40I10D100K} sont générées par un
programme développé dans le cadre du projet
db\textsc{Quest} $^{(}$\footnote{Le générateur des bases
synthétiques est disponible à l'adresse suivante
\textsl{http://www.almaden.ibm.com/software/quest/Resources/datasets/data/}.}$^{)}$.
Ces bases simulent le comportement d'achats des clients dans des
grandes surfaces \cite{hippsurvey}. Le tableau \ref{basesynt}
présente les différents paramètres des bases synthétiques.

\begin{table}[!htbp]
\begin{center}
\small{
\begin{tabular}{|c|l|}
\hline
 $|$T$|$ &   Taille moyenne des transactions \\\hline $|$I$|$ &
  Taille moyenne des itemsets maximaux potentiellement fréquents\\\hline$|$D$|$&   Nombre de transactions générées \\
\hline
\end{tabular}
} \caption{Paramètres des bases synthétiques.} \label{basesynt}
\end{center}
\end{table}

\begin{table}[!htbp]
\begin{center}

  \small{
    \begin{tabular}{||l||r|r|r|r||}
  \hline \hline
\textsc{Nom de}& \textsc{Type de}&\textsc{Nombre de}&\textsc{Nombre}& \textsc{Taille moyenne}\\
\textsc{la base} &\textsc{la base} &\textsc{ transactions} &\textsc{d'items} &\textsc{des transactions}\\
  \hline \hline
       \textsc{Connect} & Dense& 67 557 & 129 & 43\\
  \hline
    \textsc{Chess}& Dense &3 196&75& 37\\
  \hline
   \textsc{Mushroom}& Dense&8 124  & 119 &23 \\
    \hline \hline
  \textsc{T10I4D100K}& Epars&100 000&1 000 & 10\\
   \hline
   \textsc{T40I10D100K}& Epars&100 000&1 000 &40\\
  \hline \hline
\end{tabular}
} \caption{Caractéristiques des bases benchmark.}
\label{Carac_BDTs}
\end{center}
\end{table}

\section{Étude quantitative de l'ensemble des motifs rares}
Dans cette section, nous nous proposons de mener une étude sur la cardinalité des deux représentations concises exactes des motifs rares, ainsi que des différents ensembles des constituants. Tout d'abord, nous commençons par une comparaison expérimentale de la cardinalité des représentations concises par rapport à celle des méthodes existantes. Par la suite, nous entamons la quantification des ensembles formant nos représentations concises à savoir $\mathcal{MFR}$ et $\mathcal{GMR}$ et nous terminons notre étude par une comparaison entre nos deux représentations concises exactes.\\
Dans ce qui suit, nous allons diviser l'étude expérimentale de la cardinalité des ensembles selon la nature des bases benchmarks utilisées.
\subsection{Étude de la taille de la représentation basée sur les générateurs minimaux pour des bases denses}
Le tableau \ref{tab_mgr_mfr_den} et la figure \ref{fig_card_mush} présentent les cardinalités de la représentation concise basée sur $\mathcal{GMR}$ ainsi que les méthodes proposées dans la littérature. Il faut mentionner que quelques valeurs de l'ensemble $\mathcal{MRM}$ $^{(}$\footnote{$\mathcal{MRM}$ est l'ensemble des motifs rares minimaux.}$^{)}$ sont amoindries puisque l'algorithme (\textsc{Apriori-rare}), qui fournit cet ensemble, échoue parfois à cause des problèmes de gestion de la mémoire. Le tableau \ref{tab_mgr_mfr} illustre les facteurs multiplicatifs des cardinalités de notre représentation concise exacte par rapport aux méthodes existantes. Les colonnes 4 et 5 de ce tableau indiquent respectivement les facteurs multiplicatifs de la taille de l'ensemble $\mathcal{MRM}$ et $\mathcal{MR}$, qui sont les résultats de deux algorithmes \textsc{Apriori-rare} et \textsc{Minit}, par rapport à l'ensemble $\mathcal{GMR}$ (le résultat de notre première représentation concises exacte).\\

- \textbf{\textsc{Mushroom} :}
En analysant le facteur multiplicatif $\frac{|\mathcal{MRM}|}{|\mathcal{GMR}|}$ (colonne 4 du tableau \ref{tab_mgr_mfr_den}), nous remarquons que la cardinalité de $\mathcal{MRM}$ est inférieure à notre représentation concise exacte. L'explication découle directement de la définition de l'ensemble $\mathcal{MRM}$, premièrement cet ensemble ne représente qu'une frontière entre les motifs rares et ceux fréquents, cette frontière est incluse dans l'ensemble $\mathcal{GMR}$. Deuxièmement, l'algorithme s'est achevé dans quelques tests et donc ces nombres peuvent être plus grandes. Cependant, l'algorithme \textsc{Apriori-rare} nécessite un deuxième algorithme pour pallier la totalité de l'ensemble des motifs rares. Pour des \emph{minsupp} inférieurs à 1\%, la cardinalité de notre représentation $\mathcal{GMR}$ est de l'ordre de 2 par rapport à celle de $\mathcal{MRM}$ cependant à partir des valeurs de \emph{minsupp} plus élevées, l'ordre de multiplicité atteint 7 et ça peut être expliqué par le nombre très élevé des générateurs minimaux.\\
Nous focalisons notre étude sur l'ensemble $\mathcal{MR}$ qui représente la totalité des motifs rares. L'observation de la $5^{\grave{e}me}$ colonne montre le gain apporté par notre représentation concise exacte par rapport à l'extraction totale des motifs rares, le facteur multiplicatif atteint pour des valeurs de \emph{minsupp} la valeur 2. Il faut mentionner que l'algorithme \textsc{Minit} n'a pas pu fournir des résultats pour des valeurs de \emph{minsupp} supérieures à 40\% et le message d'erreur suivant a été affiché : ``process KILLED'', l'explication découle de la taille énorme de l'ensemble des motifs rares à générer.\\

- \textbf{\textsc{Chess} :}
Tout comme la base \textsc{Mushroom}, le facteur multiplicatif de notre représentation concise par rapport à l'ensemble $\mathcal{MRM}$ est nettement supérieur. L'ensemble $\mathcal{MRM}$ est inclus dans l'ensemble $\mathcal{GMR}$ puisque les minimaux rares appartiennent à l'ensemble des générateurs minimaux ce qui explique la différence au niveau de la taille entre les \textsc{MRM}s et les \textsc{GMR}s. Nous remarquons que les facteurs multiplicatifs entre l'ensemble $\mathcal{GMR}$ et l'ensemble $\mathcal{MRM}$ sont inférieurs à ceux de la base \textsc{Mushroom} ceci implique une réduction du nombre des générateurs minimaux dans cette base. La constatation des facteurs multiplicatifs entre l'ensemble total des motifs rares et l'ensemble des générateurs minimaux (colonne 5 tableau \ref{tab_mgr_mfr_den}) prouve l'infériorité des générateurs minimaux dans cette base.\\

- \textbf{\textsc{Connect} :}
En examinant le facteur multiplicatif de la colonne 5, nous remarquons que la cardinalité de notre représentation concise exacte est largement inférieure à l'ensemble total des motifs rares. Les mêmes constatations et les mêmes explications, déjà citées pour les autres bases, sont valides pour cette base. À travers une observation horizontale du tableau \ref{tab_mgr_mfr_den}, nous constatons que toutes les bases ont les mêmes spécifications et pour les mêmes explications. Cependant, une observation verticale du tableau \ref{tab_mgr_mfr_den} nous amène à constater que, contrairement aux bases précédentes, le facteur multiplicatif diminue pour une augmentation du seuil \emph{minsupp}. L'explication de cette diminution découle du fait que le nombre des générateurs minimaux, dans la base \textsc{Connect} augmente pour des \emph{minsupp} très élevés. La preuve peut être déduite à partir d'une analyse du tableau \ref{tab_mgr_mfr_den}. Ainsi, pour \emph{minsupp}=90\%, la taille de la bordure vaut 102 348 motifs et l'ensemble des générateurs minimaux contient 192 317 motifs.

\subsection{Étude de la taille  de la représentation basée sur les générateurs minimaux pour des bases éparses}
Les cardinalités des méthodes existantes ainsi que notre représentations concises exactes sont présentées dans le tableau \ref{tab_mgr_mfr_den}. Les facteurs multiplicatifs des cardinalités des méthodes existantes par rapport à nos représentations concises exactes sont résumés dans le tableau \ref{tab_mult_mgr_mfr}.\\

- \textbf{\textsc{T40I10D100K} :}
En examinant les colonnes 2, 5 et 6 du tableau \ref{tab_mgr_mfr}, nous remarquons que la cardinalité de l'ensemble $\mathcal{GMR}$ est toujours plus réduite que celle de l'ensemble $\mathcal{MR}$ et supérieur à la cardinalité de l'ensemble $\mathcal{MRM}$. Les facteurs multiplicatifs entre ces ensembles, illustrés par le tableau \ref{tab_mult_mgr_mfr}, peuvent atteindre des valeurs supérieures à 2 pour des \textit{minsupp} dépassant le $4 \%$.\\

- \textbf{\textsc{T10I4D100K}:}
Nous remarquons que l'ensemble des maximaux fréquents, dans les bases éparses à l'instar de T10I4D100K et T40I10D100K, est plus réduit que l'ensemble des minimaux rares. L'explication peut découler du fait que ces bases contiennent plusieurs motifs rares et un nombre réduit des fréquents. En analysant les colonnes 2 et 3 du tableau \ref{tab_mgr_mfr}, nous remarquons la diminution du nombre des motifs rares ainsi que le nombre des motifs fermés rares. Ceci peut être lié aux caractéristiques internes de cette base \textit{cf}. tableau \ref{basesynt}, taille moyenne de transaction 10 pour 100 000 transactions.

\subsection{Étude de la taille de la représentation basée sur les fermés pour des bases denses}
Les cardinalités des ensembles $\mathcal{MR}$, $\mathcal{MFM}$, $\mathcal{MRM}$ et $\mathcal{MFR}$ sont données dans le tableau \ref{tab_mgr_mfr_den}. Ces ensembles représentent respectivement l'ensemble des motifs rares, l'ensemble des motifs fréquents maximaux, l'ensemble des motifs rares minimaux et l'ensemble des fermés rares. Le tableau \ref{tab_mult_mfr} présente les facteurs multiplicatifs des représentations concises exactes $ \mathcal{MFR} \cup \mathcal{MFM}$ \texttt{OU} $ \mathcal{MFR} \cup \mathcal{MRM}$ par rapport aux autres ensembles déjà cités.\\

- \textbf{\textsc{Mushroom} :}
En analysant les colonnes 3 et 4 du tableau \ref{tab_mgr_mfr_den}, nous constatons une très grande différence entre les deux bordures à savoir la bordure positive pour les fréquents et celle négative pour les rares. Pour des seuils \emph{minsupp} très élevés, le facteur multiplicatif entre les deux bordures peut atteindre 2. L'ensemble des motifs rares minimaux est nettement inférieur à celui des motifs maximaux fréquents. D'une part, le nombre des motifs maximaux fréquents est très grand puisqu'il s'agit d'une base dense. D'autre part, l'ensemble des minimaux rares sont les sur-ensembles des motifs maximaux fréquents. D'où l'augmentation de la taille d'un ensemble entraîne la diminution de l'autre. Nous prenons un exemple, soit les motifs AB, AC et BC font partie de l'ensemble des maximaux fréquents, alors l'unique motif ABC fait partie de l'ensemble des minimaux rares.
Nous focalisons notre étude sur la cardinalité de notre représentation concise exacte qui est égale à la somme des cardinalités de l'ensemble $\mathcal{MFR}$ et de cardinalité de l'une de deux ensembles $\mathcal{MFM}$ ou $\mathcal{MRM}$. La constatation du tableau \ref{tab_mult_mfr} montre que la cardinalité de notre représentation est nettement meilleure pour tous les seuils \emph{minsupp} par rapport aux autres méthodes lorsque nous utilisons l'ensemble des minimaux rares au lieu de celui des maximaux fréquents dans l'ensemble final de $\texttt{R C E M R}$.\\

- \textbf{\textsc{Chess} :}
Cette base est caractérisée par la présence d'une forte corrélation entre ces motifs ce qui implique le nombre réduit des motifs rares. Par exemple, pour \emph{minsupp} égal à 10\%, nous avons 154 motifs rares alors que pour le même seuil nous avons 21 901 motifs dans la base \textsc{Mushroom}. Nous remarquons dans le tableau \ref{tab_mgr_mfr_den} dans les colonnes 3 et 4, qui représentent respectivement les motifs rares minimaux et les motifs maximaux fréquents, une importante variation des facteurs multiplicatifs. L'explication de cette baisse des valeurs des rares découle de l'existence d'un nombre très élevé des fréquents. Ainsi, l'augmentation du nombre des motifs maximaux dû à une corrélation entre les motifs.\\

- \textbf{\textsc{Connect} :}
Cette base est caractérisée par un nombre ainsi qu'une taille moyenne des transactions relativement élevées. Ces caractéristiques influent sur les facteurs multiplicatifs (\textit{cf}. les colonnes 4 et 6 du Tableau \ref{tab_mult_mfr}). Ainsi, les valeurs de ces colonnes sont les plus élevées par rapport aux autres bases. La déduction découle du fait que la majorité des motifs rares est incluse dans l'ensemble des minimaux rares.
À cause de la taille de cette base, nous avons remarqué à partir du tableau \ref{tab_mgr_mfr_den} l'existence d'un nombre très élevé des motifs rares (plus que 289 893 motifs pour un seuil \emph{minsupp }90\%)

\subsection{Étude de la taille de la représentation basée sur les fermés pour des bases éparses}
- \textbf{\textsc{T40I10D100K} :}
En analysant les facteurs multiplicatifs des colonnes du tableau \ref{tab_mult_mgr_mfr}, nous remarquons que dans la plupart des cas, notre représentation concise exacte est inférieure aux méthodes existantes. Tout de même, notre représentation basée sur l'ensemble $\mathcal{MFR}$ enrichie par l'ensemble des motifs fréquents maximaux demeure la plus réduite pour des valeurs faibles de \emph{minsupp}.
La lecture des colonnes 4 et 5, nous conduit à conclure que l'ensemble $\mathcal{MFM}$ est inférieur à celui  $\mathcal{MRM}$ quelque soit les seuils et surtout pour des seuils très élevés où le facteur multiplicatif peut atteindre 3.\\

- \textbf{\textsc{T10I4D100K}:}
Les colonnes 3 et 4 du tableau \ref{tab_mgr_mfr} montrent la différence entre la cardinalité de notre représentation concise par rapport aux autres méthodes. Cette différence apparaît pour des seuils supérieurs à 2\% où les facteurs multiplicatifs atteignent des valeurs proches de 2,5 et 3.
%

\begin{sidewaystable}
\center
\setlongtables
  \begin{longtable}{|l||r|r|r|r|r||r|r|r||r|r|r|r|c|}
    \hline
\emph{minsupp} & $|\mathcal{MR}|$ & $|\mathcal{GMR}|$&$|\mathcal{MFM}|$& $|\mathcal{MRM}|$& $|\mathcal{MFR}|$&$\frac{|\mathcal{MR}|}{|\mathcal{MRM}|}$&$\frac{|\mathcal{GMR}|}{|\mathcal{MRM}|}$&$\frac{|\mathcal{MR}|}{|\mathcal{GMR}|}$
&$\frac{|\mathcal{MRM}|}{|\mathcal{MFR}|+|\mathcal{MFM}|}$&\multirow{22}{*}{\rotatebox{270}{\textit{la page suivante contient le reste du tableau} ...}}
\\
\cline{1-10}\cline{1-10}
\multicolumn{10}{|c|}{\textsc{Mushroom}}&\\
\cline{1-10}\cline{1-10}
0.01 \% & 802& 378&8 124& 104& 221&7,71&3,63&2,12&0,32&\\
\cline{1-10}
0.1 \% & 4 721& 3 281& 7 122& 568&2561&8,31&5,77&1,43&0,18&\\
\cline{1-10}
1\% & 7 377& 4791&6768& 994&3009&7,42&4,81&1,53&0,24&\\
\cline{1-10}
5 \% & 12 322&  6315&1442&  1662& 4996&7,41 &3,79&1,95&0,24&\\
\cline{1-10}
10 \% & 21 901& 11 312&547& 3077& 8372&7,11 &3,67&1,93&0,26&\\
\cline{1-10}
20 \% & 18 001& 17 752&158& 1004& 11 981&17 &17,68&1,01&0,07&\\
\cline{1-10}
40 \% &-& 37 613& 41& 254& 28 866&-&60,24&-&0,08&\\
\cline{1-10}
\cline{1-10}
\multicolumn{10}{|c|}{\textsc{Chess}}&\\
\cline{1-10}
\cline{1-10}

10 \% & 154&  67&2 339 255& 62& 81&2,4&1,08&2,29&0,43&\\
\cline{1-10}
20 \% & 513&  313&509 355& 312&298&1,64&1,00&1,63&0,51&\\
\cline{1-10}
30 \% & 2461& 1173&134 624& 1013& 878&2,42&1,15&2,09&0,53&\\
\cline{1-10}
50 \% & 55 799& 24 313&38 050& 1671& 23 671&33,00&14,54&2,29&0,06&\\
\cline{1-10}
70 \% & 66 313& 39 112&11 463& 4097& 31 675&16,18&9,54&1,69&0,11&\\
\cline{1-10}
80 \% & -&  28 771&3323& 18 464&28 601&435&0,86&-&0,39&\\
\cline{1-10}
90 \% & -&  48 613&891& 32 671& 47 993&435&0,85&-&0,40&\\
\cline{1-10}
\cline{1-10}
\multicolumn{10}{|c|}{\textsc{Connect}}&\\
\cline{1-10}
\cline{1-10}
5 \% & 1 549& 622&743& 557&312&1&0,89&2,49&0,64&\\
\cline{1-10}
10 \% & 3 657& 2117&1 695& 1109&998&1 &0,52&1,72&0,52&\\
\cline{1-10}
30 \% & 6 246& 3742&2575& 1595&1364&3  &0,42&1,66&0,53&\\
\cline{1-10}
50 \% & 24 008& 15 231&5 808&6 621&10 317&4 &0,38&1,57&0,39&\\
\cline{1-10}
70 \% & 75 786& 42 312&9 240& 11461&33642&7 &0,21&1,79&0,75&\\
\cline{1-10}
90 \% & 289 893& 192 317&102 348& 66281&98 986&13 &0,53&1,50&0,40&\\
\cline{1-11}
\end{longtable}
\caption{Tableau comparatif entre les méthodes existantes et nos représentations $\mathcal{GMR}$ et $\mathcal{MFR}$ pour des bases denses.} \label{tab_mgr_mfr_den}
\end{sidewaystable}

\begin{table}[tb]
  \centering
  \small{
  \begin{tabular}{|l||r|r|r|r|}
    \hline
 \multicolumn{5}{|c|}{\textit{Suite du tableau} \ref{tab_mgr_mfr_den}}\\
 \hline \hline
Base&\textit{minsupp}& $\frac{|\mathcal{MR}|}{|\mathcal{MFR}|+|\mathcal{MFM}|}$ & $\frac{|\mathcal{MRM}|}{|\mathcal{MFR}|+|\mathcal{MRM}|}$ & $\frac{|\mathcal{MR}|}{|\mathcal{MFR}|+|\mathcal{MRM}|}$\\
\hline
\hline\textsc{Mushroom}
&0.01\% &0,09&0,320&2,46\\
&0.1\% &0,15 &0,180&1,50\\
&1\% &0,75&0,240&1,84\\
&5\% & 1,91&0,240& 1,85 \\
&10\% &24,5&0,260& 1,91 \\
&20\% &1,48&0,070 &1,38  \\
&40\% &-&0,008 & -  \\

\hline \textsc{Chess}
&10\%   &0,14&0,43 &1,07\\
&20\%   &0,19&0,51&0,84  \\
&30\%   &0,62&0,53 &1,30\\
&50\%   &3,46&0,06& 1,20\\
&70\%   &1,54&0,11& 1,85   \\
&80\%   &-&0,39&-  \\
&90\%   &-&0,40&-   \\

\hline\textsc{Connect}
&5\%&  1,40&0,35&1,78\\
&10\%& 1,35&0,52&1,73\\
&30\%&1,58&0,53&2,11\\
&50\%& 1,48&0,39&1,41\\
&70\% & 1,76&0,67&5,01\\
&90\% & 1,43&0,40&1,75\\
\hline
\end{tabular}
  }
\caption{Facteurs multiplicatifs des cardinalités des méthodes existantes par rapport à nos représentations $\mathcal{GMR}$ et $\mathcal{MFR}$ pour des bases denses.} \label{tab_mult_mfr}
\end{table}


\begin{figure}[!htbp]
\begin{center}
\parbox{5.cm}{\includegraphics[scale=0.7]{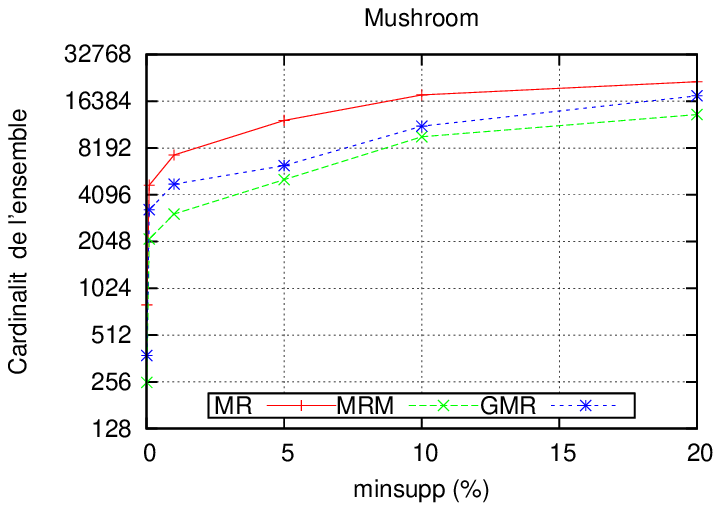}}
\parbox{5.cm}{\includegraphics[scale=0.7]{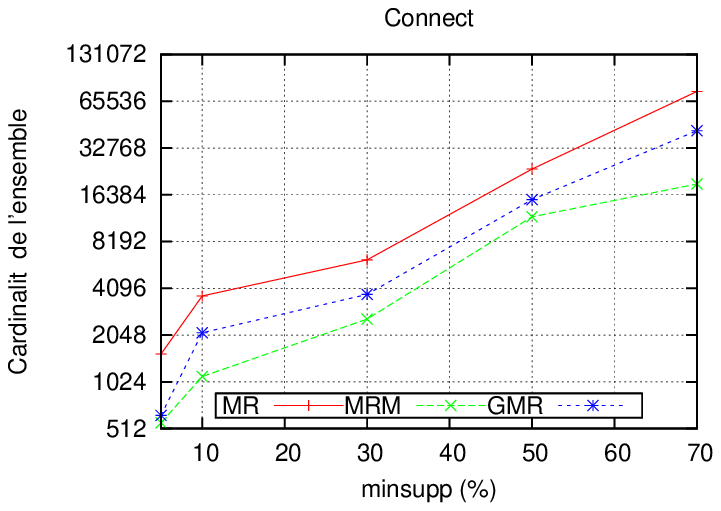}}
\parbox{5.cm}{\includegraphics[scale=0.7]{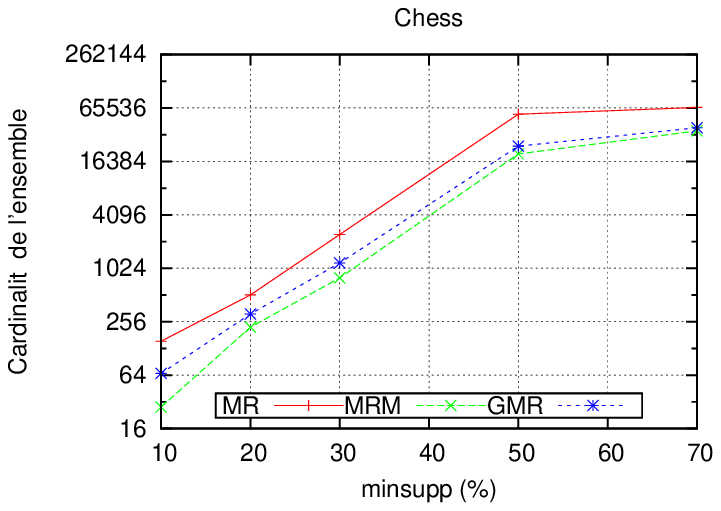}}\\
\parbox{5.cm}{\includegraphics[scale=0.7]{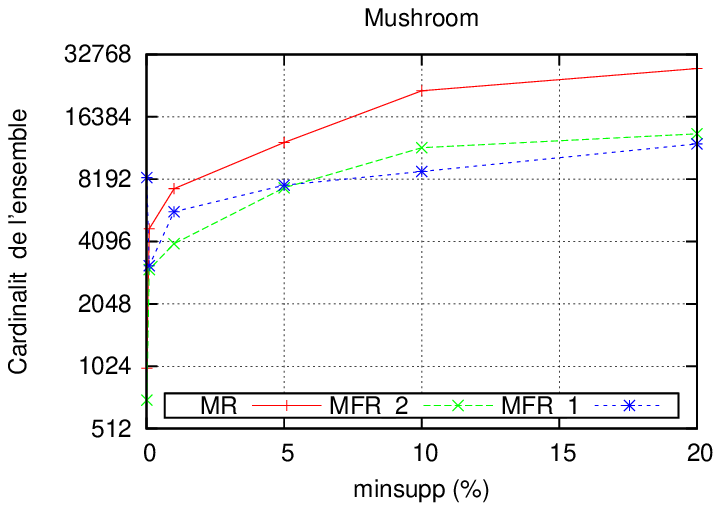}}
\parbox{5.cm}{\includegraphics[scale=0.7]{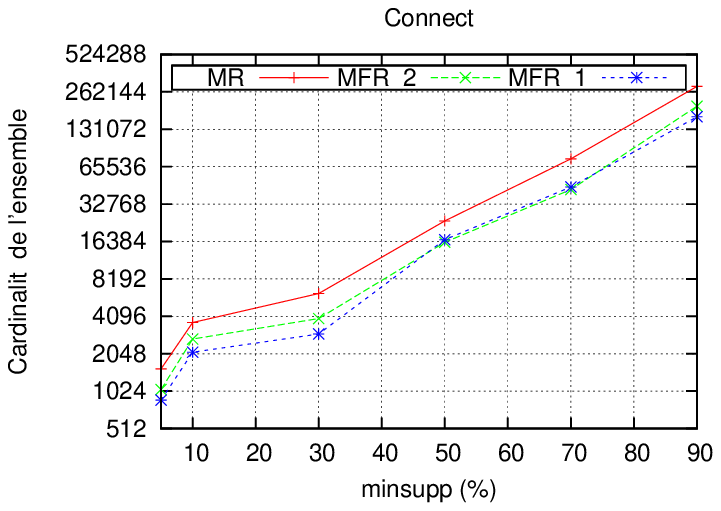}}
\parbox{5.cm}{\includegraphics[scale=0.7]{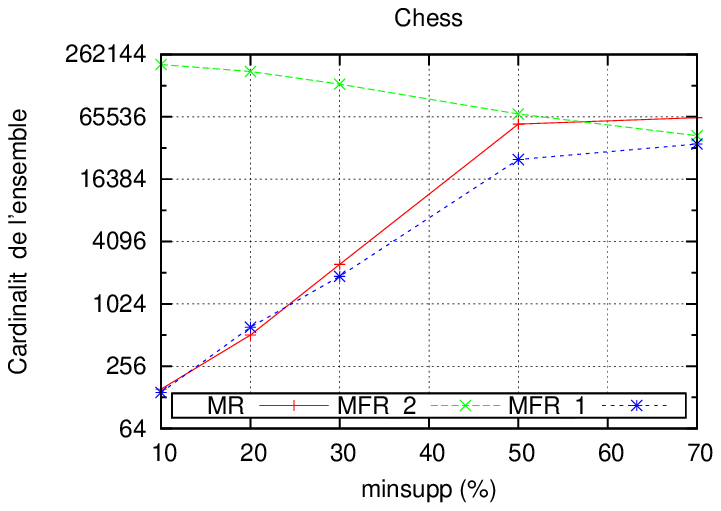}}

\end{center}
\caption{Comparaison de la cardinalité des ensembles $\mathcal{GMR}$, $\mathcal{MFR}_1$, $\mathcal{MFR}_2$, $\mathcal{MR}$ et $\mathcal{MRM}$ pour des bases denses.}
\label{fig_card_mush}
\end{figure}


\begin{table}[tb]
  \centering
  \small{
  \begin{tabular}{|l||r|r|r|r|r|}
    \hline
\emph{minsupp}& $\mathcal{MR}$ & $|\mathcal{MFR}|$ & $\mathcal{MFM}$ & $\mathcal{MRM}$ & $|\mathcal{GMR}|$\\
\hline \hline
\multicolumn{6}{|c|}{\textsc{T40I10D100K}}\\
\hline \hline
1 \% &7 347& 1 783& 21 692& 4969& 6443\\
\hline
2 \% & 23 931& 12 647& 2015& 16 230& 20455\\
\hline
3 \% & 38 426& 21 651& 700& 21 407& 36887\\
\hline
4 \% & 49 965& 33 441& 405&  22 009& 48992\\
\hline
5 \% & 58 901& 39 894& 302& 28 665& 57901\\
\hline
10 \% & 87 146& 42 537& 82& 42 321& 84 244\\
\hline
\hline
\multicolumn{6}{|c|}{\textsc{T10I4D100K}}\\
\hline \hline
0.4 \% & 2 633& 1284& 761& 841& 1783\\
\hline
0.5 \% & 3 298& 2 483&  585& 978&2843\\
\hline
1 \% & 5 427& 4 795& 370& 893&4712\\
\hline
2 \% & 7 922& 5 786& 155& 1126& 6295\\
\hline
3 \% & 12 901& 8 921& 60& 3004&10274\\
\hline
4 \% & 18 501&10 556& 26& 5265&17 924\\
\hline
5 \% & 48 121& 21 306& 10& 13 261& 44 703\\
\hline
\end{tabular}
  }
\caption{Tableau comparatif entre les méthodes existantes et nos représentations $\mathcal{GMR}$ et $\mathcal{MFR}$ pour les bases éparses.}
\label{tab_mgr_mfr}
\end{table}

\begin{sidewaystable}
  \center
  \begin{tabular}{|l||r|r||r|r|r|r|}
    \hline
\emph{minsupp}& $\frac{|\mathcal{MR}|}{|\mathcal{GMR}|}$ & $\frac{|\mathcal{MRM}|}{|\mathcal{GMR}|}$ & $\frac{|\mathcal{MR}|}{|\mathcal{MFR}|+|\mathcal{MFM}|}$ & $\frac{|\mathcal{MR}|}{|\mathcal{MFR}|+|\mathcal{MRM}|}$ & $\frac{|\mathcal{MRM}|}{|\mathcal{MFR}|+|\mathcal{MFM}|}$& $\frac{|\mathcal{MRM}|}{|\mathcal{MFR}|+|\mathcal{MRM}|}$\\
\hline \hline
\multicolumn{7}{|c|}{\textsc{T40I10D100K}}\\
\hline \hline
1 \% &1,14& 0,06& 0,30& 1,08&0,2& 0,70\\
\hline
2 \% & 1,16& 0,10& 1,63& 0,82&1,1& 0,50\\
\hline
3 \% & 1,04& 0,10& 1,70& 0,89&0,9& 0,49\\
\hline
4 \% & 1,01& 0,11& 1,47& 0,90 &0,6& 0,39\\
\hline
5 \% & 1,01& 0,20& 1,70& 0,86&0,7& 0,41\\
\hline
10 \% & 1,01& 0,41& 2,04& 1,02&0,9& 0,49\\
\hline
\hline
\multicolumn{7}{|c|}{\textsc{T10I4D100K}}\\
\hline \hline
0.4 \% & 1,47& 0,05& 1,10& 1,20&0,50& 0,30\\
\hline
0.5 \% & 1,16& 0,30&1,07& 0,96&0,39&0,20\\
\hline
1 \% & 1,15& 0,30&1,07& 0,96&0,17&0,15\\
\hline
2 \% & 1,25& 0,36& 1,30& 1,14&0,19& 0,16\\
\hline
3 \% & 1,25& 0,32& 1,30& 1,08&0,30&0,25\\
\hline
4 \% & 1,03&0,32& 1,70& 1,16&0,40&0,33\\
\hline
5 \% & 1,07& 0,37& 2,23& 1,39&0,62& 0,38\\
\hline
\end{tabular}

\caption{Facteurs multiplicatifs des cardinalités des méthodes existantes par rapport à nos représentations concises $\mathcal{GMR}$ et $\mathcal{MFR}$ pour des bases éparses.}

\label{tab_mult_mgr_mfr}
\end{sidewaystable}

\begin{figure}[!htbp]
\begin{center}
\parbox{8.cm}{\includegraphics[scale=0.95]{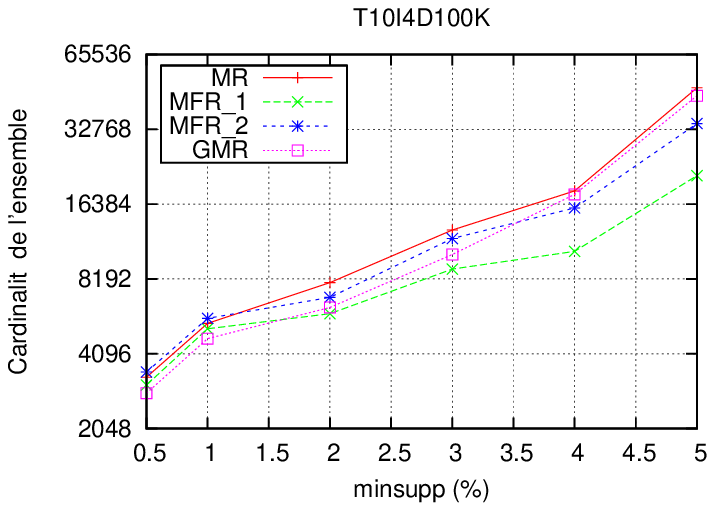}}
\parbox{7.cm}{\includegraphics[scale=0.95]{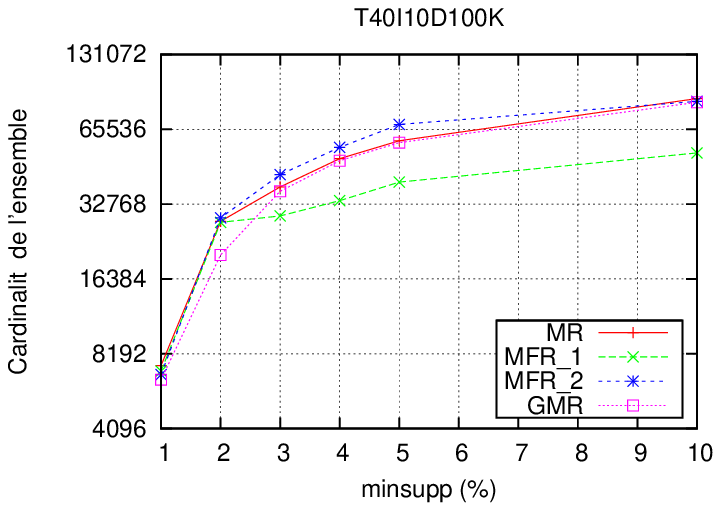}}
\end{center}
\caption{Comparaison de la cardinalité des ensembles $\mathcal{GMR}$, $\mathcal{MFR}_1$, $\mathcal{MFR}_2$ et $\mathcal{MR}$ pour des bases éparses.}
\label{fig_card_mush}
\end{figure}

\subsection{Comparaison des deux représentations $\mathcal{MFR}$ et $\mathcal{GMR}$}
La figure \ref{fig_card_mgr_mfr} illustre les variations de cardinalité des représentations concises exactes ainsi que le tableau \ref{tab_mgr vs mfr} résume dans ces colonnes 2, 3 et 4 les cardinalités de nos représentations concises exactes à savoir celle basée sur les générateurs minimaux notés $\mathcal{GMR}$ ou bien celle basée sur les fermés rares notés $\mathcal{MFR}$ tout en enrichissant cette dernière par l'une de deux ensembles suivants :
\begin{itemize}
  \item L'ensemble des maximaux fréquents noté $\mathcal{MFM}$,
  \item L'ensemble des minimaux rares noté $\mathcal{MRM}$.
\end{itemize}
Dans le tableau \ref{tab_mgr vs mfr}, nous avons utilisé une base de type dense et une autre éparse puisque nous avons remarqué la même variation sur toutes les bases de même type. En examinant les deux colonnes 3 et 4 du tableau \ref{tab_mgr vs mfr}, nous remarquons que la cardinalité de l'ensemble $|\mathcal{MFR} \cup \mathcal{MRM}|$ est légèrement inférieure à l'ensemble $|\mathcal{MFR} \cup \mathcal{MFM}|$ pour les bases denses à l'instar de \textsc{Mushroom}. En analysant les colonnes 2 et 3 du même tableau \ref{tab_mgr vs mfr}, nous remarquons que l'ensemble des générateurs minimaux rares noté $\mathcal{GMR}$ est inférieur à celui des fermés rares enrichi par les minimaux rares noté $|\mathcal{MFR} \cup \mathcal{MRM}|$ pour de telles bases.
L'analyse de la deuxième partie du tableau \ref{tab_mgr vs mfr}, qui concerne les bases éparses à l'exemple de \textsc{T10I4D100K}, montre que la cardinalité de l'ensemble $|\mathcal{MFR} \cup \mathcal{MFM}|$ est inférieure de point de vue cardinalités par rapport aux autres ensembles à savoir l'ensemble $|\mathcal{MFR} \cup \mathcal{MRM}|$ et l'ensemble $\mathcal{GMR}$.\\
Il est intéressant de noter qu'après cette étude nous avons remarqué une forte dépendance entre les contextes de données et les représentations concises adéquates à utiliser pour minimiser le nombre de motifs à extraire.\\
Il est à noter que les représentations $\mathcal{MFR}_1$ et $\mathcal{MFR}_2$ de la figure \ref{fig_card_mgr_mfr} représentent respectivement la représentation concise basée sur les fermés rares augmentée par l'ensemble $\mathcal{MRM}$ et la représentation concise basée sur les fermés augmentée par l'ensemble $\mathcal{MFM}$.
\begin{figure}[!htbp]
\begin{center}
\parbox{8.cm}{\includegraphics[scale=0.95]{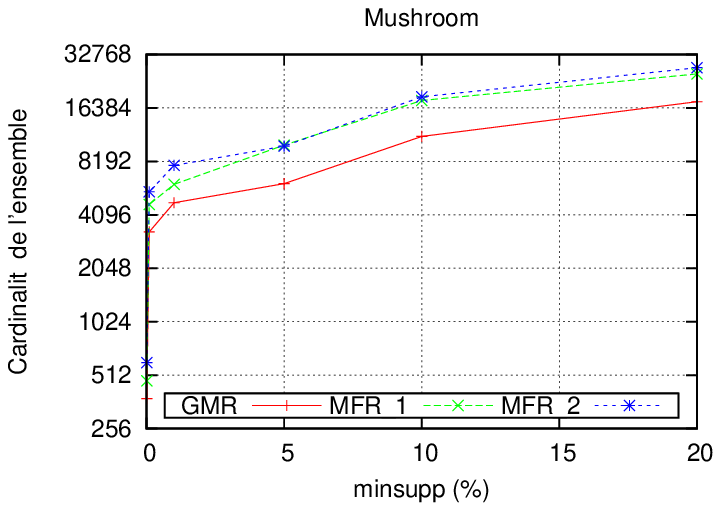}}
\parbox{7.cm}{\includegraphics[scale=0.95]{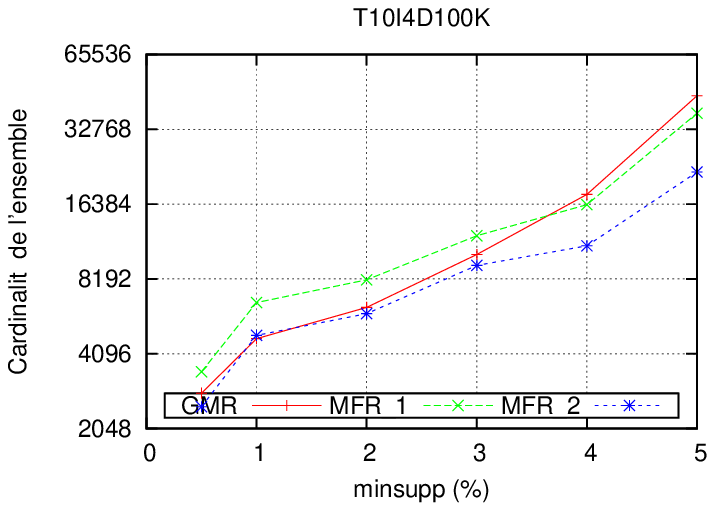}}
\end{center}
\caption{Comparaison de la cardinalité des ensembles $\mathcal{GMR}$, $\mathcal{MFR}_1$ et $\mathcal{MFR}_2$ pour une base éparse et pour une autre dense.}
\label{fig_card_mgr_mfr}
\end{figure}

\begin{table}[tbh]
  \centering
  \small{
  \begin{tabular}{|l||r|r|r|}
    \hline
\emph{minsupp} & $|\mathcal{GMR}|$ & $|\mathcal{RCEFM}|$=$|\mathcal{MFR}|$ + $|\mathcal{MRM}|$ & $|\mathcal{RCEFM}|$=$|\mathcal{MFR}|$ + $|\mathcal{MFM}|$\\
\hline \hline
\multicolumn{4}{|c|}{\textsc{Mushroom}}\\
\hline \hline
0.01 \% & 378 & 325 = 221 + 104& 8345 = 221 + 8124\\
\hline
0.1 \% & 3 281& 3129 = 2 561 + 568&9683 = 2 561 + 7122\\
\hline
1\% & 4 791& 4003= 3 009 + 994&9777 = 3 009 + 6768\\
\hline
5 \% & 6 135& 6658= 4 996 + 1662&6438 = 4 996 + 1442\\
\hline
10 \% & 11 312& 11449= 8 372 + 3077& 8919 = 8 372 + 547\\
\hline
20 \% & 17 752& 12985= 11 981 + 1004&12139 = 11 981 + 158\\
\hline
\hline
\multicolumn{4}{|c|}{\textsc{T10I4D100K}}\\
\hline
\hline
0.5 \% & 2 843& 3324= 2 483 + 841&3244 = 2 483 + 761\\
\hline
1 \% & 4 712& 5773= 4 795 + 978&5380 = 4 795 + 585\\
\hline
2 \% & 6 295& 6679= 5 786 + 893&6156 = 5 786 + 370  \\
\hline
3 \% & 10 274& 10047= 8 921 + 1126&9076 = 8 921 + 155\\
\hline
4 \% & 17 942& 13560= 10 556 + 3004&10616 = 10 556 + 60\\
\hline
5 \% & 44 703& 26571= 21 306 + 5265&21332 = 21 306 + 26\\
\hline
\end{tabular}
  }
\caption{Tableau comparatif des tailles des deux représentations concises $\mathcal{GMR}$ et $\mathcal{MFR}$.}
\label{tab_mgr vs mfr}
\end{table}

\section{Performances d'extraction de l'ensemble des motifs rares }
Dans la section précédente, nous avons vérifié expérimentalement que dans la plupart des cas, la cardinalité des représentations concises exactes, surtout celle basée sur les générateurs minimaux rares, est plus compacte que la taille de l'ensemble total des motifs rares quelque soit les bases et les seuils. Cependant, l'avantage de compacité peut être insignifiant lorsque le temps d'extraction des représentations concises est excessivement grand. Pour cette raison, nous nous proposons dans cette section la comptabilisation des temps d'extraction des deux représentations concises exactes grâce aux deux algorithmes \textsc{MFRare} et \textsc{GMRare}. Nous comparons les performances de deux algorithmes par rapport aux algorithmes \textsc{Minit} et \textsc{Apriori-rare}, tout en ajoutant à ce dernier un deuxième algorithme \textsc{Arima} qui sert à extraire l'ensemble total des motifs rares à partir d'un ensemble des motifs rares minimaux. Les codes sources de l'algorithme \textsc{Arima} et \textsc{Apriori-rare} sont disponibles sur la plate forme \textsc{CORON} à l'adresse \textsl{http://coron.loria.fr/site/index.php}, ainsi que le code source de \textsc{Minit} est disponible à \textsl{http://mavdisk.mnsu.edu/haglin/}. Il faut signaler que les deux premiers algorithmes, à savoir \textsc{Apriori-rare} et \textsc{Arima}, sont écrits en \textsc{Java} tandis que \textsc{Minit} est écrit en \textsc{C++}. Dans ce qui suit, le seuil \emph{maxc} de l'algorithme \textsc{Minit} 
est fixé à la taille maximale des transactions du contexte.\\
Dans toutes les expérimentations, nous avons utilisé la version de l'algorithme \textsc{MFRare} qui fournit en résultat l'ensemble $\mathcal{MFR}$ et l'ensemble des minimaux rares $\mathcal{MRM}$.
Tout comme la section précédente, nous divisons les résultats expérimentaux collectés selon la nature des bases considérées.\\
Il est à noter que dans l'ensemble des résultats présentés dans cette section, nous désignons par ``-'' une exécution qui n'a pas pu arrivé à terme.
\subsection{Performances de l'algorithme \textsc{GMRare} et \textsc{MFRare} \emph{vs.} celles des algorithmes \textsc{Apriori-rare} et \textsc{Minit} pour des bases denses}
Le tableau \ref{tab_tmps_mgr} illustre le temps consommé par les algorithmes \textsc{GMRare}, \textsc{MFRare}, \textsc{Apriori-rare} et \textsc{Minit}. Ainsi, les facteurs multiplicatifs des temps d'exécution de \textsc{Apriori-rare} et \textsc{Minit} par rapport à \textsc{GMRare} et \textsc{MFRare} sont présentés par la  table \ref{tab_tmps_mgr}.
En analysant le tableau \ref{tab_tmps_mgr}, nous remarquons que \textsc{Minit} est l'algorithme le plus long dans l'ensemble des bases denses considérées et pour la plupart des seuils \emph{minsupp}. Cette consommation excessive en temps d'exécution est dûe principalement au traitement récursif de l'algorithme, qui ralentit le processus d'extraction surtout dans des environnements où l'espace mémoire est réduit.

\subsection{Performances de l'algorithme \textsc{GMRare} et \textsc{MFRare} \textit{vs.} celles des algorithmes \textsc{Apriori-rare} et \textsc{Minit} pour des bases éparses}

Le tableau \ref{tab_tmps_mgr} illustre le temps consommé par les algorithmes \textsc{GMRare}, \textsc{MFRare}, \textsc{Apriori-rare} et \textsc{Minit}. Le même tableau présente également les facteurs multiplicatifs des temps d'exécution des algorithmes \textsc{Apriori-rare} et \textsc{Minit} par rapport à \textsc{GMRare} et \textsc{MFRare}.
En examinant le tableau  \ref{tab_tmps_mgr}, nous remarquons que \textsc{Minit} est l'algorithme le plus long dans l'ensemble des bases éparses considérées et pour la plupart des seuils \emph{minsupp}.

\begin{sidewaystable}
  \centering
  \begin{tabular}{|r||r|r|r|r||r|r||r|r||r|}
    \hline
\textit{minsupp}&\textsc{Apriori-rare} & \textsc{Minit} & \textsc{GMRare} &\textsc{MFRare}& $\frac{\textsc{Apriori-rare}}{\textsc{MFRare}}$& $\frac{\textsc{Minit}}{\textsc{MFRare}}$& $\frac{\textsc{Apriori-rare}}{\textsc{GMRare}}$& $\frac{\textsc{Minit}}{\textsc{GMRare}}$&$\frac{\textsc{GMRare}}{\textsc{MFRare}}$\\
&(sec)&(sec)&(sec)&(sec)&&&&&\\
\hline
\hline
\multicolumn{10}{|c|}{\textsc{Mushroom}}\\
\hline
0.01\% & 87&933&81&96    &0,90&9,71&1,07&11,5&9,71\\
0.1\% & 189&1 237&102&172 &1,09&7,10&1,85&11,2&7,19\\
1\% & 522&3 509&213&349 &1,49&10,05&2,40&15,4&10,19\\
5\% & 2 572&8 673&1818& 2 121 &1,21&4,08&1,40&4,0&1,74\\
10\% &6 701&11 439&4356&6 552  &1,02& 1,74&1,50&2,0&1,49\\
20\% &7 874 &13 539&7789&9 034 &0,87&1,49& 1,01&1,7&- \\
40\% &9 097 & - &9621& 11 325 &0,80&-&  0,94&-&\\

\hline
\multicolumn{10}{|c|}{\textsc{Chess}}\\
\hline
10\%   &102&2 551&154&198&1,04&26,03&1,21&30,36&0,85\\
20\%   &547 &5 522&579&576&1,14 &11,60&1,14&11,52&1,00\\
30\%   &1 789&10 767&1384&1 980&1,62&9,80&1,81&10,94&0,89\\
50\%   &3 201&13 882&2205&2 874& 1,11&4,83&1,45&6,29&0,76\\
70\%   &5 321&14 321&4 337&4 582& 1,47&3,34&1,45 &3,30&1,01  \\
80\%   &5 992&-&4 521&4 993&1,40 &-&1,54&-&0,90\\
90\%   &7 793&-& 5 398&6 648&1,73 &-&1,81&-&0,95\\

\hline
\multicolumn{10}{|c|}{\textsc{Connect}}\\
\hline
5\%& 821&2 982&951&1 022&0,80&2,91&0,86&3,11&0,93\\
10\%&977 &3 137&1 023&1 231&0,79&2,54&0,95&3,06&0,83\\
30\%&2 312&9 138&1 854&1 986&1,16&4,60&1,24&4,92&0,93\\
50\%& 9 278&27 887&7 883&8 231&1,12&3,38&1,17&3,53&0,95\\
70\% & 18 891&52 111&14 873&17 208&0,36&3,02&1,27&3,50&0,86\\
90\% &43 529 &98 334&41 457&50 009&0,87&1,96&1,04&2,31&0,82\\
\hline
\end{tabular}
\caption{Tableau comparatif des temps d'exécution de \textsc{GMRare}, \textsc{MFRare}, \textsc{Apriori-rare} et \textsc{Minit} pour des bases denses.}
\label{tab_tmps_mgr}
\end{sidewaystable}

\begin{table}[tb]
  \centering
  \small{
  \begin{tabular}{|l||r||r|r|r||r|r|}
    \hline
Base&\textit{minsupp}&\textsc{Apriori-rare} & \textsc{Minit} & \textsc{GMRare} & $\frac{\textsc{Apriori-rare}}{\textsc{GMRare}}$& $\frac{\textsc{Minit}}{\textsc{GMRare}}$\\
\hline
\hline\textsc{T10I4D100K}
&1\% &  4 327&5 127&4 194&1,03 &1,22\\
&2\% & 7 349&9 271&6 982&1,05&1,32\\
&3\% & 11 871&14 412&10 321& 1,15&0,13 \\
&4\% &15 722 &18 732&15 978& 0,98&1,17 \\
&5\% &27 887&36 812&30 643&1,91&1,20  \\
&10\%& 32 597& - &37 887 & 0,86&-  \\

\hline \textsc{T40I10D100K}
&0.4\%   &3 218&4 213&1 788&1,79&2,35\\
&0.5\%   &5 223&6 112&4 989&1,04 &1,22\\
&1\%   &13 767&15 566&9 231&1,49&1,68\\
&2\%   &16 253&20 331&14 557& 1,11&1,39\\
&3\%   &27 437&32 226&23 377& 1,15&1,37   \\
&4\%   &38 113&-&39 759&0,95 & -\\
&5\%   &57 223&-&57 889&0,98& -\\
\hline
\end{tabular}
  }
\caption{Tableau comparatif des temps d'exécution de \textsc{GMRare}, \textsc{Apriori-rare} et \textsc{Minit} pour des bases éparses.}
\label{tab_tmps_mgr_epar}
\end{table}

\begin{table}[tb]
  \centering
  \small{
  \begin{tabular}{|l||r||r|r|r||r|r|}
    \hline
Base&\textit{minsupp}&\textsc{Apriori-rare} & \textsc{Minit} & \textsc{MFRare} & $\frac{\textsc{Apriori-rare}}{\textsc{MFRare}}$& $\frac{\textsc{Minit}}{\textsc{MFRare}}$\\
\hline
\hline\textsc{T10I4D100K}
&1\% & 4 327&5 127&3 467&1,24 &1,47\\
&2\% & 7 349&9 271&5 719&1,28&1,62\\
&3\% & 11 871&14 412&9 741& 1,21& 1,47\\
&4\% & 15 722 &18 732&13 878& 1,13& 1,34\\
&5\% & 27 887&36 812&25 384&1,09&1,45  \\
&10\%& 32 597& - &31 861 & 1,02&-  \\

\hline \textsc{T40I10D100K}
&0.4\%   &3 218&4 213&1 129&2,85&373,00\\
&0.5\%   &5 223&6 112&3 931&1,32&1,55\\
&1\%   &13 767&15 566&9 001&1,52&1,72\\
&2\%   &16 253&20 331&12 332& 1,31&1,64\\
&3\%   &27 437&32 226&21 912& 1,25&1,47   \\
&4\%   &38 113&-&32 112&1,18 & -\\
&5\%   &57 223&-&49 972&1,14  & -\\
\hline
\end{tabular}
  }
\caption{Tableau comparatif des temps d'exécution de \textsc{MFRare}, \textsc{Apriori-rare} et \textsc{Minit} pour des bases éparses.} \label{tab_tmps_mfr_epar}
\end{table}

\begin{figure}[!htbp]
\begin{center}
\parbox{5.cm}{\includegraphics[scale=0.75]{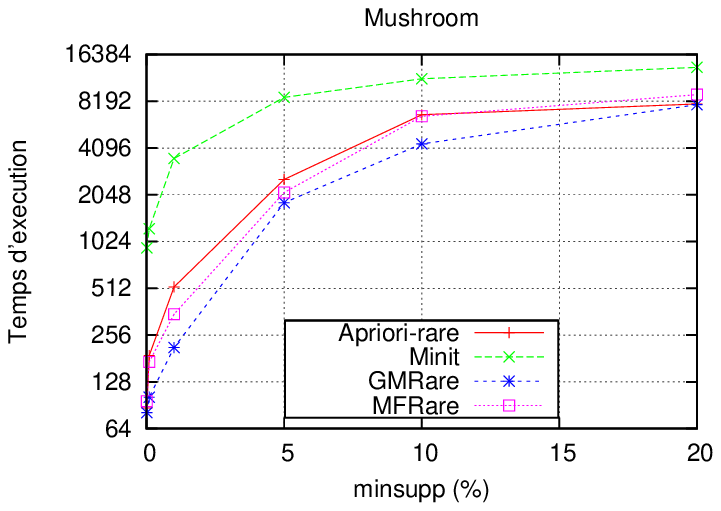}}
\parbox{5.cm}{\includegraphics[scale=0.75]{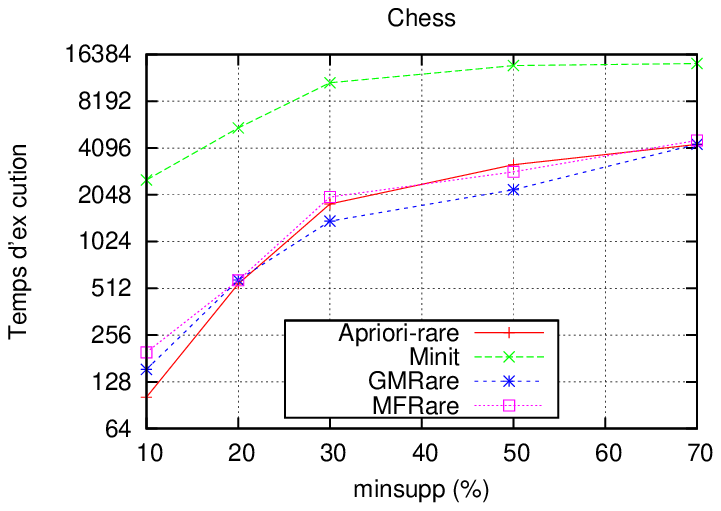}}
\parbox{5.cm}{\includegraphics[scale=0.75]{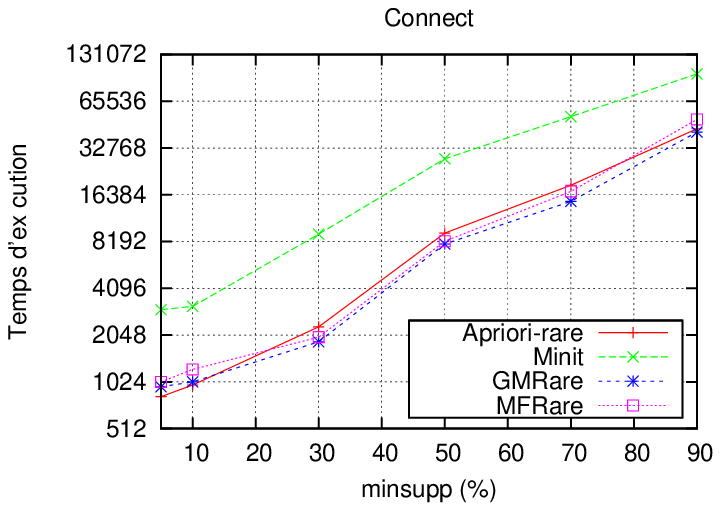}}\\
\parbox{5.5cm}{\includegraphics[scale=0.75]{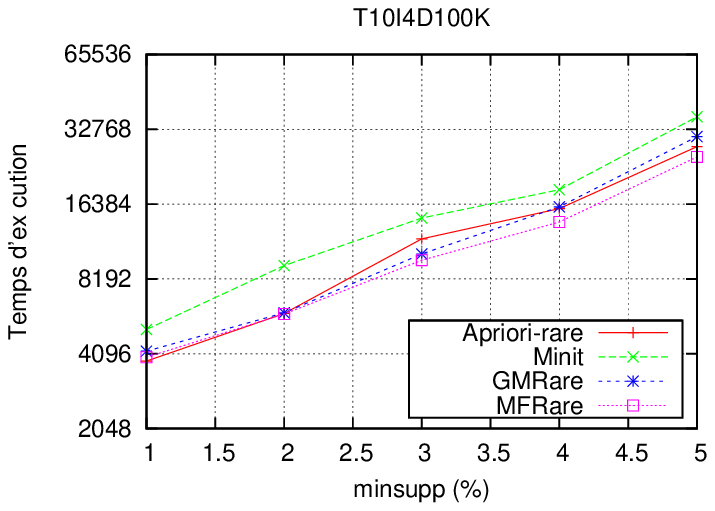}}
\parbox{5.cm}{\includegraphics[scale=0.75]{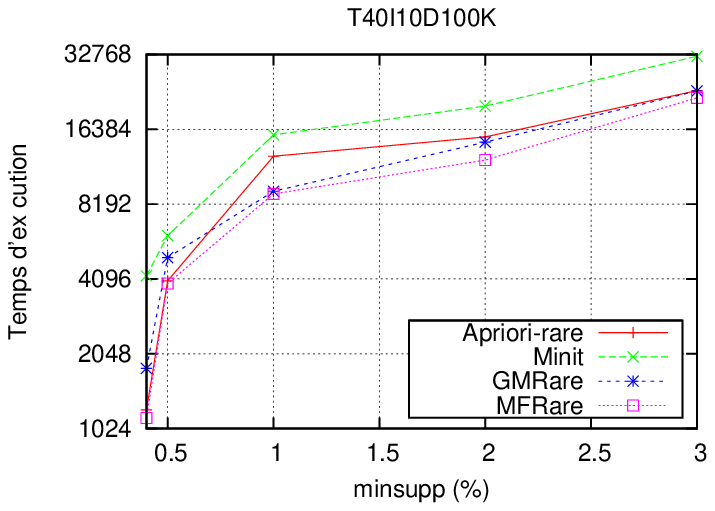}}
\end{center}
\caption{Comparaison des temps d'exécutions des algorithmes $\textsc{MGRare}$, $\textsc{MFRare}$, $\textsc{Apriori-rare}$ et $\textsc{Minit}$.}
\label{fig_card_mgr_mfr}
\end{figure}

\section{Conclusion}
Dans ce chapitre, nous avons mené une étude expérimentale ayant comme objectif la quantification de la cardinalité des représentations concises exactes ainsi que le temps d'extraction de ces représentations dans différentes bases et pour plusieurs  seuils \emph{minsupp}.\\
À la lumière de cette étude, nous avons constaté que nos représentations concises exactes, et surtout celle basée sur les générateurs minimaux, sont dans la plupart des cas de taille inférieure à celle des méthodes proposées dans la littérature. De plus, outre le fait que le temps d'extraction de nos deux représentations concises soient respectables par rapport à l'information extraite, nous constatons qu'il est inférieur au temps d'extraction consommé par les algorithmes d'extraction de l'ensemble total des motifs rares proposés dans la littérature et un peu plus supérieur au temps d'extraction consommé par quelques algorithmes d'extraction d'une partie de l'ensemble des motifs rares (un ensemble qui lui manque l'information support pour ces motifs).
\addcontentsline{toc}{chapter}{Conclusion générale}
\markboth{Conclusion générale}{Conclusion générale}

\chapter*{Conclusion générale}
Dans ce travail, nous avons présenté les notions de bases relatives à l'extraction des motifs rares et celles relatives à l'extraction des motifs fréquents. Par suite, nous avons fait une étude exhaustive des méthodes permettant l'extraction des motifs rares. Nous avons également mené une étude critique de ces méthodes afin de déceler les limites et les avantages de chacune d'elles. Nous avons constaté qu'aucune de ces méthodes ne permet d'extraire une représentation concise exacte des motifs rares. Tout de même, nous avons présenté les principales représentations concises exactes des motifs fréquents. Afin de pallier ces lacunes, nous avons introduit deux représentations concises exactes des motifs rares l'une basée sur les fermés rares et l'autre sur les générateurs minimaux rares. Afin d'extraire ces deux nouvelles représentations concises exactes, nous avons introduit deux nouveaux algorithmes, à savoir \textsc{MGRare} et \textsc{MFRare}. Tout de même, nous avons prouvé la correction et la complétude de deux algorithmes. De plus, nous avons évalué sa complexité au pire des cas. Afin de s'assurer que les cardinalités des deux représentations concises exactes sont bien réduites dans la plupart des cas, nous avons mené une étude expérimentale dans laquelle nous avons évalué leurs cardinalités sur plusieurs bases benchmarks et pour des seuils différents \textit{minsupp}. Enfin, Nous avons évalué l'aspect temporel associé aux algorithmes d'extractions de nos représentations concises.\\

À la lumière de cette étude, nous avons constaté que dans la plupart des bases benchmarks et pour des seuils \emph{minsupp} différents, la taille de nos représentations concises exactes à savoir celle basée sur les générateurs minimaux ou bien celles basée sur les fermés rares sont plus réduites que celles des méthodes de la littérature. Nous avons prouvé expérimentalement que nos objectifs définis d'avance ont été atteints, à savoir pallier les problèmes liés à l'extraction des motifs rares. Ces problèmes peuvent se résumer dans les points suivants :
\begin{itemize}
  \item Un nombre très important des motifs rares implique une taille très importante de l'ensemble résultat.
  \item La nécessité de deux algorithmes pour extraire l'ensemble total des motifs rares munis de leurs supports pour la plupart des méthodes existantes.
\end{itemize}
Ces problèmes ont été résolus, puisque nous avons offert un taux de compacité meilleur que la plus pertinente des méthodes de la littérature.

La fouille des représentations concises exactes des motifs rares constitue une piste de recherche intéressante. Au meilleur de notre connaissance, ces deux représentations sont les deux premières représentations concises exactes pour les motifs rares. Cet axe de recherche est plein de promesses puisque plusieurs pistes sont non encore explorées. Nous citons comme perspectives de recherche :
\begin{enumerate}
  \item Notre algorithme est moins efficace pour les contextes épars. \`A cet égard, il faut concevoir d'autres algorithmes permettant d'extraire plus efficacement notre représentation concise dans des bases éparses moyennant de nouvelles stratégies d'élagage ainsi que l'adoption de nouvelles techniques d'exploration de l'espace de recherche (en profondeur d'abord par exemple au lieu de la technique basée sur une exploration en largeur actuellement utilisée).
  \item D'après le chapitre 4, il est clair que les performances des représentations concises exactes, à savoir celle basée sur les motifs fermés rares ou bien celle basée sur les générateurs rares, varient selon les types des benchmarks. Afin de remédier ce problème, il faut penser à mettre un système hybride, qui selon le type de la base de transaction, extrait la représentation la plus adéquate.
 \item Explorer d'éventuelles relations entre les support conjonctif et disjonctif. Grâce à telles relations, nous pouvons bénéficier des algorithmes d'extraction des motifs fréquents pour les adapter dans le domaine des motifs rares.
 \item Prendre le problème de l'extraction des motifs rares noté $\mathcal{MR}$ d'un contexte $\mathcal{K}$ à l'envers et chercher la possibilité  d'extraire un ensemble des motifs fréquents, noté $\mathcal{MF}$, à partir d'un contexte $\mathcal{K'}$ telle que $\mathcal{MR}$ soit égal à $\mathcal{MF}$. Le problème est de trouver la relation de transformation de $\mathcal{K}$ pour avoir la nouvelle base de transactions $\mathcal{K'}$.
\end{enumerate}

\nocite{ref1} \nocite{ref2} \nocite{ref3} \nocite{ref4} \nocite{ref5} \nocite{ref6}
\nocite{ref7} \nocite{ref8} \nocite{ref9} \nocite{ref10} \nocite{ref11} \nocite{ref12} \nocite{ref13} \nocite{ref14} \nocite{ref15} \nocite{ref16} \nocite{ref17} \nocite{ref18} \nocite{ref19} \nocite{ref20}

\bibliography{biblio}
\bibliographystyle{plain}
\end{document}